\documentclass[10pt,a4paper,twocolumn,english,prb,aps,showpacs,floatfix,groupedaddress,superscriptaddress]{revtex4-1}
\usepackage{graphicx}
\usepackage{epsfig}
\usepackage[english]{babel}
\usepackage{amsmath}
\usepackage{amssymb}
\usepackage{amsfonts}
\usepackage{longtable}
\usepackage{dcolumn}
\usepackage{bm}
\usepackage{bbm}
\usepackage{nicefrac}
\usepackage{color,array}
\usepackage{colortbl}
\usepackage{dsfont}
\usepackage{mathtools}
\usepackage{subfigure}  
\usepackage[breaklinks,colorlinks=true,citecolor=blue,linkcolor=blue]{hyperref}
\usepackage{breakurl}

\def\bra#1{\mathinner{\langle{#1}|}}           
\def\ket#1{\mathinner{|{#1}\rangle}}           

\begin{document}

\newcommand{\be}{\begin{equation}}
\newcommand{\ee}{\end{equation}}
\newcommand{\bearr}{\begin{eqnarray}}
\newcommand{\eearr}{\end{eqnarray}}
\newcommand{\nn}{\nonumber}
\newcommand{\reqn}{\eqref}
\newcommand{\bs}{\begin{subequations}}
\newcommand{\es}{\end{subequations}}

\title{Dispersive Excitations in the One-Dimensional Ionic Hubbard Model}

\author{M. Hafez Torbati}
\email{mohsen.hafez@tu-dortmund.de}
\author{Nils A. Drescher}
\email{nils.drescher@tu-dortmund.de}
\author{G\"{o}tz S. Uhrig}
\email{goetz.uhrig@tu-dortmund.de}
\affiliation{Lehrstuhl f\"ur Theoretische Physik I, Technische Universit\"at Dortmund,
Otto-Hahn-Stra\ss e 4, 44221 Dortmund, Germany}

\date{\rm\today}

\begin{abstract}
A detailed study of the one-dimensional ionic Hubbard model with interaction $U$ is presented. 
We focus on the band insulating (BI) phase and the spontaneously  dimerized insulating (SDI) 
phase which appears on increasing $U$. By a recently introduced continuous unitary transformation 
[Krull et al.\ Phys.~Rev. B {\bf 86}, 125113 (2012)] we are able to describe the system
even close to the phase transition from BI to SDI although the 
bare perturbative series diverges before the transition is reached.
First, the dispersion of single fermionic quasiparticles is determined in the full
Brillouin zone. Second, we describe the binding phenomena between two fermionic quasiparticles
leading to an $S=0$ and to an $S=1$ exciton. The latter corresponds to the lowest spin excitation
and defines the spin gap which remains finite through the transition from BI to SDI.
The former becomes soft at the transition indicating that the SDI corresponds to a condensate
of these $S=0$ excitons. This view is confirmed by a BCS mean field theory for the SDI phase.
\end{abstract}



\maketitle

\section{Introduction}
\label{sec:introduction}

Electrons in solids can imply metallic, i.e., conducting behavior. But
there are also several mechanisms leading to insulating
behavior. The most common one is realized in band insulators (BI) which
are characterized by filled bands. A variant of this scenario consists
in the occurrence of spontaneous breaking of the symmetry in the ground state. The reduction
of the symmetry splits the bands into sub-band such that fractionally  filled bands 
become filled sub-bands so that insulating behavior results again. 
This happens for instance in antiferromagnets on bi-partite lattices 
with long-range order.

Another mechanism leading to insulating behavior is disorder. If the system
is strongly enough disordered the electronic states are localized so that
no extended, conducting states exist. This is called an Anderson insulator.

Strong interactions imply the third scenario of insulating behavior, the
Mott insulators (MI). It is generic for half-filled narrow-band systems. 
A strong local repulsion prevents the electrons to pass each other.

The band insulator and the Mott insulator will be the focus of the present article.
We will concentrate on one-dimensional systems where quantum fluctuations are most
strongly felt. For instance, long-range order due to the breaking of a continuous symmetry
generically does not occur. Our particular interest lies in the description
and the understanding of the elementary excitations. This includes their
dispersion and their interaction which partly induces bound states, i.e., excitons.
The softening of the energies of these bound states signals phase transitions.

Besides the conceptual, theoretical interest there are also many 
experimental systems for which our investigation is relevant.
Mixed stacked organic charge-transfer compounds are composed of alternating
donor and acceptor molecules. These materials are either nominally ionic or 
nominally neutral. They are insulating due to the double periodicity of the lattice.
If the compounds are situated close to the boundary between neutral and ionic behavior, such as 
TTF-chloranil, a reversible phase transition from the neutral phase to 
the ionic phase can be induced by changing pressure \cite{Torrance81a} or temperature \cite{Torrance81b}. 
The transition from the neutral ground state to the ionic ground state
appears  not to be of first order. Yet there is an intermediate region  where both neutral and 
ionic molecules coexist \cite{Torrance81b}.

The observation of the neutral-ionic phase transition has been the subject of
various experimental \cite{Tokura82,Ayache83,Mitani84,Tokura84,Kobayashi12} and theoretical 
investigations \cite{Hubbard81,Rice82,Horovitz83,Nagaosa86a,Nagaosa86b,Nagaosa86c,Giovannetti09}. 
In theory, the chain 
of alternating doner and acceptor molecules of the mixed stacked organic compounds 
is described by the ionic Hubbard model (IHM) \cite{Nagaosa86a}. 
This model consists of three terms: A nearest-neighbor (NN) hopping,
an on-site Hubbard interaction, and an ionic potential which describes the 
on-site energy difference between the donor and acceptor molecules. The effect
of additional terms such as electron-electron interaction 
on NN sites~\cite{Nagaosa86b} or electron-lattice interaction \cite{Nagaosa86c}
was also considered in order to make the Hamiltonian describe the experimental
situation more closely.

The one-dimensional (1D) IHM in the electron-hole symmetric form reads
\begin{eqnarray}
H=\frac{\delta}{2}\sum_{i\sigma} (-1)^i n_{i,\sigma} 
&+&U\sum_{i} \left( n_{i,\uparrow}-\frac{1}{2} \right) \left(n_{i,\downarrow}-\frac{1}{2} \right) \nn \\
&+&t\sum_{i\sigma} (c^\dagger_{i,\sigma}c^{\phantom{\dagger}}_{i+1,\sigma} + {\rm h.c.}),
\label{eq:IHM}
\end{eqnarray}
where $c_{i,\sigma}$ and $c^\dagger_{i,\sigma}$ create and annihilate
an electron at site $i$ with spin $\sigma$, respectively. The density operator  
$n_{i,\sigma} := c^\dagger_{i,\sigma}c^{\phantom{\dagger}}_{i,\sigma}$ counts the number of electrons
with spin $\sigma$ at site $i$. It is convenient to choose 
$\delta$ as unit of energy. The IHM for $U=0$
at half-filling describes a BI with equal spin gap and charge gap. 
The density of particles on odd sites is larger than the density on the even sites so that
the phase is nominally ionic. In reciprocal space, the lower band is completely filled, the upper one
is empty.

In the opposite limit $U-\delta \gg t$, the IHM at half-filling 
can be mapped to the Heisenberg model whose ground state in known be a 
Mott insulator (MI) with zero spin gap. Although the IHM~\reqn{eq:IHM} has two sites per 
unit cell, it has been shown that the effective Heisenberg model has
the full translational symmetry in all orders of the hopping term $t$ \cite{Nagaosa86a}. 
In this MI phase, the densities of particles
of even and odd sites are close to each other, $n_{i,\sigma} \approx 1$ so 
that the phase is nominally neutral.

The IHM attracted further interest as a model to describe ferroelectric 
perovskites \cite{Egami93}. Since then, various analytical and 
numerical methods have been employed to find the phase diagram and excitation spectrum
of the IHM. In one dimension, Fabrizio {\it et al.} 
showed by using bosonization techniques that a spontaneously dimerized insulator (SDI) 
represents a stable intermediate phase between BI  and MI. The transition from BI to SDI
at a critical value $U_{c1}(t)$ was 
recognized as Ising type and the transition from SDI to MI at a critical value $U_{c2}(t)$ 
found to be of Kosterlitz-Thouless type\cite{Fabrizio99}. 

Using exact diagonalization techniques on finite size clusters, 
it was found that the BI and MI are separated
by a transition point where both spin and charge gaps 
vanish \cite{Gidopoulos00,Anusooya-Pati01}. In Ref.\ \onlinecite{Gidopoulos00}, however, 
it could not be decided whether the spin and charge gaps
close exactly at the same value or at slightly different values due to limitations in the
finite size scaling. Contrary to this finding,
an exact diagonalization study of the Berry phase by Torio {\it et~al.} indicates 
that the BI and MI are separated by an intermediate SDI region \cite{Torio01}.
The results of approximations such as self-consistent mean-field theory \cite{Resta95,Ortiz96},
renormalization group complemented by a mean-field analysis \cite{Gupta01}, and
the slave-boson approach \cite{Caprara00} are in favor of a single transition point 
between the BI and the MI without intermediate phase.

A variational quantum Monte Carlo study only found a single transition from the BI to the SDI phase
without a second transition to the MI phase.
It was argued that the  MI phase only stabilizes
for $\delta = 0$ \cite{Wilkens01}. Furthermore, the density-matrix renomalization group (DMRG)
method was used by various groups to investigate the phase diagram of the 
IHM \cite{Takada01,Lou03,Manmana04,Zhang03,Legeza06,Tincani09,Kampf03}. 
By extrapolating the DMRG results of finite size lattices to infinite size lattices 
most of them support the scenario of two transition 
points $U_{c1}$ and $U_{c2}$ 
\cite{Takada01,Lou03,Manmana04,Zhang03,Legeza06,Tincani09}.
But the reported behavior for charge gap and spin gap near the transition points differs\cite{Takada01,Lou03,Manmana04,Kampf03}. 
It was also deduced by Kampf {\it et~al.} that, within the 
accuracy of DMRG and the accessible chain lengths, it is not possible
to establish the second transition from SDI to MI beyond doubt \cite{Kampf03}.

In two dimensions, the phase diagram of the IHM has been discussed controversely. 
Although the existence of the BI at small values and of the MI at large values of the
Hubbard interaction is established \cite{Garg06,Paris07,Bouadim07,Craco08,Kancharla07,Chen10}, 
the nature of the intermediate phase is not clear. 
A single-site dynamic mean field theory (DMFT) indicates a metallic phase between BI 
and MI \cite{Garg06} which is confirmed by determinant quantum Monte Carlo 
method \cite{Paris07,Bouadim07}. In another single-site DMFT study, 
a parameter range with coexistence of MI, metallic behavior, and BI is found 
in addition to the pure metallig phase \cite{Craco08}. 
Cluster-DMFT, however,
indicates that the intermediate phase is a SDI similar to the case in one dimension \cite{Kancharla07}. 
In the variational cluster approach, the intermediate phase is a bond-located spin density wave with
magnetic order which produces the lowest energy between BI and MI \cite{Chen10}.

The excitation spectrum of the IHM has attracted much less attention so far. 
The low-energy spectrum and the dynamic spin and charge structure
factors in the BI phase of the model are investigated in 1D using perturbative continuous
unitary transformations \cite{Hafez10b,Hafez11}. 
The expansion parameters of these studies are the hopping $t$ and the interaction $U$.
In the reduced Brillouin Zone (BZ), one singlet bound state
and two triplet bound state modes are found in the two-fermion sector \cite{Hafez10b}. 
But due to the perturbative nature of the approach, the authors were not
able to approach the transition point and the results for the two-particle excitations 
are obtained deep in the BI phase \cite{Hafez10b}.

In the present paper, the phase transitions of the 1D IHM and its
excitation spectrum in the BI phase are investigated in the vicinity of the transition point $U_{c1}$. 
We use the recently formulated method of {\it directly
evaluated enhanced perturbative continuous unitary transformations} (deepCUT) \cite{Krull12}
in two subsequent steps to derive simpler effective Hamiltonians which allow a quantitative
analysis of the dynamics of the excitations.

In the first step, we employ the deepCUT method to obtain an 
effective Hamiltonian describing the low-energy physics of the system for $\delta \approx U \gg t$. 
This corresponds to eliminating doubly occupied states on even sites and empty states on odd sites. 
This reduces the relevant energy scale from $U$ to $t$.

In the next step, the resulting low-energy Hamiltonian is mapped to various effective
Hamiltonians using various generators in the deepCUT method. In this step, the processes
creating particle-hole pairs from the vacuum or in addition to existing fermionic
excitations are eliminated. The one-particle dispersion and 
the dispersion of two-particle bound states are obtained by the deepCUT in the BI phase. 
In addition, we aim at improving
the accuracy of the results by analyzing the effective Hamiltonians obtained from the deepCUT by
using exact diagonalization (ED) techniques valid in  the thermodynamic limit. 
For charge, spin, and exciton gaps, we compare our results 
with the extrapolated DMRG results of Ref.~\onlinecite{Manmana04}.

Finally, we use a BCS-type mean-field theory to describe the phase
beyond the transition point $U_{c1}$. We can show that the SDI phase is indeed stable
for $U> U_{c1}$.

The paper is organized as follows: In the next section \ref{sec:method}, we introduce the
various employed methods. In Sect.~\ref{sec:direct_evaluation}, we present the results of 
the application of deepCUT alone. In Sect.~\ref{sec:exact_diagonalization}, the ED method is described and the results
obtained by combining deepCUT and ED are discussed. The last-but-one section~\ref{sec:mean_field} is devoted
to the analysis of the effective Hamiltonian in the mean-field level.
Finally, the paper is concluded.

\section{Method}
\label{sec:method}

In this section, the employed 
deepCUT \cite{Krull12} and the ED methods are presented. 
The deepCUT is  based on the continuous unitary transformations 
(CUT) \cite{Wegner94,Glazek93}. 
First, the general concepts of CUT are briefly presented. Finally, the deepCUT and the ED
are illustrated.

\subsection{The CUT method} 

The CUT or flow equation approach was proposed by Wegner \cite{Wegner94} and independently
by G\l{}azek and Wilson \cite{Glazek93} in 1994.
In this approach, a given Hamiltonian $H$ is mapped
by a unitary transformation to a diagonal or block-diagonal effective Hamiltonian 
in a systematic fashion \cite{Kehrein06}. The unitary transformation $U(\ell)$ depends on an auxiliary continuous
parameter $\ell$ which defines the flow under which the Hamiltonian transforms from its 
initial form $H=H(\ell)\big{|}_{\ell=0}$ to its final effective form $H_{\rm eff}=H(\ell)\big{|}_{\ell=\infty}$.
A related approach is the projective renormalization (PRG), 
which maps a given Hamiltonian to an effective Hamiltonian by iteration of discrete steps~\cite{Becker2002,Phan2011}.

In CUT, the transformed Hamiltonian $H(\ell)=U^\dagger(\ell)HU(\ell)$ is determined from
an ordinary differential equation, called flow equation, 
\begin{equation}
\partial_\ell H(\ell) = [ \eta(\ell),H(\ell) ] \quad ,
\label{eq:flow_equation}
\end{equation}
where the anti-hermitian operator $\eta(\ell)=-U^{\dagger}(\ell) \partial_\ell U(\ell)$ is 
the infinitesimal generator of the flow. It is seen from Eq.~\reqn{eq:flow_equation} that we can directly deal
with the generator $\eta(\ell)$ instead of the unitary transformation $U(\ell)$. 

Wegner suggested to define the generator as $\eta_W(\ell)=[H_d(\ell),H(\ell)]$ where
$H_d(\ell)$ is the diagonal part of the Hamiltonian $H(\ell)$. It can be shown that
for $\ell \rightarrow \infty$, Wegner's choice of generator brings the Hamiltonian into a
diagonal form except for degenerate states \cite{Wegner94,dusue04a}.

A disadvantage of Wegner's generator is that it spoils certain simplifying features
of the initial Hamiltonian $H(\ell=0)$. 
If  the initial Hamiltonian has a band-diagonal
structure, this property will be lost during the flow. Mielke introduced a modified 
generator which preserved the band-diagonality for matrices \cite{Mielke98}. 
In the context of second quantization, Stein \cite{stein98} effectively used the
analogous generator. Knetter and Uhrig \cite{Uhrig98,Knetter00} 
realized the importance of the sign function in the proper generalization for second quantization.
This generator is efficient in  deriving an effective block-diagonal Hamiltonian that preserves the number 
of excitations, also called quasiparticles (QPs), in the system. Thus we call this generator the
 particle-conserving generator (pc) which reads
\be
\eta_{\rm pc}(\ell)=\sum_{i,j=0} {\rm sign}(i-j) H_{i:j}(\ell)
\label{eq:mku_generator}
\ee
where $H_{i:j}$ is the part of the Hamiltonian which creates $i$ and annihilates 
$j$ quasiparticles. It is defined that ${\rm sign}(0)=0$.

The pc generator makes the Hamiltonian block-diagonal in the sense that 
the final effective Hamiltonian conserves the number of QPs. It is desirable to 
reach this goal. But for many properties, it is unnecessarily ambitious. For the 
energetically low-lying excitation spectrum there is no
need to block-diagonalize the sectors with large numbers of QPs. 
It is sufficient to decouple only the sectors with low numbers 
of QPs from the remaining Hilbert space.
The corresponding reduced generator, which allows us to decouple the first $n \ge 0$
quasiparticle sectors, reads \cite{Fischer10} 
\be
\eta_{\rm p:n}(\ell)=\sum_{i=0}^{n}\sum_{j>n} \left( H_{j:i}(\ell) - H_{i:j}(\ell) \right) .
\label{eq:reduced_generator}
\ee
In comparison to Eq.~\reqn{eq:mku_generator},
one sees that only terms that act on the first $n$ quasiparticle sectors 
and link them to other sectors contribute to the reduced generator. 
Note that contributions from the sectors with up to $n$ QPs to sectors
with arbitrarily large numbers of QPs may occur in the generator.
It is especially useful 
to describe the decay of QPs due to the energetic degeneracy of
eigen states with different number of QPs, the so-called overlap of continua,
 in the framework of CUTs \cite{Fischer10}. 
The reduced generator allows us not only to avoid divergences  
that may occur in the flow due to overlapping continua, but also to increase the speed of calculations 
significantly because only less terms need to be considered \cite{Fischer10,Hamerla10}.

\subsection{The deepCUT method}

In the sequel, we present the deepCUT in real space. But we emphasize that locality is not
needed but only an appropriate small expansion parameter and a sufficiently simple
unperturbed Hamiltonian $H_0$ \cite{Krull12}. In addition, a truncation scheme is
needed to obtain a closed set of equations. The guiding idea is
to keep all operators and their prefactors in the flow equation which
contribute to the quantities of interest, for instance the dispersion,
up to a given order in the expansion parameter.

To put the deepCUT in real space to use, we assume that the initial Hamiltonian can
be decomposed into a local part ($H_0$) and a nonlocal part ($V$)
\be
H=H_0+xV,
\label{eq:general_hamiltonian}
\ee
where $x$ is an expansion parameter on which we base the 
truncation of the flow equations \cite{Krull12}. 
Targeting the first sectors with a few QPs allows us to use
simplification rules which highly accelerate the calculations by
eliminating unnecessary contributions early on. For 
the details about the truncation scheme and the simplification rules, we 
refer to Ref.\ \onlinecite{Krull12}.

In order to use second quantization in terms of the QPs, the Hamiltonian
\reqn{eq:general_hamiltonian}
is written in terms of creation and annihilation operators \cite{Knetter03b}. 
In this representation, $H_0$ simply counts the number of excitations present in the system.
For $x=0$, these excitations are the 
{\it true} QPs of the system. Their vacuum is the ground state. For any finite
value of $x$, however, these QPs become dressed and the initial Hamiltonian \reqn{eq:general_hamiltonian} 
does not necessarily conserve the number of these excitations. 

The Hamiltonian in QP representation can be denoted as a sum of
monomials of operators $\{A_i\}$ which describe specific interactions
in real space. These monomials create and annihilate a certain number
of QPs. Hence the transformed Hamiltonian $H(\ell)$ can be expressed generally as
\be
H(\ell) = \sum_i h_i(\ell) A_i,
\label{eq:ham_monomial_rep}
\ee
where the coefficients $h_i(\ell)$ carry the $\ell$-dependence of the Hamiltonian.
Similarly, the generators \reqn{eq:mku_generator} and \reqn{eq:reduced_generator} 
are written as
\be
\eta(\ell) = \sum_i \eta_i(\ell) A_i := \sum_i h_i(\ell) \hat{\eta}[ A_i],
\label{eq:gen_monomial_rep}
\ee
where the super-operator $\hat{\eta}$ defines how a given monomial enters the 
generator. Using the above representations for Hamiltonian and generator the flow 
equations \reqn{eq:flow_equation} read
\be
\partial_\ell h_i(\ell) = \sum_{jk} D_{ijk} h_j(\ell) h_k(\ell)
\label{eq:deepCUT_FE}
\ee
where the contributions $D_{ijk}$ results from the re-expansion of the 
commutator in terms of the monomials
\be
[\hat{\eta}[ A_j], A_k] = \sum_i D_{ijk} A_i.
\label{eq:contributions}
\ee

\noindent
Summarizing, the solution of the flow equation \reqn{eq:flow_equation} requires two 
major steps:
\begin{itemize}
 \item Finding the contributions $D_{ijk}$ from Eq.~\reqn{eq:contributions},
 \item Solving the set of ordinary differential Eqs.~\reqn{eq:deepCUT_FE}.
\end{itemize}
The first step is algebraic work and we assume that this can be done up 
to a certain number of terms.

The second step is the integration of the flow equations which can be done
in two ways. The first one is perturbative and relies on an expansion
of the coefficients in powers of $x$. These coefficients can be determined
from the integration of the flow equation yielding a perturbative evaluation
of the effective Hamiltonian. In contrast to the original perturbative CUT (pCUT) method \cite{Knetter00,Knetter03a,Knetter03b}, this approach also works for cases where the unperturbative part 
has a non-equidistant spectrum. This approach, which generalizes the pCUT, 
is called {\it enhanced perturbative} CUT (epCUT) \cite{Krull12}.

The second approach consists in the direct numerical integration of
Eq.~\reqn{eq:deepCUT_FE} once only those contributions are kept which
would be required to yield the correct epCUT result in a fixed order in $x$. 
This procedure is called deepCUT method \cite{Krull12} and has been introduced
for spin ladders and successfully applied
to the transverse-field Ising model in 1D yielding to systematically controlled multi-particle 
excitation spectra and dynamical correlation functions \cite{Fauseweh13}.

In the deepCUT as in other non-perturbative CUTs, see for instance Ref.\ \onlinecite{Fischer10},
one generically has to check if the flow equation converges reliably. In the perturbative CUTs the
hierarchical structure of the differential equations guarantees convergence.
In order to track the convergence quantitatively we use 
the {\it residual off-diagonality} (ROD) defined by
\be
{\rm ROD}(\ell) = \sqrt{\sum_i |\eta_i(\ell)|^2},
\label{eq:ROD}
\ee 
where the sum runs over all the monomials appearing in the generator. The coefficient
$\eta_i(\ell)$ is the prefactor of the monomial $A_i$ as defined
in Eq.~\reqn{eq:gen_monomial_rep}. In the deepCUT analysis, the ROD can diverge due to the
energetic overlap of continua with different number of QPs. 
In this case, a less ambitious decoupling of sectors with lower numbers of QPs may restore
convergence of the flow as we will show in the following. 

\subsection{Exact Diagonalization}

The ED can be applied in two ways. In the first
way, the size of the lattice is limited to a finite number of sites and 
the corresponding Hamiltonian matrix is constructed and diagonalized. The major problem
in this approach is the effect of the finite size of the system. This is the
most commonly used ED scheme.

A second approach by ED is possible {\it if} the ground state is decoupled from the other
parts of the Hilbert space. Such a decoupling can be obtained, for instance, by 
the deepCUT method, see above or Ref.\ \onlinecite{Krull12}. 
In this case, the ground state is given by the vacuum of QPs
and  states with a few QPs describe the low-energy spectrum of the system. Because the 
ground state is already decoupled, it is possible to work directly in the thermodynamic limit.
The Hilbert space is restricted by limiting the maximum number of QPs considered
and the maximum relative distances between them.
This approach is employed in Ref.~\onlinecite{Fischer10} to describe QP decay in the asymmetric 
two-leg Heisenberg ladder. If we use the term ED in the remainder of this article
we refer to this second approach valid in the thermodynamic limit.

\section{Direct Evaluation Analysis of the Band Insulator Phase}
\label{sec:direct_evaluation}

In this section, the low-lying excitation spectrum of the IHM including 1-QP 
dispersion, 2-QP continuum, and possible singlet and triplet bound
states are discussed using the deepCUT.
The results for charge gap, exciton gap, and spin gap are compared to the 
available results obtained by DMRG~\cite{Manmana04}.

\subsection{Preliminary Considerations}

To apply the deepCUT method in the BI phase of the IHM, we put
the local staggered potential and the Hubbard interaction in the IHM~\reqn{eq:IHM} 
into $H_0$ and consider the hopping term as perturbation $V$. 
The unperturbed Hamiltonian $H_0$ has a unique ground state only for $U<\delta$. 
The energy gap of inserting a single fermion takes the value $\delta-U$
so that the dimensionless expansion parameter is $t/(\delta - U)$
In the limit $U \to \delta$, any purely perturbative analysis breaks down. 
Below, however, we will show that in the deepCUT approach the on-site energy 
is renormalized to larger values so that the BI phase is stabilized beyond 
$U=\delta$ and one can obtain $H_\text{eff}$ for $U>\delta$ as well.

\begin{figure}[tb]
  \centering
  \includegraphics[width=0.95\columnwidth]{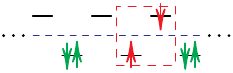}
  \caption{Schematic representation of excitations of the ionic Hubbard model 
  in the band insulator phase. In the absence of hopping terms, the ground state 
  for $U<\delta$ is characterized by occupied odd sites and empty even sites. 
  A pair of excitations appears when an electron hops from an odd to an even site.}
  \label{fig:IHM}
\end{figure}

In the ground state of $H_0$ all odd sites are occupied and 
all even sites are empty. An electronic hop
from an odd site to an even sites excites the system, see Fig.~\ref{fig:IHM}. 
In order to make the fermionic vacuum the ground state of $H_0$, 
we apply an electron-hole transformation to the odd sites. 
To be specific, we define
\be
c_{i,\sigma} = h^\dagger_{i,\sigma}.
\label{eq:ph_trans}
\ee
Due to this  transformation the spin operators on odd sites change
\bs
\bearr
S^z_i &=& \sum_\sigma \sigma c^\dagger_{i,\sigma} c^{\phantom{\dagger}}_{i,\sigma}
=-\sum_\sigma \sigma h^\dagger_{i,\sigma} h^{\phantom{\dagger}}_{i,\sigma}
:= -\tilde{S}_i^z, \\
S_i^+ &=& c^\dagger_{i,\uparrow} c^{\phantom{\dagger}}_{i,\downarrow} 
= -h^\dagger_{i,\downarrow} h^{\phantom{\dagger}}_{i,\uparrow}
:= -\tilde{S}_i^-, \\
S_i^- &=& c^\dagger_{i,\downarrow} c^{\phantom{\dagger}}_{i,\uparrow} 
= -h^\dagger_{i,\uparrow} h^{\phantom{\dagger}}_{i,\downarrow}
:= -\tilde{S}_i^+,
\eearr
\es
and the spin operators on even sites remain unchanged. 
Hence the spin states 
that include a mixture of electrons and holes are different
from the usual definitions. For instance, the singlet state of an 
electron-hole pair on adjacent sites reads
\be
 | {\rm e-h} \rangle^{S=0} = 
 \frac{1}{\sqrt{2}} \left(|\uparrow\uparrow \rangle + |\downarrow\downarrow \rangle \right),
\ee
On an even site the arrow refers to the spin of an electron and on an odd site 
it refers to the spin of a hole.
This point must be kept in mind in the considerations below.

In order to unify all electron and hole operators, we define the fermionic 
operator  
\be
f_{i,\sigma} = 
\begin{cases}
c^{\phantom{\dagger}}_{i,\sigma} & {\rm for} ~ i \in {\rm even} \\
h^{\phantom{\dagger}}_{i,\sigma} & {\rm for} ~ i \in {\rm odd}
\end{cases}.
\label{eq:f_operator}
\ee
According to these definitions  the Hamiltonian \reqn{eq:IHM} reads
\bearr
H = \frac{U\!-\!2\delta}{4} \sum_i \mathds{1} 
&+& \frac{\delta - U}{2} \sum_{i,\sigma} f^\dagger_{i,\sigma} f_{i,\sigma}^{\phantom{\dagger}} \nn \\
&+& U \sum_i f^\dagger_{i,\uparrow}f^\dagger_{i,\downarrow} 
f_{i,\downarrow}^{\phantom{\dagger}}f_{i,\uparrow}^{\phantom{\dagger}} \nn \\
&+& t \sum_{i,\sigma} (-1)^i ( f^\dagger_{i,\sigma} f^\dagger_{i+1,\sigma} + {\rm h.c.} ) ,
\label{eq:IHM_int}
\eearr
it still has two sites per unit cell as the original Hamiltonian \reqn{eq:IHM}.
 
It is possible to restore full translational symmetry by applying a suitable
local transformation on the fermionic operators
\be
f^\dagger_{j,\sigma} \longrightarrow 
e^{-i\frac{\pi}{4}} e^{i \frac{\pi}{2} j } f^\dagger_{j,\sigma}. 
\label{eq:local_trans}
\ee
This transformation
leaves the first three terms in Eq.~\reqn{eq:IHM_int} unchanged and eliminates
the prefactor $(-1)^i$ from the last term. Thereby, we reach
\bearr
H = \frac{U\!-\!2\delta}{4} \sum_i \mathds{1} 
&+& \frac{\delta - U}{2} \sum_{i,\sigma} f^\dagger_{i,\sigma} f_{i,\sigma}^{\phantom{\dagger}} \nn \\
&+& U \sum_i f^\dagger_{i,\uparrow}f^\dagger_{i,\downarrow} 
f_{i,\downarrow}^{\phantom{\dagger}} f_{i,\uparrow}^{\phantom{\dagger}}  \nn \\
&+& t \sum_{i,\sigma} ( f^\dagger_{i,\sigma} f^\dagger_{i+1,\sigma} + {\rm h.c.} ) .
\label{eq:IHM_fin}
\eearr
The last term of this Hamiltonian is a Bogoliubov term which creates and annihilates a 
pair of QPs (originally an  electron and a hole) with total spin {\it zero} on 
neighboring sites. In the following,
the deepCUT method will be applied to this Hamiltonian. 

The conservation of the original  {\it electron} number in the representation \reqn{eq:IHM_fin} 
is not manifest. Thus we write down 
the operator of the total electron number $\widehat{N}$ in terms of $f$-operators 
\bearr
\widehat{N} := \sum_{i,\sigma} c^{\dagger}_{i,\sigma} c^{\phantom{\dagger}}_{i,\sigma}
= L + \sum_{i,\sigma} (-1)^i f^{\dagger}_{i,\sigma} f^{\phantom{\dagger}}_{i,\sigma},
\label{eq:pn_operator}
\eearr
where $L$ is the number of sites in the chain. The difference between the number of QPs 
on even sites and on odd sites is a constant of motion. Thus they are always created
or annihilated in pairs with an odd distance between them.

\subsection{Low-Energy Effective Hamiltonian}

The interesting physics of the IHM happens at large values of the Hubbard interaction $U$
approaching the first transition at $U_{c1}$. We focus on the case $U, \delta \gg t$ where
the states with finite number of double occupancies (DOs) lie very high in energy. Thus the low-energy 
physics of the Hamiltonian \reqn{eq:IHM_fin} is governed by the  Hilbert
subspace without DO.  But the subspaces with and without DOs are linked by the Bogoliubov term.

In a first step, we decouple the low- and the high-energy 
parts of the Hilbert space. The same idea was first realized 
by Stein perturbatively for the Hubbard model on the square lattice \cite{Stein97}. 
Extended calculations using self-similar CUTs (sCUT)
were carried out  at and away from half-filling \cite{Reischl04, Hamerla10} 
to investigate the range of validity of the 
mapping from the Hubbard model to the $t$-$J$ model.
High-order perturbative calculations for the Hubbard model on the triangular lattice 
at half-filling have been performed by Yang and co-workers \cite{Yang2010}.

\begin{figure}[tb]
  \centering
  \includegraphics[width=0.9\columnwidth]{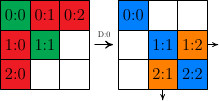}
  \caption{(Color online) Schematic representation of the application of the generator $D\!:\!0$ 
  to the initial Hamiltonian \reqn{eq:IHM_class}. Each part $H_{i:j}$ of the Hamiltonian
  is depicted by a block; the notation $i\!:\!j$ stands for the number of DOs
  which is first annihilated ($j$) and then created ($i$). 
   The blank blocks 
  indicate the absence of the correspond interaction in the Hamiltonian. In the final effective Hamiltonian 
  the sector with zero number of DOs is decoupled and the coefficients in other blocks are renormalized as
  indicated by the change of color/shading.}
  \label{fig:D:0_generator}
\end{figure}

In the fermionic representation \reqn{eq:IHM_fin} of the IHM it is not evident
how many DOs are created or annihilated by a term because this depends on the
state to which the terms are applied. 
Thus, we introduce a representation (Hubbard operators \cite{Hubbard1964}) of hard-core particles defined by
\bs
\bearr
g^\dagger_{i,\sigma}& := & \ket{\sigma}_i \prescript{}{i}{\bra{0}} 
= (1-n_{i,\bar{\sigma}} ) f^\dagger_{i,\sigma} \\
g^\dagger_{i,d}& := & \ket{\uparrow \downarrow}_i \prescript{}{i}{\bra{0}} 
= f^\dagger_{i,\uparrow} f^\dagger_{i,\downarrow}
\eearr
\es
where $\bar{\sigma}=-\sigma$. The fermionic hard-core  operator 
$g^\dagger_{i,\sigma}$ creates a fermion with spin $\sigma$ at site $i$ 
from the vacuum and the bosonic operator $g^\dagger_{i,d}$ creates a DO at
site $i$ from the vacuum. They obey the hard-core (anti-)commutation relation
\be
\left[ g^{\phantom{\dagger}}_{i,\alpha},g^\dagger_{j,\beta} \right]_{\pm} = \delta_{i,j} 
\left( \delta_{\alpha,\beta} \pm g^\dagger_{i,\beta} g^{\phantom{\dagger}}_{i,\alpha} 
\!-\delta_{\alpha,\beta} \!\!\!\sum_{\gamma=\uparrow,\downarrow,d} \!\!\!
g^\dagger_{i,\gamma}g^{\phantom{\dagger}}_{j,\gamma} \right)
\ee
where the anticommutation $[~,~]_+$ is to be used if both operators are fermionic, otherwise
the commutation $[~,~]_-$ is to be used. 
The above representation can be reversed to express the $f$-operators
in terms of the $g$-operators
\bearr
f^\dagger_{i,\sigma} = g^\dagger_{i,\sigma} 
+ {\rm sign}(\sigma) g^\dagger_{i,d} ~ g^{\phantom{\dagger}}_{i,\bar{\sigma} } .
\eearr

The IHM~\reqn{eq:IHM_fin} in terms of the $g$-operators can be split into
different parts which create and annihilate a specific number of DOs. Explicitly
one has
\be
H=H_{0:0}+H_{1:1}+H_{1:0}+H_{0:1}+H_{2:0}+H_{0:2},
\label{eq:IHM_class}
\ee
where $H_{i:j}$ creates $i$ and annihilates $j$ DOs. These parts are given by
\bs
\label{eq:H_nm}
\bearr
\label{eq:H_00}
H_{0:0} &=& \frac{U\!-\!2\delta}{4} \sum_i \mathds{1}
+ \frac{\delta\!-\!U}{2} \sum_{i,\sigma} g^\dagger_{i,\sigma} g_{i,\sigma}^{\phantom{\dagger}}  \nn \\
&\quad&+t \sum_{i,\sigma} ( g^\dagger_{i,\sigma} g^\dagger_{i+1,\sigma} +{\rm h.c.}) , \\
\label{eq:H_11}
H_{1:1} &=& \delta \sum_i g^\dagger_{i,d} g_{i,d}^{\phantom{\dagger}} , \\
\label{eq:H_10}
H_{1:0} &=&  t \sum_{i,\sigma} {\rm sign}(\sigma) 
( g^\dagger_{i,d} g_{i,\bar{\sigma}}^{\phantom{\dagger}} g^\dagger_{i+1,\sigma} 
+ g^\dagger_{i,\sigma} g^\dagger_{i+1,d} g_{i+1,\bar{\sigma}}^{\phantom{\dagger}} ) \nn\\  
&\quad&= (H_{0:1})^\dagger, \\
\label{eq:H_20}
H_{2:0} &=& t \sum_{i,\sigma}
g^\dagger_{i,d} g_{i,{\sigma}}^{\phantom{\dagger}} g^\dagger_{i+1,d} g_{i+1,{\sigma}}^{\phantom{\dagger}} 
=(H_{0:2})^\dagger .
\eearr
\es
These expressions indicate that for $U \approx \delta \gg t$, the low-energy
physics takes place in the subspace without DOs.
The reduced generator $\eta_{D:0}$ is applied to \reqn{eq:IHM_class} 
to disentangle the subspace without any DOs from the remaining Hilbert space.
The process is schematically shown in Fig.~\ref{fig:D:0_generator}. 
The final low-energy effective Hamiltonian acts on a  three-dimensional local Hilbert space (no
fermion present or an $\uparrow$ or $\downarrow$ fermion is present).
The fermionic hard-core QP can hop and they interact with one another. 
In table \ref{tab:effective_interactions}, the relevant monomials  $A_j$ up to 
the minimal order $O_{\rm min} \leq 2 $ are given. The expression ``minimal order''
refers to the lowest power in the expansion parameter $x$ in which this term appears.
Together with the prefactors
$h_j(\infty)$ the monomials define the low-energy effective Hamiltonian after the first CUT
\be
H^{\rm eff}_{0:0} = \sum_{j=0}h_j(\infty)A_j.
\label{eq:LE_hamiltonian}
\ee

\begin{table}[t]
 \begin{tabular}{|ccc|}
\hline
$j\hspace{1cm}$	&	$A_j$	&	$\hspace{1cm}O_{\rm min}$	\\
\hline
$0\hspace{1cm}$	&	$\sum\limits_{i}\mathds{1}$	&	$\hspace{1cm}0$	\\
$1\hspace{1cm}$	&	$\sum\limits_{i,\sigma} g^\dagger_{i,\sigma} g_{i,\sigma}^{\phantom{\dagger}}$	&	$\hspace{1cm}0$	\\
$2\hspace{1cm}$	&	$\sum\limits_{i,\sigma} \left(g^\dagger_{i,\sigma} g_{i+1,\sigma}^{\dagger}+{\rm h.c.} \right)$	&	$\hspace{1cm}1$	\\
$3\hspace{1cm}$	&	$\sum\limits_{i,\sigma} g^\dagger_{i,\sigma} g_{i,\sigma}^{\phantom{\dagger}}g^\dagger_{i+1,\sigma} g_{i+1,\sigma}^{\phantom{\dagger}}$	&	$\hspace{1cm}2$	\\
$4\hspace{1cm}$	&	$\sum\limits_{i,\sigma} g^\dagger_{i,\sigma} g_{i,\sigma}^{\phantom{\dagger}}g^\dagger_{i+1,\bar{\sigma}} g_{i+1,\bar{\sigma}}^{\phantom{\dagger}}$	&	$\hspace{1cm}2$	\\
$5\hspace{1cm}$	&	$\sum\limits_{i,\sigma} g^\dagger_{i,\sigma} g_{i,\bar{\sigma}}^{\phantom{\dagger}}g^{\dagger}_{i+1,\sigma} g_{i+1,\bar{\sigma}}^{\phantom{\dagger}}$	&	$\hspace{1cm}2$	\\
$6\hspace{1cm}$	&	$\sum\limits_{i,\sigma} \left(g^\dagger_{i,\sigma} g_{i+1,\bar{\sigma}}^{\dagger}g^{\phantom{\dagger}}_{i+1,\bar{\sigma}} g_{i+2,\sigma}^{\phantom{\dagger}} + {\rm h.c.} \right)$	&	$\hspace{1cm}2$	\\
$7\hspace{1cm}$	&	$\sum\limits_{i,\sigma} \left(g^\dagger_{i,\sigma} g_{i+1,\sigma}^{\dagger}g^{\phantom{\dagger}}_{i+1,\bar{\sigma}} g_{i+2,\bar{\sigma}}^{\phantom{\dagger}} + {\rm h.c.} \right)$	&	$\hspace{1cm}2$	\\
\hline
\end{tabular}
\caption{The operators $A_j$ up to the minimal order $O_{\rm min} = 2$ 
present in the low-energy effective Hamiltonian \reqn{eq:LE_hamiltonian}.
Note that we combined certain monomials which must have the same prefactor due to
symmetries or hermitian conjugation.}
\label{tab:effective_interactions}
\end{table}

\begin{figure}[t]
  \centering
  \includegraphics[width=1.15\columnwidth,angle=-90]{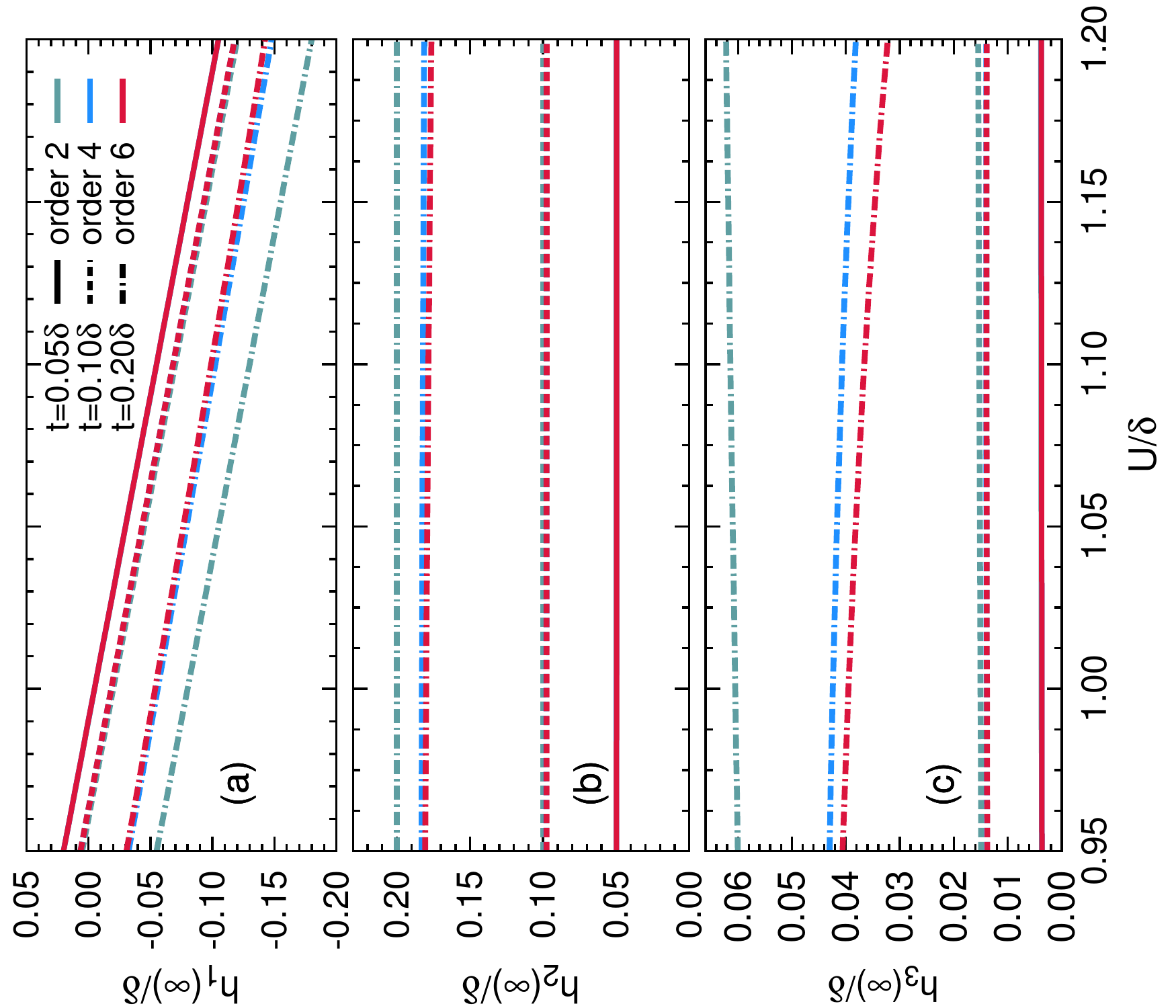}
  \caption{(Color online) The coefficiens $h_1(\infty)/\delta$ (a), $h_2(\infty)/\delta$ (b), and 
  $h_3(\infty)/\delta$ (c)
  defined in Eq.~\reqn{eq:LE_hamiltonian} plotted versus the Hubbard 
  interaction $U/\delta$. The results are obtained by applying the $D\!:\!0$
  generator to the Hamiltonian~\reqn{eq:H_nm}. Each panel includes the results 
  for the hopping parameters $t=0.05\delta$ (solid line), $t=0.10\delta$ (dashed line),
  and $t=0.20\delta$ (dotted-dashed line) in three diferent orders $2$ (green/light gray), $4$ (blue/dark
  gray),  and $6$ (red/gray).}
  \label{fig:effective_interactions}
\end{figure}

In order to verify  the convergence of the results, the prefactors of the
monomials $A_1$, $A_2$, and $A_3$ are plotted versus $U$ in panels (a), (b), and (c) 
of Fig.~\ref{fig:effective_interactions}, respectively. In each panel, the results for 
the hopping parameters $t=0.05$ (solid line), $t=0.10$ (dashed line), and $t=0.20$ (dotted-dashed line) 
in three different orders $2$ (green/light gray), $4$ (blue/dark gray), and $6$ (red/gray) are depicted. 
For $t=0.05$ the results in the different orders agree nicely
for all the three prefactors. Fig.~\ref{fig:effective_interactions} also
shows that for $t=0.10$, order $4$ and $6$ still coincide. But for 
$t=0.20$ we need to go to higher orders to obtain the effective Hamiltonian quantitatively.
In the following, we fix the order of deepCUT in this first step 
to $4$ in the hopping parameter $t$. This appears to be sufficient 
as long as we focus on low values of $t$ in the following.

The underlying idea to eliminate processes changing the number of DOs is similar to the one
used in the well-known derivation of the $t$-$J$ model from the Hubbard model \cite{Hirsch85,Fazekas99}.
We stress that the obtained
effective Hamiltonian is a {\it renormalized} one and that can be 
{\it systematically} improved by including higher orders in $t/\delta$, see also
Refs.\ \onlinecite{Reischl04,Hamerla10,Yang2010}. In Ref.\ \onlinecite{Tincani09}, Tincani {\it et al.}
investigated the IHM by restricting the local Hilbert space to the three states. They deal directly
with the Hamiltonian \reqn{eq:H_00} 
omitting the other processes completely.
Their findings for the transition points tend towards the results of 
the IHM in the limit $U,\delta \gg t$ \cite{Tincani09}.

\subsection{The One-Quasiparticle Sector}

The effective Hamiltonian derived in the previous subsection is
still complicated.
It includes various interactions between different 
QP sectors. These QPs are created and annihilated by the $g$-operators
of spin $\uparrow$ and $\downarrow$. To determine the dispersion of a single quasiparticle
(1QP), we need to decouple at least the zero- and one-QP sectors from the sectors with more QPs. 
The reduced generator $\eta_{g:1}$ is required for this goal.
Various symmetries and simplification rules are used 
in order to decrease the runtime and the memory requirement in the deepCUT algorithm so that
the high orders can be reached.  

We use the symmetries of reflection, the rotation about the $z$-axis of
the spins, and the self-adjointness of the Hamiltonian to reduce the
number of representative terms by about  a factor  $8$.
The various  simplification rules we use are analogous 
to those introduced in the first paper on  deepCUT \cite{Krull12}. 
In addition, we exploit the conservation of the particle number for each spin separately.
For details about the implementation of the simplification rules we refer
the reader to the Appendix \ref{app:simpl_rules}. In this way, we were able to reach order $20$ 
in the hopping parameter $t$ in the calculations for the 1QP dispersion. Up to this order 
no divergence in the numeric evaluation of the flow equations occurred in
the investigated parameter regime.

The final effective Hamiltonian is translationally invariant
so that the one-QP sector is diagonalized by a Fourier transformation. 
The resulting one-QP dispersion reads
\be
\omega(k) = h_0 + 2 \sum_{d=1}^{n} h_{2n} \cos(2nk)
\label{eq:disp}
\ee
where the prefactors $h_d$ is the hopping element from site $i$ to $i\pm d$. 
Only hopping elements over {\it even} distances occur because odd hops would
violate the conservation of the total particle number of original particles, see Eq.~\reqn{eq:pn_operator}. All 
bilinear terms acting on odd distances are of Bogoliubov type.

\begin{figure}[tb]
\includegraphics[width=0.85\columnwidth,angle=-90]{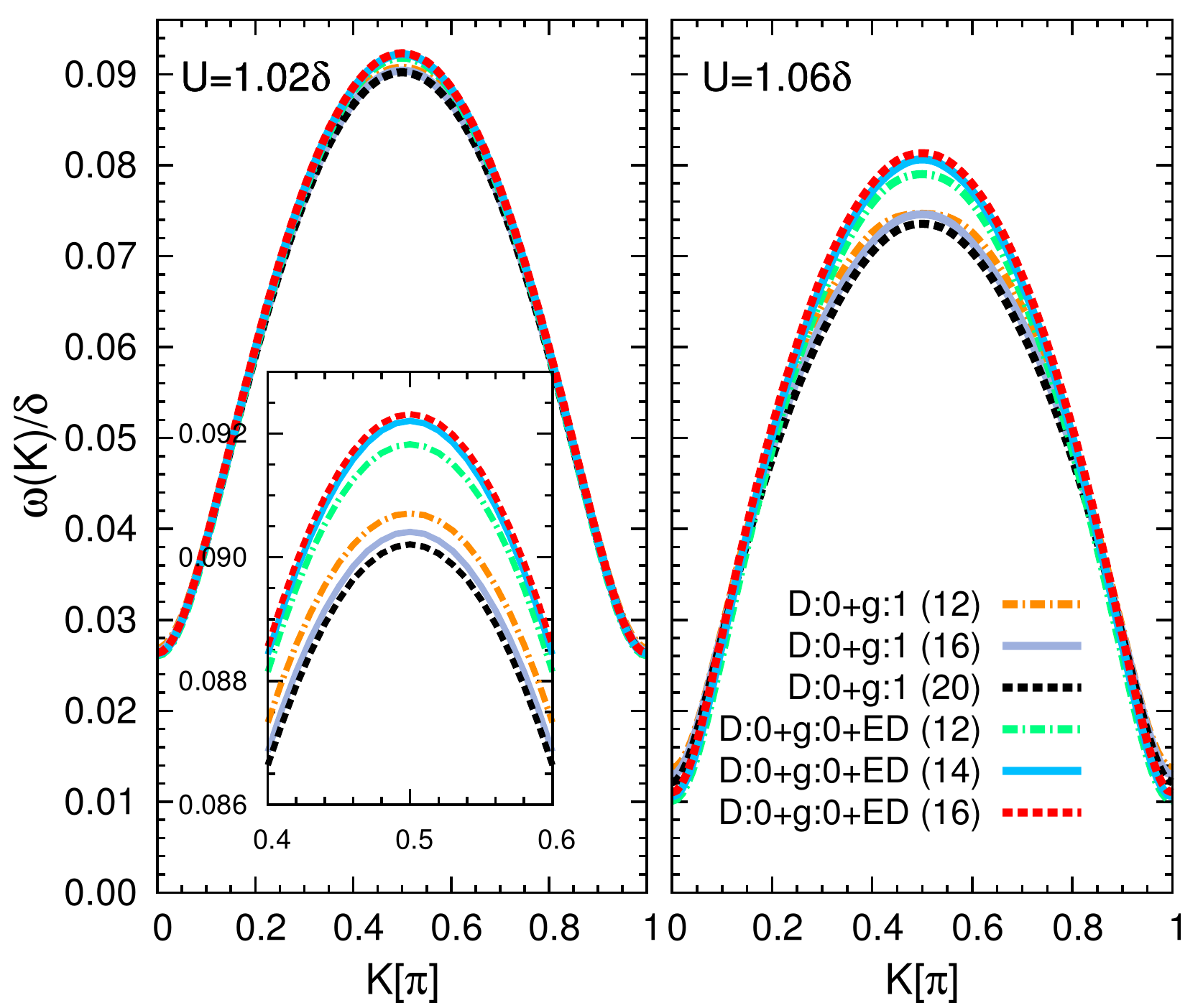}
  \caption{(Color online) The one-QP dispersion of the IHM for 
   $t=0.05\delta$ and $U=1.02\delta$ (left panel) and $U=1.06\delta$ 
  (right panel). The one-QP dispersion 
  is obtained by a successive application of the generators $D\!:\!0$ and $g\!:\!1$,
  denoted by $D\!:\!0\!+\!g\!:\!1$. The order in $D\!:\!0$ step is fixed to $4$ and
  the step $g\!:\!1$ is realized in order $12$ (dotted-dashed line), $16$ (solid line), 
  and $20$ (dashed line). 
  The deviations between different orders are maximum close to $K=\frac{\pi}{2}$.
  In addition, the dispersion resulting from the  combination of the deepCUT with ED,
  denoted $D\!:\!0\!+\!g\!:\!0\!+\!{\rm ED}$, see also main text below, is depicted. 
  The largest deviation between the two approaches
  occurs around $K=\frac{\pi}{2}$, i.e., at the maximum value of the dispersion.}
  \label{fig:disp}
\end{figure}

The one-QP dispersion \reqn{eq:disp} resulting from the consecutive 
application of the generators $D\!:\!0$ and $g\!:\!1$, denoted by $D\!:\!0+g\!:\!1$,
 is depicted in Fig.~\ref{fig:disp}. The shorthand $D\!:\!0+g\!:\!1$ stand for a
 first application by applying the generator $\eta_{D:0}$. Then, the resulting
 effective Hamiltonian is block-diagonalized by applying the generator $\eta_{g:1}$. 
The left panel of Fig.~\ref{fig:disp} is for $U=1.02\delta$ and the right panel is for $U=1.06\delta$.  
The hopping prefactor $t$ in Eq.~\reqn{eq:IHM_fin} is fixed to $0.05\delta$. 
The one-QP dispersion is presented for order $12$ (dotted-dashed line), $16$ (solid line), 
and $20$ (dashed line) in the hopping prefactor $t$.

The left panel in Fig.~\ref{fig:disp} shows that
the results of different orders $12$, $16$, and $20$ accurately coincide in the whole 
range of momenta $0 \leq K < \pi$ demonstrating a good convergence of the deepCUT method. 
The largest deviation occurs around the momentum $K=\frac{\pi}{2}$
where the dispersion is maximum as is shown in the inset.

But the convergence for increasing order is  worse for $U=1.06\delta$ because we approach the 
transition point, $U_c \approx 1.07\delta$.
Again, the largest deviation between different orders
occurs near the total momentum $K=\frac{\pi}{2}$. The one-QP dispersion $\omega(k)$ shows a tendency
to decrease on increasing the order of calculations.

\begin{figure}[t]
\includegraphics[width=0.73\columnwidth,angle=-90]{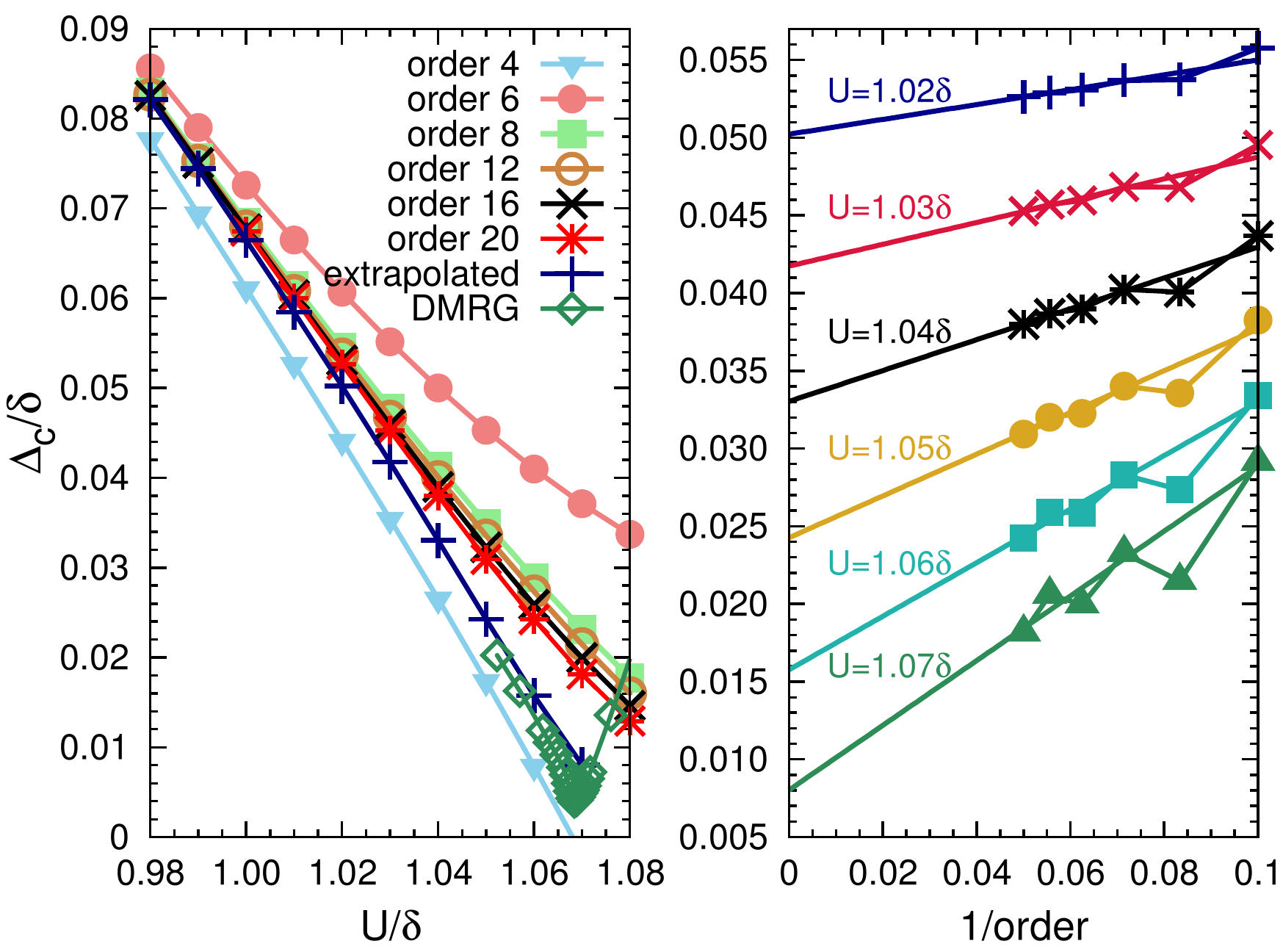}
\caption{(Color online) The charge gap of the ionic Hubbard model for the hopping parameter $t=0.05\delta$. 
	  Left panel: The charge gap as a function of $U/\delta$ in various orders. The deepCUT
	  results extrapolated to infinite order are also depicted. The 
	  extrapolated DMRG results \cite{Manmana04} are 
	  shown for comparison. The deviation in our analysis at finite orders becomes large close to the 
	  transition point, $U_c=1.07\delta$. The difference between extrapolated deepCUT results and 
	  DMRG results is about $0.002\delta$.
	  Right panel: The charge gap versus the inverse of the order for different values of 
	  Hubbard interaction $U$. The deepCUT results are extrapolated to infinite order 
	  by a linear fit to the last four points in $1/{\rm order}$.}
  \label{fig:gap_deepCUT}
\end{figure}

The charge gap ($\Delta_c$) is defined as the energy necessary to add an electron 
plus the energy for taking an electron from the system
\be
\Delta_c = E_0(N+1)+E_0(N-1)-2E_0(N),
\label{eq:charge_gap}
\ee
where $E_0(N)$ is the ground state energy of the system with $N$ particles.
For our electron-hole symmetric Hamiltonian \reqn{eq:IHM}, it is twice the 
minimum of the dispersion $\Delta_c = 2\omega_{\rm min}$.

Besides the charge gap, the following gaps are relevant in the IHM as well.
The exciton gap $\Delta_e$ is defined as the first excitation energy in the sector
with the same particle number as in the ground state and with total spin zero
\be
\Delta_e := E_1(N,S=0)-E_0(N,S=0),
\label{eq:exciton_gap}
\ee
where $E_1$ stands for the first excited state in the corresponding sector. 
Similarly, the spin gap is defined as the first excitation energy in the sector with
the same particle number, but with total spin one
\be
\Delta_s := E_1(N,S=1)-E_0(N,S=0).
\label{eq:spin_gap}
\ee
In our formalism, the exciton gap is given by the lowest energy 
of the first singlet bound state and the spin gap by the first triplet bound state, if binding occurs.
Otherwise, the lowest scattering states matter. Excited states of two QPs will be considered in
detail below in Sect.\ \ref{ssc:2qp}.

The charge gap for different orders of the hopping prefactor $t$ is plotted in the left
panel of Fig.~\ref{fig:gap_deepCUT} as function  of Hubbard interaction $U$. 
The deepCUT results extrapolated to infinite
order by a linear fit in $1/{\rm order}$ are also depicted. 
The DMRG results, rescaled to the present units, are shown for comparison \cite{Manmana04}. 
Data is given up to $U=1.08\delta$  because around this point 
the phase transition to the SDI takes place (see the next subsection) and 
the QP picture breaks down.
Fig.~\ref{fig:gap_deepCUT} shows that the results of 
the deepCUT at high orders coincide very well for $U<1.00\delta$.
For $U>1.00\delta$, however, especially close to the transition point, the different
orders separate due to the numerical deviations indicating a poorer convergence.
The difference between the extrapolated deepCUT results and the DMRG results is about $0.002\delta$.
We draw the reader's attention to the accuracy of such data. The energy scale of
the initial model before the renormalizing unitary transformations is $U+\delta\approx 2\delta$ so that
the transformations are still precise on energy scales reduced by three orders of magnitude.

In the right panel of Fig.~\ref{fig:gap_deepCUT}, the charge gap versus the inverse order is 
displayed for various values of $U$. The charge gap decreases on increasing  order. 
The deepCUT results are extrapolated to infinite order 
by a linear fit to the last four points. This plot illustrates how the deepCUT calculations 
converge as function of the order in the hopping $t$. The deepCUT method as used in the present work
is a renormalizing approach based on a truncation in real space. This means that processes are tracked only up to 
a certain range in real space. This range is determined by the order of the calculations
 interpreted as the maximum number of hops on the lattice.
Thus it is clear that the approach as presented here runs into difficulties upon
approaching continuous phase transitions where long-range processes become
essential.

\subsection{The Two-Quasiparticle Sector}
\label{ssc:2qp}

In the framework of deepCUT, the treatment of sectors with higher number of QPs is also possible
\cite{Knetter00,Windt2001,Knetter2001,Knetter03b,Schmidt2005,Fischer10,Krull12}.
The two-QP sector can be decoupled by using the reduced generator $\eta_{g:2}$.
This generator will yield an effective Hamiltonian that can be diagonalized for 
each combination of the total momentum $K$, total spin $S$, and total magnetic quantum number $M$. 
The two-QP states with fixed $K$, $S$, and $M$ read
\bearr
\ket{K;d}^{S,M}\!\! &=& \sum_{\sigma_1 \sigma_2} 
A^{S,M}_{\sigma_1,\sigma_2} \ket{K, \sigma_1 ; d ,\sigma_2} \nn \\
&=& \frac{1}{\sqrt{L}} \sum_r e^{iK(r+\frac{d}{2})}
\sum_{\sigma_1 \sigma_2} A^{S,M}_{\sigma_1,\sigma_2} 
\ket{r,\sigma_1; r+d, \sigma_2}, \nn \\
\label{eq:2qp_state}
\eearr
where $\sigma_1$ and $\sigma_2$ indicate the spins of the QPs, 
$A^{S,M}_{\sigma_1,\sigma_2}$ are the appropriate Clebsch-Gordon coefficients, 
$L$ is the system size, i.e, the number of sites. The sum runs over all lattice sites 
and $d>0$ is the distance between the two QPs which cannot be zero 
due to the hard-core property. Furthermore, because the two constituting
fermions are indistinguishable after the particle-hole transformation,
the Clebsch-Gordon coefficients take the contributions with negative $d$ into account.

The Hamiltonian matrix in the two-QP sector
is composed of three different submatrices refering to different total charge. 
Both QPs can be original electrons, or holes, or one is an electron and the other a hole.
Here we refer to the fermions before the particle-hole transformations. If the 
two-QP state \reqn{eq:2qp_state} contains only odd distances $d$
it consists of an electron and a hole. But if the two-QP state is made of two original electrons or
two holes, the distances between them are even, cf.\ Eq.\ \eqref{eq:pn_operator}.
Here we focus on the case of two-QP states with  one
electron and one hole and discuss the possible triplet and singlet 
bound states. 

The sector with two holes (or two electrons) is also very interesting 
in the context of superconductivity. A recent investigation of the IHM 
including next-nearest neighbor (NNN) hopping terms on the honeycomb
lattice found evidence for superconducting behavior upon hole doping \cite{Watanabe13}. 
A dynamic mean field theory 
study of the model also indicates an interesting half-metallic behavior on doping away 
from half-filling \cite{Garg13}. But these issues are beyond the scope of present article.

The Hamiltonian matrix can be constructed by applying the Hamiltonian parts
$H_{1:1}$ and $H_{2:2}$ to the state $\ket{K, \sigma_1 ; d ,\sigma_2}$. For $H_{1:1}$, we
obtain
\bearr
&H_{1:1}& \ket{K, \sigma_1 ; d ,\sigma_2} = \nn \\
&&+ \sum_{n>-d} \sum_{\beta_1} e^{iK\frac{n}{2}} 
\prescript{\beta_1 \!\!}{\sigma_1 \!\!}{\Big{[}}{ \mathcal{C}_{1}^{1}\Big{]}}^{\!n}
\ket{K, \beta_1 ; +d+n ,\sigma_2} \nn \\
&&- \sum_{n<-d} \sum_{\beta_1} e^{iK\frac{n}{2}} 
\prescript{\beta_1 \!\!}{\sigma_1 \!\!}{\Big{[}}{ \mathcal{C}_{1}^{1}}{\Big{]} }^{\!n}
\ket{K, \sigma_2 ; -d-n ,\beta_1} \nn \\
&&+ \sum_{n<d} \sum_{\beta_2} e^{iK\frac{n}{2}} 
\prescript{\beta_2 \!\!}{\sigma_2 \!\!}{\Big{[}}{ \mathcal{C}_{1}^{1}}{\Big{]} }^{\!n}
\ket{K, \sigma_1 ; +d-n ,\beta_2} \nn \\
&&- \sum_{n>d} \sum_{\beta_2} e^{iK\frac{n}{2}} 
\prescript{\beta_2 \!\!}{\sigma_2 \!\!}{\Big{[}}{ \mathcal{C}_{1}^{1}}{\Big{]} }^{\!n}
\ket{K, \beta_2 ; -d+n ,\sigma_1},
\label{eq:H11_on_2qp}
\eearr
where the appearance of the minus signs is due to the fermionic nature of the problem.
We use the shorthand
\be
\prescript{\beta \!\!}{\sigma \!\!}{\Big{[}}{ \mathcal{C}_{1}^{1}}{\Big{]} }^{\!n}
:=\bra{r\!-\!n,\beta} H_{1:1} \ket{r, \sigma} .
\label{eq:c11_def}
\ee
Similarly, for $H_{2:2}$ we have
\bearr
&H_{2:2}& \ket{K, \sigma_1 ; d ,\sigma_2} = \nn \\ 
&&-\sum_{n} \sum_{d' > 0} \sum_{\beta_1 \beta_2} e^{iK \left(n+\frac{d-d'}{2} \right)} 
\prescript{\beta_1 \! \beta_2 \!\!}{\sigma_1 \! \sigma_2 \!\!}{\Big{[}}{ \mathcal{C}_{2}^{2}}{\Big{]} }^{\!n d'}_{d}
\ket{K, \beta_1 ; d' ,\beta_2}, \nn \\
\label{eq:H22_on_2qp}
\eearr
with the definition
\be
\prescript{\beta_1 \! \beta_2 \!\!}{\sigma_1 \! \sigma_2 \!\!}{\Big{[}}{ \mathcal{C}_{2}^{2}}{\Big{]} }^{\!n d'}_{d}
\!\! :=\! \bra{r\!-\!n,\beta_1; r\!-\!n\!+\!d',\beta_2} H_{2:2} \ket{r, \sigma_1;r+d,\sigma_2} .
\label{eq:c22_def}
\ee
In order to fix the fermionic sign in the definition \reqn{eq:c22_def} uniquely
we assume from now on that in each monomial of $H_{2:2}$ the creation operators
are placed in front of the annihilation operators and the annihilation and the creation 
parts are separately site-ordered.

\begin{figure}[t]
  \centering
  \includegraphics[width=0.8\columnwidth,angle=-90]{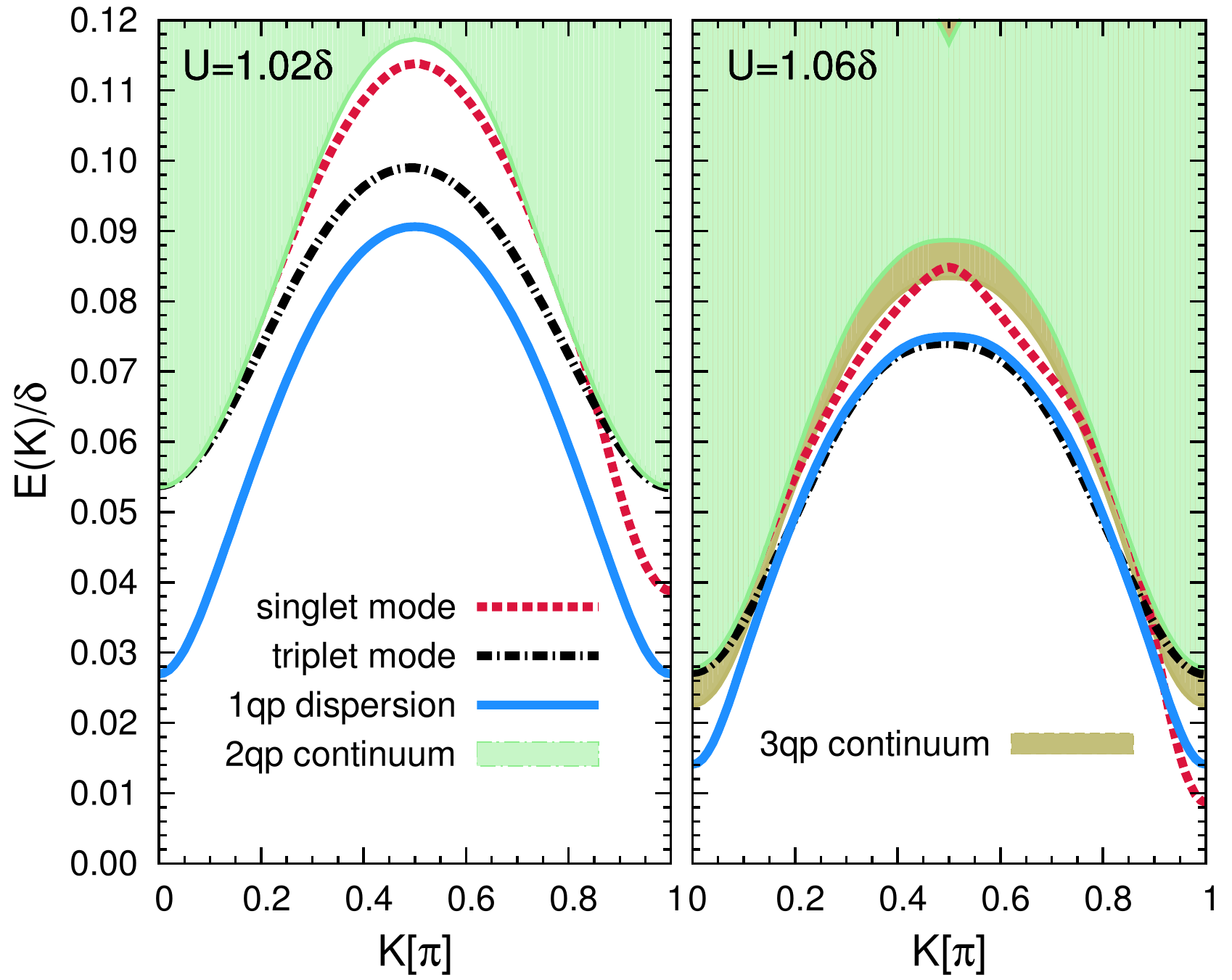}
  \caption{(Color online) Low-lying excitation spectrum including one-quasiparticle dispersion (solid line), 
	  two-quasiparticle and three-quasiparticle continuum (colored/shaded regions), 
	  singlet (dashed line) and triplet (dotted-dashed line) 
	  bound states. The results are obtained by consecutive
	  application of the generators $D\!:\!0$ and $g\!:\!2$. 
	  The order of the transformation for $D\!:\!0$ is $4$ and for $g\!:\!2$ 
	  it is  $12$ which is the highest converging order. The hopping 
	  element is $t=0.05\delta$.
	  The Hubbard interaction $U$ is fixed to $1.02\delta$ for 
	  the left panel and to $U=1.06\delta$ for the right panel. There
	  are two singlet bound states near the total momentum
	  $K=\pi$ and around $K=\frac{\pi}{2}$. The triplet bound 
	  state is almost symmetric around $K=\frac{\pi}{2}$
	  and exists in the whole BZ. For $U=1.06\delta$,
	  the two-quasiparticle continuum lies completely within the 
	  three-quasiparticle continuum.}
  \label{fig:spectrum}
\end{figure}

The low-lying excitation spectra for $U=1.02\delta$ and $U=1.06\delta$ are 
depicted in the left and in the right panel of Fig.~\ref{fig:spectrum}, respectively. The
hopping $t$ is fixed to $0.05\delta$. The order of the second CUT is $12$. 
We cannot go beyond this order because the flow equations for the two-QP sector
do not converge in higher orders. As can be seen 
in the right panel, the two-QP continuum lies within
the three-QP continuum. The lower edge of the four-QP continuum (not shown) 
lies also close in energy to the lower edge of two-QP continuum. This large overlap 
between continua of different number of QPs is the major reason of divergence
of the flow equations \cite{Fischer10}.

For both values
of $U$ in Fig.~\ref{fig:spectrum} there are two singlet and one triplet bound states. 
The singlet bound modes occur only near the total momentum $K=\pi$ and around $K=\frac{\pi}{2}$.
The triplet bound mode becomes more and more symmetric about $K=\frac{\pi}{2}$ as the 
Hubbard interaction $U$ is increased and approaches the transition point $U_{c1}\sim 1.07\delta$.
We attribute the wiggling of the 
singlet mode for $U=1.06\delta$ around the total momentum $K=\frac{\pi}{2}$ 
to the truncation of the flow equations. For $U=1.06\delta$, the lowest excited state
is the singlet bound state that appears at the total momentum $K=\pi$. 
This mode becomes soft, i.e, its energy vanishes, upon increasing the Hubbard interaction further
indicating the first phase transition at $U_{c1}$ from the BI to the SDI phase.

\begin{figure}[t]
\includegraphics[width=0.85\columnwidth,angle=-90]{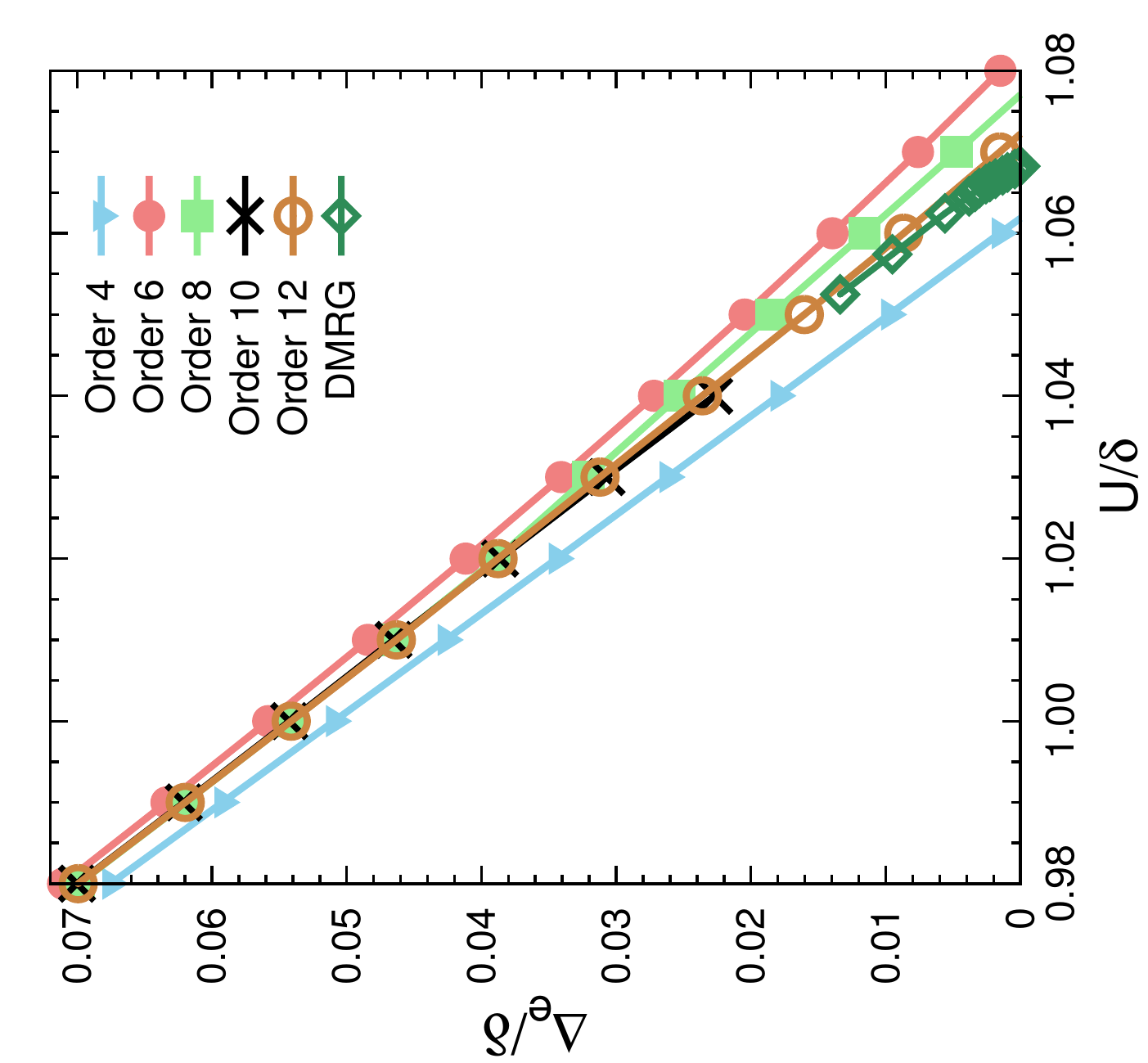}
  \caption{(Color online) The exciton (or singlet) gap $\Delta_e$ versus the Hubbard interaction
	  in various orders. The hopping is fixed to $t=0.05\delta$. The exciton gap at becomes soft for $U_c=1.072\delta$
	  in order $12$. For comparison, the DMRG prediction
	   of the first transition is $U_c=1.069\delta$ \cite{Manmana04}.}
  \label{fig:exciton_gap}
\end{figure}

The exciton gap is plotted versus the interaction $U$ in Fig.~\ref{fig:exciton_gap}
for different orders. Due to divergence of the flow equations no values are reported in order $10$ for $U>1.04\delta$.
For the same reason, orders higher than $12$ were not accessible. 
The extrapolated DMRG results extracted from Ref.\ \onlinecite{Manmana04} and rescaled to the present units
are also shown. 
The deepCUT results at high orders are very close to the DMRG results. The DMRG prediction
of the transition point is $1.069\delta$. In our analysis in order $12$ the exciton gap 
vanishes at $U_c=1.072\delta$. The deepCUT results for the exciton gap $\Delta_e$ converge better
upon increasing order than the deepCUT results for the charge gap $\Delta_c$ shown in Fig.~\ref{fig:gap_deepCUT}.
We attribute this to the larger separation in energy from the closest continuum.

We studied the energy difference between the spin and the charge gap. While this difference is
finite in any finite order, its extrapolation in the inverse order is consistent with a zero difference
in the BI phase, i.e., for $U\le U_{c1}$. In view of the definitions \eqref{eq:exciton_gap}
and \eqref{eq:spin_gap} this implies that no binding between two QPs occurs
in the two-QP sector with total $S=1$. These findings are to be compared to
previous DMRG data. Takada and Kido extrapolated the 
DMRG results to infinite system size and deduced that the spin gap and charge gap
become different before the first transition point $U_{c1}$ \cite{Takada01}. 
The equality of spin and charge gaps up to the first transition point 
is supported by other extrapolated DMRG calculations \cite{Manmana04,Lou03,Kampf03}.

The deepCUT approach realized in real space, the range of processes taken into account 
is proportional to the order of the calculation. Thus we expect 
the deepCUT method to provide accurate results as long as the
order is larger than the correlation length $\xi$ in units of the lattice spacing of the system.
The correlation length can be estimated as \cite{Sorensen93,Okunishi01}
\be
\xi \approx \frac{v}{\Delta},
\label{eq:correl_length}
\ee
where $v$ is the velocity for vanishing gap and $\Delta$ is
the gap present in the system. The relation \eqref{eq:correl_length}
stems from the assumption that the low-energy physics of the model
fulfills an (approximate) Lorentzian symmetry.

In the IHM, the exciton gap is the smallest
gap and hence we set $\Delta=\Delta_e$. The fermionic velocity can be obtained by 
fitting  $\omega(k)=v\sin(k)$ to the one-QP dispersion.
We find $\xi = \frac{0.09}{0.038} \approx 2.4$ for $U=1.02\delta$, 
$\xi = \frac{0.07}{0.008} \approx 8.8$ for $U=1.06\delta$, and 
$\xi = \frac{0.07}{0.0015} \approx 47$ for $U=1.07\delta$. The rapid
increase of $\xi$ on approaching the transition point $U_{c1}$ 
reflects the vanishing exciton gap $\Delta_e$. This implies that
the deepCUT approach parametrized in real space naturally becomes
inaccurate on approaching $U_{c1}$.

\section{Exact Diagonalization in the Thermodynamic Limit}
\label{sec:exact_diagonalization}

The deepCUT results close to
the transition point is not quantitative, especially for the charge gap
for reasons given above. In this section, we aim at improving the results
by following the route used previously in Ref.\ \onlinecite{Fischer10}.
The goal of the deepCUT is chosen less ambitious, i.e., less terms
are rotated away. This makes the deepCUT step less prone to inaccuracies
and convergence can be achieved more easily. But the disadvantage
is that the resulting effective Hamiltonian is not yet diagonal or
block-diagonal so that the subsequent analysis becomes more demanding.
Here we will employ exact diagonalization in restricted subspaces for this purpose.

\subsection{Construction of the Hamiltonian Matrix}

In order to take into account  processes of longer range for the important {\it excited states}, we only decouple 
the ground state from the subspaces with finite number of QPs.
This is achieved by applying the reduced generator $g\!:\!0$.
This generator keeps interactions and transitions between different excited states.
Because the system under study is fermionic, there are only terms in the Hamiltonian 
with even number of fermionic operators. Thus there is no process linking
one QP and two QPs:  $H_{2:1} = 0$. Therefore, the major 
off-diagonal interaction for one-QP states is $H_{3:1}+H_{1:3}$ and for two-QP states 
it is $H_{4:2}+H_{2:4}$. 

After applying the generator $g\!:\!0$, the effective Hamiltonian 
has the following structure
\bearr
H_{\rm eff} &=& H_{0:0} + H_{1:1} + H_{2:2} + H_{3:3} + H_{4:4} \nn \\
&& + (H_{3:1} + {\rm h.c.} ) + (H_{4:2} + {\rm h.c.}) \nn \\
&& + {\rm  less~important~terms},
\label{eq:ED_ham}
\eearr
where the less important terms include the parts which involve states with more than four QPs. 
These interactions have much less effect than $H_{3:1}$ and $H_{4:2}$ on the low-energy spectrum
given by the eigenvalues in the one-QP and in the two QP sectors.

The effect of off-diagonal interactions between one- and three-QP states and
between two- and four-QP states can be 
considered by restricting the Hilbert space to four-QP states and 
performing an exact diagonalization (ED) within this restricted
Hilbert space. The effect of the Hamiltonian is stored 
in two separate matrices, one for
the states that are built from one and three QPs and the other for the states built from
 two and four QPs. We stress that  also states  with four QPs have to be considered
to be able to address modifications in the two-QP spectrum.

Because the ground state is decoupled in the deepCUT step,
we can work directly in the thermodynamic limit by introducing the states with specific 
total momentum $K$, total spin $S$, and total magnetic number $M$
\bs
\label{eq:ED_states}
\begin{align}
\ket{K}^{S,M} \!&=\! \frac{1}{\sqrt{L}} \sum_r e^{iKr} \ket{r}^{S,M}, 
 \label{eq:ED_states_1} \\
\ket{K;d}^{S,M} \!&=\! \frac{1}{\sqrt{L}} \sum_r e^{iK \left(r+\frac{d}{2} \right) } \ket{r;r\!+\!d}^{S,M}, 
\label{eq:ED_states_2} \\
\ket{K;d_1;d_2}^{S,M}_{\alpha} \!&=\! \frac{1}{\sqrt{L}} 
\sum_r e^{iK \left(r+\frac{2d_1+d_2}{3} \right)} \times \nn \\ 
&\hspace{0.8cm}\times \ket{r;r\!+\!d_1;r\!+\!d_1\!+\!d_2}^{S,M}_{\alpha}, 
\label{eq:ED_states_3} \\
\hspace{-0.2cm} \ket{K;d_1;d_2;d_3}^{S,M}_{\alpha} \!&=\! \frac{1}{\sqrt{L}} 
\sum_r e^{iK \left(r+\frac{3d_1+2d_2+d_3}{4} \right)} \times \nn \\ 
&\hspace{-1.5cm}\times \ket{r;r\!+\!d_1;r\!+\!d_1\!+\!d_2;r\!+\!d_1\!+\!d_2\!+\!d_3}^{S,M}_{\alpha},
\label{eq:ED_states_4}
\end{align}
\es
where $d_1$, $d_2$, and $d_3$ are the distances between the QPs, and $\alpha$
is an additional quantum number that specifies the spin configuration. 
The quantum number $\alpha$ is required for distinction because there is more than
one spin configuration with three and four QPs for given total spin and total $S_z$.

The Hamiltonian matrix is constructed for each fixed set of
$K$, $S$, and $M$. The action 
of the parts of the Hamiltonian $H_{i:j}$ for $i,j \leq 4$ on the 
states~\reqn{eq:ED_states} is calculated analytically. The effect
of $H_{1:1}$ and $H_{2:2}$ on the two-QP state is already reported in 
Eqs.~\reqn{eq:H11_on_2qp} and \reqn{eq:H22_on_2qp}. The effect of 
$H_{1:1}$ on two-QP state has $4$ contributions while it has $9$
and $16$ contributions for three- and four-QP states, respectively. 
The application of $H_{2:2}$  on three- and four-QP states 
lead to  $9$ and $36$ different contributions. 
The numbers of contributions can be understood easily. For instance, 
$H_{1:1}$ has three different possibilities to annihilate a QP when it
acts on a three-QP state and it can also create a QP in three distinct
positions, namely to the left, between, and to the right of the two QPs on the chain, leading to $9$ contributions. 
The process is schematically shown in Fig.~\ref{fig:schematic_H11_on_3qp}.

\begin{figure}[t]
  \centering
  \includegraphics[width=0.99\columnwidth,angle=0]{fig8.pdf}
  \caption{(Color online) Schematic representation of the application of
   $H_{1:1}=\sum_{i,j} t_{i,j} g_{i}^{\dagger}g_{j}^{\protect\phantom{\dagger}}$ on the three-QP
  state $\ket{K;d_1;d_2}$ defined in~\reqn{eq:ED_states_3}. There are three different 
  possibilities for the operator $g_{j}^{\protect\phantom{\dagger}}$ to annihilate
  a QP and the operator $g_{i}^{\dagger}$ can create a QP in three distinct positions: 
  to the left, between, and to the right of the two QPs already present. 
  This leads to $9$ different contributions.
  }
  \label{fig:schematic_H11_on_3qp}
\end{figure}

The explicit expressions for 
the action of different parts of the Hamiltonian~\reqn{eq:ED_ham} on the states~\reqn{eq:ED_states}
are calculated and reported in the supplementary electronic material. These expressions 
are general and can be used for all hardcore fermionic or bosonic problems.
The two-, three-, and four-QP states with total 
spin $S=0$ and $S=1$, total magnetic number $M=0$ and $M=1$, and the additional label $\alpha$
are also given in the supplementary electronic part.

The idea that we have applied here is similar to what had been introduced in 
Ref.\ \onlinecite{Fischer10} to describe QP decay with CUT.
The main difference is that we have to  take care of the fermionic minus sign 
and  to consider also the states with four QPs. Including the 
four-QP states not only leads to large analytic expressions, but also 
limits the maximum relative distances that can be treated numerically. 
For the following results, the Hamiltonian matrix has been constructed with maximum
distances $d_1^{\rm max}=d_2^{\rm max}=d_3^{\rm max}=24$.

\subsection{Low-lying Excitation Spectrum}

\begin{figure}[t]
\includegraphics[width=0.8\columnwidth,angle=-90]{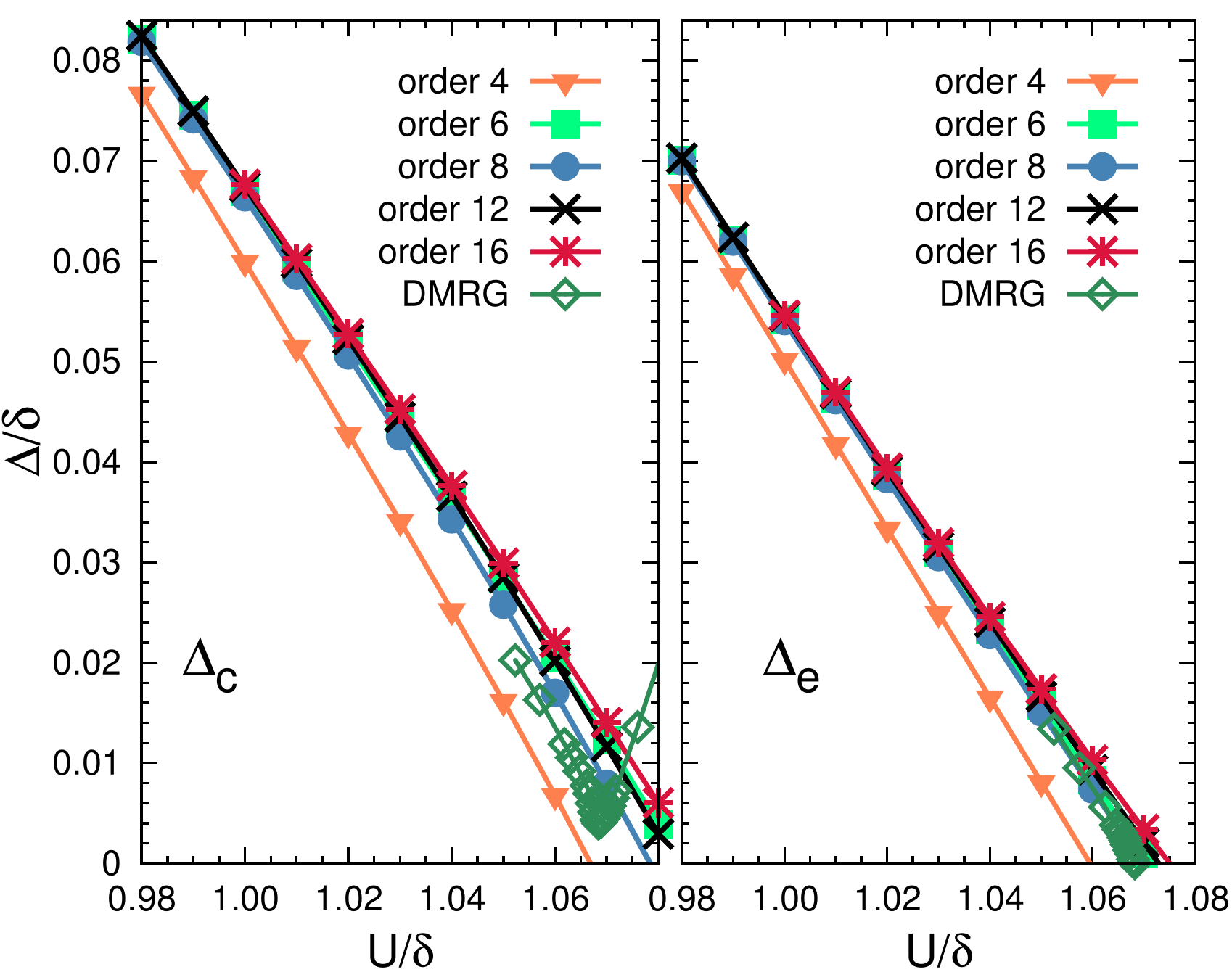}
  \caption{(Color online) The charge gap (left panel) and the 
  exciton gap (right panel) vs.\ $U$ obtained by $D\!:\!0\!+\!g\!:\!0\!+\!{\rm ED}$, see main text. 
  The order of the $D\!:\!0$ step is $4$ and $g\!:\!0$ is carried out in various orders, see legend.
  The data for charge gap appears to be more robust in $D\!:\!0\!+\!g\!:\!0\!+\!{\rm ED}$ than in the pure 
  deepCUT analysis $D\!:\!0\!+\!g\!:\!2$.}
  \label{fig:gap_ED}
\end{figure}

The charge and exciton gaps obtained by the combination of deepCUT and ED  are depicted
in Fig.~\ref{fig:gap_ED}. We denote this approach by $D\!:\!0+g\!:\!0+{\rm ED}$
which means that the effective Hamiltonian is derived by the consecutive application
of the generators $\eta_{D:0}$ and $\eta_{g:0}$. Then this effective Hamiltonian is 
analyzed by ED method as described above. Due to the restriction
of the Hilbert space in the ED, its results {\it overestimate} the eigen values 
of the {\it effective Hamiltonian}, i.e., they provide upper bounds to them.
But note that the effective Hamiltonian has only a limited accuracy due to
the truncations in the course of the deepCUT $D\!:\!0+g\!:\!0$ so that 
the ED results cannot be taken as rigorous upper bounds. If we, however, assume
that the inaccuracies introduced in the derivation of the 
effective Hamiltonian are of minor importance, the ED results can be taken as 
an {\it upper} bound for the {\it correct} eigen values.

The left panel of Fig.~\ref{fig:gap_ED} shows that the difference between
the data obtained by $D\!:\!0\!+\!g\!:\!0\!+\!{\rm ED}$ and the DMRG results is smaller than 
the difference of the data of the pure application of the deepCUT to the DMRG results, 
cf.\ Fig.~\ref{fig:gap_deepCUT}. For the charge gap close to the phase transition, 
the deviation between our results and the DMRG data is decreased from about $1\%$ for the pure deepCUT to about $0.5\%$ for 
the combination of deepCUT and ED.

In the right panel of Fig.~\ref{fig:gap_ED}, the exciton gap is plotted vs.\
the Hubbard interaction $U$. The results agree nicely with the DMRG results
for all orders higher than $4$. Inspecting the trend of the results for
increasing order they appear to converge to values slightly higher
than the DMRG results. We attribute this fact to the restriction 
of the Hilbert space in the ED treatment making it an upper bound.

\begin{figure}[t]
  \includegraphics[width=0.79\columnwidth,angle=-90]{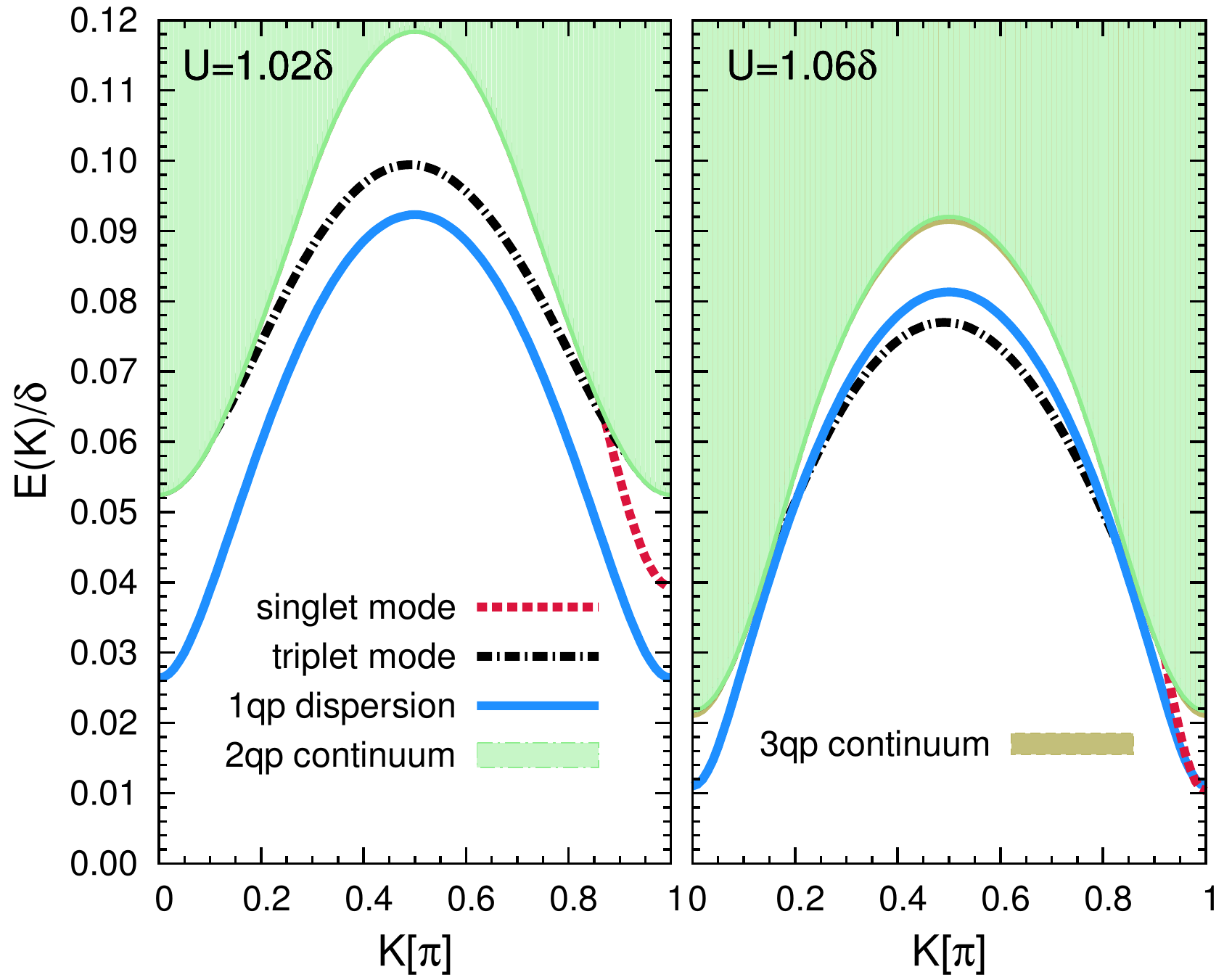}
  \caption{(Color online) The low-energy spectrum of the IHM including
  one-quasiparticle dispersion (solid line), two- and three-quasiparticle 
  continuum (solid region), and singlet (dashed line) and
  triplet (dotted-dashed line) bound states. The results are obtained by $D\!:\!0\!+\!g\!:\!0\!+\!{\rm ED}$. 
  The deepCUT steps $D\!:\!0$ and $g\!:\!0$ are  performed in orders $4$ and $16$, respectively. 
  For the ED, the Hamiltonian matrix is constructed for maximum relative distances of $24$. The hopping is fixed to $t=0.05\delta$ 
  for both panels; $U=1.02\delta$  in the left panel and $U=1.06\delta$ in the right panel. For $U=1.06\delta$, the lower edge
  of two- and three-quasiparticle continuum are very close to each other. No singlet bound state is found
  near the total momentum $K=\pi/2$ in contrast to the pure deepCUT $D\!:\!0\!+\!g\!:\!2$, see Fig.~\ref{fig:spectrum}.}
  \label{fig:deepCUT_ED_disps}
\end{figure}

The one-QP dispersion obtained by the combination $D\!:\!0+g\!:\!0+{\rm ED}$ is plotted 
in Fig.~\ref{fig:disp} for the two different values of the Hubbard interaction
$U=1.02\delta$ (left panel) and $U=1.06\delta$ (right panel). 
In this figure, we compare the results of 
pure deepCUT with the results of the combination of deepCUT and ED. For $U=1.02\delta$, both
methods coincide nicely except very close to $K=\frac{\pi}{2}$ where the maximum 
deviation occurs. 
Around $K=\frac{\pi}{2}$ the results of deepCUT plus ED lie a bit
higher in energy than those by pure deepCUT, see also inset. 
It is not clear whether the small difference is due to the restriction of the Hilbert space in ED implying
 a certain overestimation or whether it is due to the effect of long-range processes 
 that are less well captured by the pure  deepCUT.

Next, we focus on the right panel of Fig.~\ref{fig:disp} where $U=1.06\delta$
close to the transition point. Here the difference between the two 
methods is larger. For the momenta near $0$ and $\pi$
the combination $D\!:\!0+g\!:\!0+{\rm ED}$ yields a dispersion with lower energy
while for the momenta around $\frac{\pi}{2}$ the result from $D\!:\!0+g\!:\!2$ 
is the lower one. From the comparison with the extrapolated DMRG results 
for the charge gap we deduce that
the dispersion of $D\!:\!0+g\!:\!0+{\rm ED}$ is more accurate near $K=0$ and $K=\pi$.
Thus we presume that also around $K=\frac{\pi}{2}$ the $D\!:\!0+g\!:\!0+{\rm ED}$
data is more accurate, but there is no  data from alternative approaches 
available to corroborate this  conclusion.

Let us turn to  Fig.~\ref{fig:deepCUT_ED_disps} which shows the low-energy spectrum of the IHM
obtained by $D\!:\!0+g\!:\!0+{\rm ED}$ with the orders $4$ and $16$ for the deepCUT steps
 $D\!:\!0$ and $g\!:\!0$, respectively.
 The left and right panels are again for $U=1.02\delta$ and 
$U=1.06\delta$. The lower edge of the three-QP continuum for $U=1.06\delta$ lies close to
the lower edge of the two-QP continuum. The major difference between this figure
and the pure deepCUT results plotted in Fig.~\ref{fig:spectrum} is the 
absence of the singlet bound state around the total momentum $K=\pi/2$.
This difference may arise from the restricted relative distances of
QPs in the ED treatment. The singlet bound state mode near $K=\pi/2$ 
has a \emph{small} binding energy indicating that it is weakly bound and thus 
extending over large distances. Its extension is restricted due to computational limitations
and the binding may be suppressed in the ED spuriously.

\section{Beyond the Transition Point: A Mean Field Study}
\label{sec:mean_field}

The deepCUT approach realized in the previous sections
is based on the QPs of the BI, i.e., the more complicated, dressed
excitations close to the transition to the SDI are continuously mapped
to the simple QPs of the BI. The same quantum numbers are used in analogy to
Fermi liquid theory which uses the same quantum
numbers as the Fermi gas. As long as the system is located
on the BI side of the phase transition, only 
a few-particle problem remains to be solved in a subsequent step to find
the low-lying excitation spectrum.
But this QP picture breaks down when a phase transition occurs.
Beyond the transition point, a macroscopic number of QPs of the BI
condenses forming the new phase. This new phase displays other types
of elementary excitations.

Our analysis of the BI of the IHM in the previous sections showed that the exciton gap 
decreases on increasing the Hubbard interaction and vanishes at a critical
value  $U_{c1}$. This critical interaction was found to be $U_{c1}=1.072\delta$
for $D\!:\!0+g\!:\!2$ in order $12$, see Fig.~\ref{fig:exciton_gap}. 
How can we proceed beyond the transition and still profit from the
effective Hamiltonians obtained by deepCUT?
The most systematic way would be to set up a CUT with respect to the
ground state and the elementary excitations for $U> U_{c1}$. But there are
two obstacles to this route. The first one is that one has to know and to characterize the SDI ground state 
sufficiently well to be able to set up a CUT. The second one is that this approach would require
to implement another, different CUT which is tedious.

Thus we choose a slightly modified approach and continue to use the implemented CUT
to derive an effective Hamiltonian by applying $D\!:\!0+g\!:\!2$ and then
to analyze this effective Hamiltonian by a perturbative approach. 
The guiding idea is that the terms driving the phase transition are small
and can be treated perturbatively as long as the system is considered close to the
phase transition. In this way, one continues to profit from the deepCUT implemented
to obtain effective Hamiltonians. 
We use the deepCUT $D\!:\!0+g\!:\!2$ in order $12$ to derive the effective
Hamiltonian that we analyse perturbatively in the sequel. This deepCUT is not yet
so sensitive to be spoilt by the instability towards the SDI because the latter
takes place on very low energy scales.

For simplicity, we choose here a mean-field approximation
as a first step of a perturbative treatment. Although this approach is
not able to capture the correct critical behavior in low dimensions and
underestimates the role of fluctuations, it provides us with 
an estimate which phases are lower in energy.
Since the exciton becomes soft at $U_{c1}$ the SDI can be seen as a condensate of excitons.
The particle-hole transformation that we performed maps the original exciton into
a bound state of two fermions, i.e., the exciton appears as Cooper pair.
Thus we expect a BCS-type theory to describe the SDI phase transition. 

The effective Hamiltonian is represented in terms of hardcore 
fermions $\{ g_{i,\sigma} \}$ and it includes
various interactions within and between sectors of different numbers of QPs.
In the following,  we consider this effective Hamiltonian up to quadrilinear interactions and 
ignore interaction terms acting on higher numbers of QPs. Hence
the effective Hamiltonian takes the general form
\begin{subequations}
\be
H_{\rm eff} = H_{0:0}+H_{1:1}+H_{2:2},
\label{eq:D0_g2}
\ee
where
\bearr
H_{0:0} &=&  E_0 \mathds{1} , \allowdisplaybreaks[4] 
\\
H_{1:1} &=& \sum_{ij} \Gamma_{j;i}~ g^{\dagger}_j g^{\phantom{\dagger}}_i, \allowdisplaybreaks[4] 
\\
H_{2:2} &=& \sum_{ijkl} \Gamma_{kl;ij}~ g^{\dagger}_k g^{\dagger}_l g^{\phantom{\dagger}}_i g^{\phantom{\dagger}}_j.
\eearr
\end{subequations}

The prefactors $\Gamma_{j;i}$ and $\Gamma_{kl;ij}$ are nonzero up to an interaction range 
proportional to the order of calculations. The processes of longer range are all zero.
Because the effective Hamiltonian~\reqn{eq:D0_g2} is obtained 
by applying the reduced generator $g\!:\!2$ no off-diagonal interactions such as $H_{3:1}$ appear.

In order to apply the Wick theorem, we neglect the hardcore property of the operators and treat
them like usual fermions. Due to this approximation two fermions with
different spin are allowed to occupy the same site. It is also possible to deal with the hardcore
property by the slave-particle techniques, see Ref.\ \onlinecite{VaeziThesis} and references therein,
or by the Brueckner approach \cite{kotov98}.
But such analyses are beyond the scope of present investigation.

For a self-consistent mean-field approximation the symmetries of the ground state are essential. 
In order to describe the SDI phase of the IHM, we take  the possibility of 
a spontaneous symmetry breaking into account with nonzero anomalous 
expectation values (see below). The broken symmetry is the parity with 
respect to reflection about a site. Thus adjacent bonds may become different
even though in the original Hamiltonian the (directed) bond from site 0 to 1 was
identical to the one from 0 to -1. This is characteristic of the SDI as
found in previous studies  based on variational quantum Monte 
Carlo \cite{Wilkens01} and DMRG \cite{Manmana04,Takada01,Lou03,Zhang03,Kampf03,Tincani09,Legeza06}.
Thus, we assume for the expectation values
\bs
\label{eq:MF_EVs}
\bearr
\langle g^\dagger_{i,\sigma} g^\dagger_{i+m,\sigma}  \rangle
&\neq& \langle g^\dagger_{i+1,\sigma} g^\dagger_{i+m+1,\sigma}  \rangle
\neq 0,
\label{eq:MF_Bog}
\\
\langle g^\dagger_{i,\sigma} g^{\phantom{\dagger}}_{i+n,\sigma}  \rangle
&=& \langle g^\dagger_{i+1,\sigma} g^{\phantom{\dagger}}_{i+n+1,\sigma}  \rangle
\neq 0,
\label{eq:MF_hop}
\eearr
\es
where $m$ and $n$ stand for odd and even distances, respectively. 
The maximum values of $m$ and $n$ depend on the order in which the deepCUT 
was performed. All the above expectation values 
are zero in the BI phase where the ground state is the vacuum of
``$g$-particles'', but they become finite as soon as the exciton begins to 
condense and the phase transition occurs.

For a transparent notation, we express the $g$-operators acting on even and 
odd sites by $a$- and $b$-operators, respectively. The resulting mean field Hamiltonian takes the BCS-form 
\bearr
H_{\rm BCS} &=& \frac{L}{2}(\epsilon^{A}_0+\epsilon^{B}_0) 
\nn \allowdisplaybreaks[4] 
\\
&&\hspace{-1.2cm}+ \hspace{-0.2cm} \sum_{r \in {\rm even},\sigma } 
\!\! \left( 
t^A_0 :a^{\dagger}_{r,\sigma} a^{\phantom{\dagger}}_{r,\sigma}:
+ \hspace{-0.2cm} \sum_{n=2,4,\cdots} \!\! 
t^A_{n} :a^{\dagger}_{r,\sigma} a^{\phantom{\dagger}}_{r+n,\sigma}+ {\rm h.c.}:
\right) \nn \allowdisplaybreaks[4] 
\\
&&\hspace{-1.2cm}+ \hspace{-0.2cm} \sum_{r \in {\rm odd},\sigma } 
\! \left( 
t^B_0 :b^{\dagger}_{r,\sigma} b^{\phantom{\dagger}}_{r,\sigma}:
+ \hspace{-0.2cm}\sum_{n=2,4,\cdots} \!\! 
t^B_{n} :b^{\dagger}_{r,\sigma} b^{\phantom{\dagger}}_{r+n,\sigma}+ {\rm h.c.}:
\right) \nn \allowdisplaybreaks[4] \\
&&\hspace{-1.2cm} +\hspace{-0.2cm} \sum_{r \in {\rm even},\sigma } 
\hspace{0.1cm} \sum_{m=1,3,\cdots} \Delta^A_{m} :a^{\dagger}_{r,\sigma} b^\dagger_{r+m,\sigma}:+ {\rm h.c.}
\nn \allowdisplaybreaks[4] \\
&&\hspace{-1.2cm} + \hspace{-0.2cm} \sum_{r \in {\rm odd},\sigma } 
\hspace{0.2cm} \sum_{m=1,3,\cdots} \Delta^B_{m} :b^{\dagger}_{r,\sigma} a^\dagger_{r+m,\sigma}:+ {\rm h.c.}
\label{eq:MF_ham_RS}
\eearr
where we have divided the lattice into the two sublattices $A$ and $B$ of even sites 
and odd sites, respectively. The prefactors 
$\epsilon^{A}_0$, $\epsilon^{B}_0$, $t^A_{d}$, $t^B_{d}$, $\Delta^A_{d}$, and 
$\Delta^B_{d}$ depend on the coefficients of the effective Hamiltonian, which 
stem from the flow equations, see Eq.~\reqn{eq:ham_monomial_rep}, and from the expectation
values introduced in Eq.~\reqn{eq:MF_EVs}. Due to the identity \reqn{eq:MF_hop}, the hopping prefactors 
of the two sublattices are identical. So we unify them omitting the sublattice index
 $t^A_{d}=t^B_{d} =: t_{d}$.

\begin{figure}[t]
\includegraphics[width=0.7\columnwidth,angle=-90]{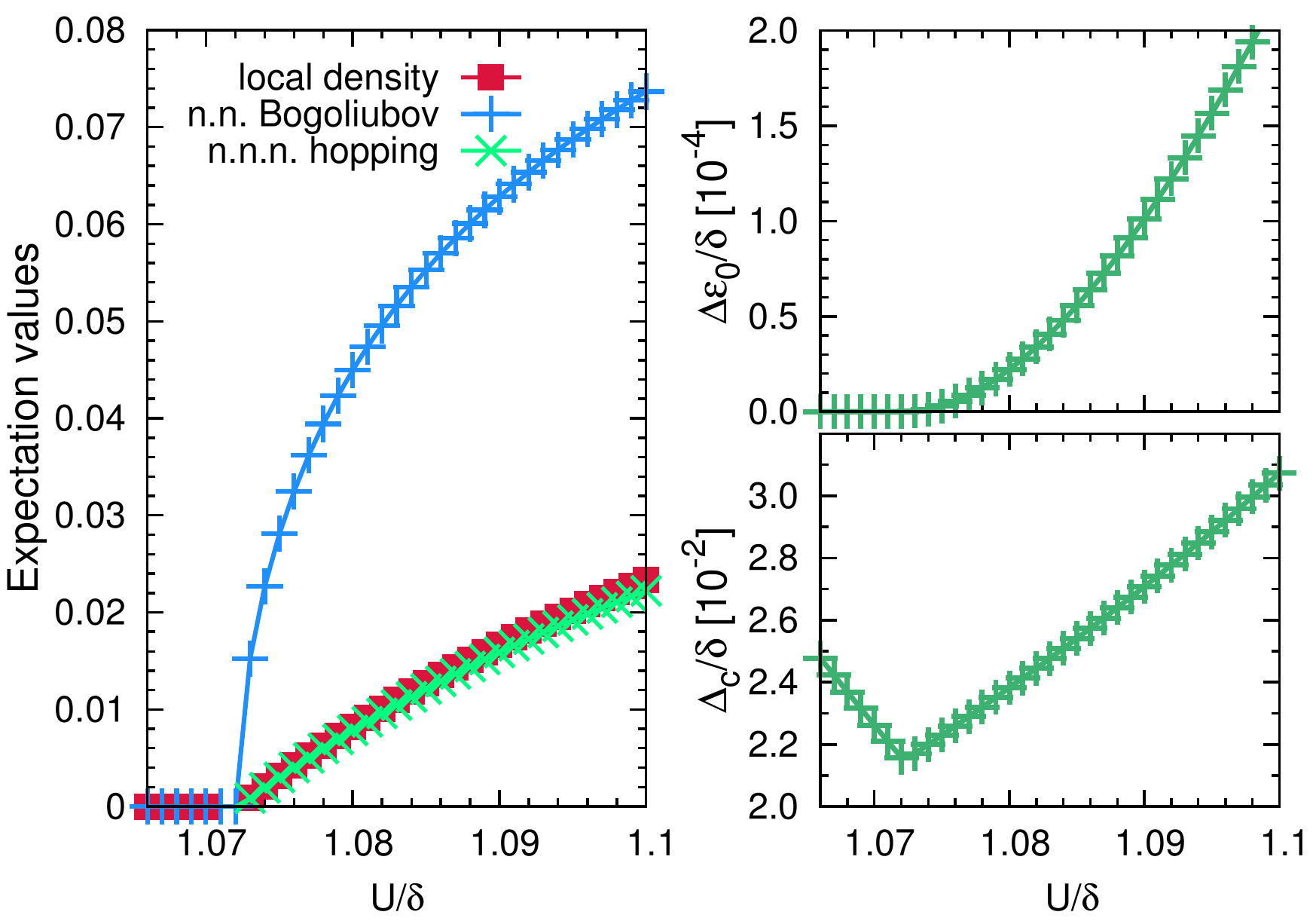}
  \caption{(Color online) Analysis of the effective Hamiltonian obtained by $D:0+g:2$ 
  in order 12 within the BCS-type approximation. The hopping prefactor $t$ is 
  $0.05\delta$. 
  Left panel: Expectation values of the local density, the nearest-neighbor (NN) 
  Bogoliubov term, and the next-nearest-neighbor (NNN) hopping term vs.\ 
  the Hubbard interaction $U$. At the phase transition at $U_c=1.072\delta$ 
  the expectation values become finite. The Bogoliubov term shows a square
  root behavior near the transition point. Right panel: The condensation 
  energy per site $\Delta \epsilon_0$ (upper panel) and the charge gap $\Delta_c$ 
  (lower panel) vs.\ Hubbard interaction $U$. The charge gap 
  starts to increase beyond the transition point $U>U_{c1}$.}
  \label{fig:MF}
\end{figure}

The BCS Hamiltonian~\reqn{eq:MF_ham_RS} is diagonalized by a Bogoliubov 
transformation in momentum space. The self-consistency equations to be solved
are found after some lengthy standard calculations
\bs
\label{eq:MF_eq}
\bearr
\langle a^\dagger_{r,\sigma} a^{\phantom{\dagger}}_{r+n,\sigma} \rangle 
&=& \langle b^\dagger_{r,\sigma} b^{\phantom{\dagger}}_{r+n,\sigma} \rangle \nn \\
&=&\frac{1}{\pi} \int_{0}^{\frac{\pi}{2}} \!\!dk ~\frac{\lambda(k)-t(k)}{\lambda(k)} \cos(n k),
~~~~~~~~~~~\allowdisplaybreaks[4] \\
\langle a^\dagger_{r,\sigma} b^\dagger_{r+m,\sigma} \rangle 
&=& \nn \\ && \hspace{-2.5cm} \frac{1}{\pi} \int_{0}^{\frac{\pi}{2}} \!\!dk~ 
\frac{ {\rm Im}(\Delta(k))\sin(m k) - {\rm Re} (\Delta(k) ) \cos(m k) }{\lambda(k)},
\allowdisplaybreaks[4] \\
\langle b^\dagger_{r,\sigma} a^\dagger_{r+m,\sigma} \rangle 
&=& \nn \\ && \hspace{-2.5cm} \frac{1}{\pi} \int_{0}^{\frac{\pi}{2}} \!\!dk~ 
\frac{ {\rm Im}(\Delta(k))\sin(m k) + {\rm Re} (\Delta(k) ) \cos(m k) }{\lambda(k)},
\eearr
\es
where $n$ and $m$ take even and odd values, respectively. The functions $t(k)$, $\Delta(k)$,
and $\lambda(k)$ are defined as
\bs
\bearr
t(k)\!&=&\!t_0 + 2\sum_{n=2,4,\cdots} t_{n} \cos(n k), 
\label{eq:MF_disp_hop} \allowdisplaybreaks[4] \\
\Delta(k)\!&=&\! \hspace{-0.3cm}\sum_{m=1,3,\cdots} \hspace{-0.3cm} 
\left( (\Delta^A_{m}\!-\!\Delta^B_{m})\cos(m k)
\!-\!i (\Delta^A_{m}\!+\!\Delta^B_{m})\sin(m k) \right),
\label{eq:MF_disp_Bog} \nn \\ \allowdisplaybreaks[4] \\
\lambda(k)\!&=&\! \sqrt{t^2(k)+|\Delta(k)|^2}.
\label{eq:MF_disp_1qp}
\eearr
\label{eq:MF_disps}
\es

Once the parameters $t$ and $U$ are specified, the mean field equations~\reqn{eq:MF_eq}
have to be solved self-consistently for the expectation values~\reqn{eq:MF_EVs}. The results
are show in Fig.~\ref{fig:MF}. The left panel displays the expectation values of the local density operator,
of the NN Bogoliubov term, and of the NNN hopping term.
For $U \le 1.072\delta$, all the expectation values are zero;
they continuously increase from zero for $U \ge 1.072\delta$. This critical Hubbard interaction 
$U_c=1.072\delta$  is precisely the value we found in our study of the 
BI phase in the previous sections based on the deepCUT $D:0+g:2$ in order 12,
see Fig.~\ref{fig:exciton_gap}. This demonstrates the overall consistency of the approach used.

The two NN Bogoliubov terms in the unit cell are related to each other by a minus sign
\be
\langle a^\dagger_{r,\sigma} b^\dagger_{r+1,\sigma} \rangle 
=-\langle b^\dagger_{r+1,\sigma} a^\dagger_{r+2,\sigma} \rangle; \quad r \in {\rm even}.
\ee
Two equivalent solutions are possible corresponding to the two ground states. In 
one of them $\langle a^\dagger_{r,\sigma} b^\dagger_{r+1,\sigma} \rangle >0$ holds and in the 
other $\langle a^\dagger_{r,\sigma} b^\dagger_{r+1,\sigma} \rangle <0$. It is seen 
from the left panel of Fig.~\ref{fig:MF} that the expectation 
values of the local density and of the NNN hopping term are 
close to each other and behave linearly in the vicinity of the transition
point. The NN Bogoliubov term displays  a square root behavior
around the transition point. This square root behavior of the order parameter near
the transition point is what one expects from a mean-field theory without spatial
fluctuations, i.e., Landau theory, for transitions
from a unique ground state to a state with spontaneously broken symmetry.

We define the condensation energy  as the energy difference between the 
vacuum of ``$g$-particles'' and the mean field ground state of the system.
In the right panel of Fig.~\ref{fig:MF} the condensation energy per site 
$\Delta \epsilon_0$ and the charge gap are plotted vs.\ $U$. Of course, the condensation 
energy is zero in the BI phase and becomes finite when the condensation starts. 

The mean field analysis shows that the charge gap starts to increase as soon as the transition
has taken place. The behavior of the charge gap beyond the first transition point $U_{c1}$ has been
discussed controversially in previous studies. 
Lou {\it et~al.} \cite{Lou03} concluded by extrapolating DMRG results to infinite chain length that
the charge gap continues to decrease beyond $U_{c1}$ up to the second transition
point $U_{c2}$. At this second transition point both charge and spin gaps vanish and
for $U> U_{c2}$ the charge gap starts to increase while the 
spin gap remains zero \cite{Lou03}. The DMRG method employed
by other groups, however, show that the charge gap starts to 
increase just from the first transition point on \cite{Manmana04,Takada01,Kampf03}.
Our findings clearly support the latter scenario.

Because the IHM can be mapped to the Heisenberg model in the limit $U-\Delta \gg t$,
we expect a MI phase in the large $U$ limit with a vanishing spin gap.
However, the effective Hamiltonian analysed on the mean field level shows no evidence for a second
transition to the MI phase. But there is 
strong evidence for a second transition to the MI phase  obtained 
by field theoretical approach \cite{Fabrizio99,Torio01} and by DMRG 
 \cite{Manmana04,Takada01,Lou03,Zhang03,Tincani09,Legeza06} even though
 it appears to be difficult to determine it unambiguously \cite{Wilkens01,Kampf03}.

The question arises why we do not see any evidence for the transition SDI to MI.
From the employed approach two sources are conceivable.
The first source consists in errors in the mapping 
of the IHM to the effective Hamiltonian using $D\!:\!0+g\!:\!2$. 
We have already seen that this effective Hamiltonian includes some inaccuracies. 
This is seen, for instance, in the charge gap
calculated from the effective Hamiltonian and compared to  DMRG results
in the left panel of Fig.~\ref{fig:gap_deepCUT}. But this is only
a quantitative discrepancy which can explain quantitative deviations
and it is unlikely that the qualitative aspect of a mechanism driving
the system from the SDI to the MI is completely missed.

The second source arises from the analysis of the effective Hamiltonian.
The mean field analysis can capture the essential aspects of the gaps
of \emph{single} fermionic excitations, but it is not powerful enough to provide information
about binding phenomena. The physics of the MI is characterized by the
massless excitations of a generalized Heisenberg model. In higher dimensions
it would display magnetic long-range order. In 1D this order is reduced
by quantum fluctuations to a quasi-long-range order with power law decay.
Still we expect that the transition SDI to MI is driven by the
softening of a magnetic $S=1$ excitation, i.e., a triplon. The condensation
of such a triplon would indicate the transition to a phase dominated
by magnetic fluctuations or with magnetic long-range order \cite{Sachdev1990}.

In terms of fermions, the triplon is an exciton with $S=1$, in contrast
to the exciton with $S=0$ which signalled the BI to SDI transition.
Thus we conclude that the second transition $U_{c2}$ can only be 
found if the binding of $S=1$ excitons formed by two fermionic
excitations above the SDI ground state is analyzed.
This is left to future research.

\section{Conclusions}

Summarizing we studied the ionic Hubbard model (IHM) with interaction $U$
at half-filling in one dimension
in detail to understand the nature of its three phases: The band insulator (BI),
the spontaneously dimerized insulator (SDI), and the Mott insulator (MI).
We employed the recently developed deepCUT approach \cite{Krull12} to 
derive an effective Hamiltonian in a systematically controlled fashion.
The obtained effective Hamiltonian describes the physics in terms of
the elementary excitations of the correlated BI, i.e., dressed electrons and holes
in an empty and a filled band, respectively.

We quantitatively determined the dispersion of single fermionic excitations 
(quasiparticles, QP) in the whole Brillouin zone in the BI phase 
almost up to the first transition point $U_{c1}$. Very good accuracy
could be reached if the system was not too close at the transition point.
This has been established (i) by comparing the results of various orders
establishing convergence for the limiting process towards infinite order,
and (ii) by comparison of the results for the charge gap to DMRG data \cite{Manmana04}.
We emphasize that our approach has the merit to address the full dispersion,
going beyond the gap.

Technically, the use of the deepCUT is essential because the bare
perturbative series contains powers of $t/(\delta-U)$ which imply its divergence
for $U\to\delta$ from below. Thus any perturbative description necessarily
breaks down at some point $U<\delta$ while the phase transition takes place
at $U_{c1}\approx 1.07\delta$. We could show that this transition 
\emph{beyond} the point $U=\delta$ is due to a renormalization of the 
local excitation energy $\delta-U$ to positive values when
higher lying excitations are integrated out. Thus  a strict
perturbative approach fails, but the renormalizing properties of the deepCUT
manages to capture this effect properly.

Moreover, we computed the binding phenomena occurring for two
fermionic excitations. Our focus was the formation of a non-magnetic $S=0$
exciton at momentum $\pi$ which becomes soft on approaching the phase
transition to the SDI. We have computed the dispersion of this collective excitation
in the BI and also very close to the phase transition. 
To support the idea that the SDI is a phase with a condensate
of these excitons we applied a straightforward mean field theory to
the effective Hamiltonian beyond the phase transition, i.e., for $U>U_{c1}$.
Indeed, we could establish that the condensation amplitude of the exciton grows from
the phase transition at $U_{c1}$ on. This condensed phase
displays the same symmetries as the SDI phase,
namely an alternating bond strength. Thereby, a consistent picture of the
BI to SDI quantum phase transition has been provided.

Furthermore, we argued that the second transition from the SDI to the MI
is signalled by the softening of an $S=1$ exciton in the SDI phase. Its 
condensation would lead to the quasi-long-range order in the MI phase.
But the computation of this binding effect and the determination
of $U_{c2}$ was beyond the present investigation.

From the above, two possible extensions suggest themselves for
future research.
The first is to develop a quantitative transformation yielding
an effective Hamiltonian for the SDI phase. This should allow for
a determination of the softening of the collective magnetic excitation
for $U\to U_{c2}$ from below, signalling the transition to the MI phase.
Of course, such an effective Hamiltonian for the SDI has also to include the instability 
towards the BI by a softening $S=0$ excitation for $U\to U_{c1}$ from above.

Second, we recall that much less is known about the IHM in two and higher
dimensions. So an application of the presented approach is called for.
The first question is whether the BI becomes unstable towards some
modulated phase similar to the SDI. This would be seen in the softening
of an $S=0$ exciton at the corresponding wave vector.
Alternatively, it is possible that no modulated phase occurs but
that the BI becomes unstable directly towards a MI phase.
This would be signalled by the softening of an $S=1$ exciton. Another scenario would 
be that no collective bosonic excitation condenses, but that the fermionic dispersion 
becomes negative leading to a strongly correlated metallic behavior. Thus further 
research is called for to clarify these issues.

\section*{Acknowledgment}
We would like to thank K.~Coester, B.~Fauseweh, H.~Krull,
S.~R.~Manmana, and M.~D.~Schulz for fruitful discussions. 
We gratefully acknowledge financial support by the NRW-Forschungsschule
``Forschung mit Synchrotronstrahlung in den Nano- und Biowissenschaften''.

\appendix

\section{Simplification Rules}
\label{app:simpl_rules}

The algebraic part of the deepCUT method requires to keeping track of many
monomials and to calculate their commutators. The number of monomials
to be tracked can be substantially reduced if we are interested 
in sectors with only a few QPs and in processes up to a specific order $n$ in
the formal expansion parameter. For the bookkeeping  \cite{Krull12}, we define
two different orders for each monomial $A_i$.  The first is the minimal order
$O_{\rm min}(A_i)$ which is the order in which the monomial $A_i$ appears. 

The second is the maximal order $O_{\rm max}(A_i)$  which gives the order up to which 
the prefactor of the monomial $A_i$ is needed 
to describe the targeted sector up to order $n$. 
By the term ``targeted'' we simply express that it is this sector that
we want to know and to compute finally. The maximal
orders of monomials can be determined from the minimal orders and the flow
equations in an iterative way, see Ref.\  \onlinecite{Krull12}
for details and examples. Finally,
if $O_{\rm max}(A_i) < O_{\rm min}(A_i)$ holds the monomial $A_i$ has no effect on 
the targeted quantities up to order $n$ and we can discard it.

This omission of unnecessary monomials is possible only {\it after}
determining the flow equations.  
The idea of {\it simplification rules} (SRs) is to find an {\it upper bound}  $\widetilde{O}_{\rm max}$  
for the maximal order of each monomial $A_i$ {\it during} the algebraic part of the 
calculations. Then this bound  $\widetilde{O}_{\rm max}$ 
is used to discard at least some of the unnecessary monomials in the algebraic 
calculations leading to an acceleration of the algorithm and reduced memory requirements.

Here, we present two different kinds of SRs: \emph{a-posteriori} and \emph{a-priori} SRs.
They are employed in our second application of the deepCUT analysis where effective 
Hamiltonians are drived which preserve the number of fermionic QPs.

\subsection{The \emph{a-posteriori} Simplification Rules}

The a-posteriori SRs are applied {\it after} the calculation of 
each commutator. They check whether a monomial can be 
discarded or not. In the sequel, it is assumed that the order of calculations is $n$.
First, we discuss the simplifications if the sector with zero QPs is targeted,
i.e., the ground state because this is the simplest case.
But we also discuss what is necessary to target sectors with $q$ QPs.

For an upper bound to the maximal 
order of the monomial $A$, let us assume that $c_{\sigma}$ and $a_{\sigma}$ 
are the number of creation and annihilation operators with spin 
$\sigma$ which occur in $A_i$. We explain the idea for the creation operators. 
The annihilation operators can be treated in the same way.

The ground state energy is just a number so that its corresponding
operator is the identity $\mathds{1}$. For the monomial $A_i$ 
to influence the ground state energy, all creation
operators have to be cancelled in the commutation process.
The generator $\eta$ comprises the monomials 
\be
\eta_{\rm eff}^{(1)} = \sum_{i,\sigma} 
\left( g^{\dagger}_{i,\sigma}g^{\dagger}_{i+1,\sigma}+ {\rm h.c.} \right)
\label{eq:FO_gen}
\ee
in first order. In commutation, this generator term can compensate two creation or two annihilation 
operators with the {\it same} spin. This is the key observation for the SR.
We point out  that the higher order terms in the generator may be able to compensate 
more than two operators, but the ratio between the number of compensated operators
and the minimal order of the generator term is always equal or less than $2$.
Thus it is sufficient to consider just the first order term of the 
generator in our analysis \cite{Krull12}. The minimal number of commutations 
needed to cancel all the creation operators reads 
\be
K_0^c = \sum_\sigma \Big{\lceil} \frac{c_\sigma}{2} \Big{\rceil}
\label{eq:basic_K0c}
\ee
where the ceiling brackets stand for the smallest integer larger than
the argument.
A lower number of commutations is necessary if sectors with more QPs
are targeted. If we want to target $q$ QPs we denote the required
minimal number of commutations by $K_q^c$.
The least number of commutations are required if these operators are chosen
from monomials with an odd number of operators. 
In this way, one can reach the sector with $q$ quasiparticles by
a minimum of
\be
K_q^c ={\max}\left( K_0^c - d_c - \left\lfloor \frac{q-d_c}{2} \right\rfloor, 0 \right)
\label{eq:Kcq}
\ee
commutations with $d_c := {\rm min}(q,\alpha_c)$.  The floor brackets stand
for the largest integer smaller than the argument.
The parameter $\alpha_c$ is zero if both the numbers of creation operators
with spin up and with spin down are even; it is one if one of them is even
and the other odd; it is two if both of them are odd.
Analogously, $K_q^a$ for the annihilation part is defined.

Because each commutation with the generator~\reqn{eq:FO_gen} increases
the order by one, we deduce from the above considerations the upper bound 
\be
\widetilde{O}_{\rm max}(A) = n - K_q^c - K_q^a
\label{eq:omax}
\ee
for the maximal order of the monomial~$A$.
The monomial $A$ is safely omitted if $\widetilde{O}_{\rm max}(A) < O_{\rm min}(A)$.
We refer to the described analysis for the maximal order as {\it basic a-posteriori SR}.

The above upper bound of the maximal order can be reduced further by 
considering the structure of the generator terms on the lattice. 
The term \reqn{eq:FO_gen} contains two creation or annihilation operators with
the same spin only on {\it adjacent} sites. This means that the compensation of 
two operators which do not act on neighboring sites needs at least two commutations with 
leading to an increase by two in the maximal order.

We point out that in the generator there are also other terms with
extended structure in real space, but they occur in higher minimal orders  \cite{Krull12} so that 
it is sufficient to focus on the first order term \reqn{eq:FO_gen}.    
To exploit this structural aspects on the lattice, the clusters of creation and 
annihilation operators with spin $\sigma$ are divided into different {\it linked} subclusters.
We denote the number of creation and annihilation operators with spin $\sigma$ in the subcluster 
labelled by $i$ by $k_i^{c,\sigma}$ and $k_i^{a,\sigma}$, respectively \cite{Krull12}. 
The number of commutations with \reqn{eq:FO_gen} needed to compensate all the creation operators reads
\be
K_0^c = \sum_{i,\sigma} \left\lceil \frac{k_i^{c,\sigma}}{2} \right\rceil.
\label{eq:extended_K0c}
\ee
This equation extends \reqn{eq:basic_K0c} by considering the 
real space structure of the monomials. In full analogy  to the basic a-posteriori SR, 
the relation~\reqn{eq:extended_K0c} can be generalized to $K_q^c$ if the sectors with  $q$ QPs 
are targeted.  In order to minimize $K_q^c$, the $q$ operators are taken at first from
 subclusters with odd number of sites saving one commutation for each operator. Then the remaining 
operators are taken from even subclusters which needs at least two operators to 
save one commutation. Eventually, we obtain
\be
K_q^c ={\max}\left( K_0^c - d_c - \left\lfloor \frac{q-d_c}{2} \right\rfloor, 0 \right)
\label{eq:extended_Kqc}
\ee
where $d_c := {\rm min}(q,\alpha_c)$ and $\alpha_c$ is the number of odd-size 
linked subclusters present in {\it both} spin up and spin down creation clusters. Similarly,
one can find the corresponding relation for annihilation yielding $K_q^a$.
Replacing them for $K_q^c$ and $K_q^a$ in Eq.~\reqn{eq:omax} 
leads to a maximal order which is lower than the estimate of the basic a-posteriori 
SR. This improved analysis for the maximal order which takes into account the real space 
structure of the monomials is called {\it extended a-posteriori SR}.

\subsection{The \emph{a-priori} Simplification Rules}

The a-priori SRs are applied {\it before} commutators are computed explicitly. 
Thus they allow for an additional speed-up.
This type of SRs checks whether the result of the commutator $\left[ T,D \right]=TD-DT$ 
leads to any monomial which can pass the a-posteriori SRs or not. 
Here $T$ stands for any monomial from the generator and $D$ for any monomial from the Hamiltonian.
If all the monomials which may ensue from the studied commutator are
unnecessary, one can ignore this commutator improving the computational speed. 
Two different basic and extended a-priori SRs can be defined 
corresponding to the basic and extended a-posteriori SRs.

In the basic a-priori SR, the minimal number of creation and annihilation operators 
with spin $\sigma$ which result from the products $TD$ and $DT$ are
estimated separately. Then the basic a-posteriori SR is employed to obtain an upper bound
for the maximal orders of $TD$ and $DT$. We explain the method for $TD$; the product $DT$ can 
be analyzed in the same way.

Let $a_T^\sigma$ and $c_T^\sigma$ be the numbers of
creation and annihilation operators with spin $\sigma$ in the monomial~$T$.
Similarly, we define $a_D^\sigma$ and $c_D^\sigma$ for the monomial~$D$. 
Next, the product $TD$ is normal-ordered,  the creation operators are sorted left to the
annihilation operators by appropriate commuations.
In the course of this normal-ordering, the number  $s^\sigma_{TD} := \min(a_T^\sigma,c_D^\sigma)$ 
from the creation and annihilation parts with spin~$\sigma$ may cancel at maximum. 
Therefore, the minimal number 
of creation and annihilation operators of the normal-ordered product~$TD$ is given by
\bearr
c_{TD}^\sigma &=& c_T^\sigma + c_D^\sigma - s^\sigma_{TD}, \\
a_{TD}^\sigma &=& a_D^\sigma + a_T^\sigma - s^\sigma_{TD}.
\eearr
Using these estimates, one can find the upper bound for the maximal order of 
the normal-ordered product~$TD$. Eventually, we conclude that the commutator 
$\left[ T,D \right]$ can be ignored if
\be
\max \left(\widetilde{O}_{\max}(TD),\widetilde{O}_{\max}(DT) \right) < 
O_{\min}(T) + O_{\min}(D).
\label{eq:apriori_cri}
\ee

In the extended a-priori SR, similar to the extended a-posteriori SR, the lattice
structure of monomials is considered as well. Using this property, one can
manage to identify more unnecessary commutators before computing them. 
Again, we describe the method for $TD$; $DT$ is treated in the same fashion.
All we have to do is to evaluate the clusters of creation and annihilation operators 
of the normal-ordered product~$TD$ for up and down spins. 
Then, the method uses the extended a-posteriori SR to find the maximal order 
of $TD$ based on the estimated clusters.

Because the operator algebra is local, only 
operators acting on the same sites can cancel in the normal-ordering. 
This means that the spin-$\sigma$ creation operators of $T$ and the spin-$\sigma$ 
annihilation operators of $D$ which are elements of the set
\be
\mathbf{S}_{TD}^\sigma := \mathbf{A}_T^\sigma \cap \mathbf{C}_D^\sigma
\ee
may cancel in the normal-ordering. The sets $\mathbf{A}_X^\sigma$ and 
$\mathbf{C}_X^\sigma$ denote the spin-$\sigma$ creation and annihilation clusters 
of operator $X$. Hence, the creation and annihilation clusters of $TD$ are 
given by
\bs
\bearr
\mathbf{C}_{TD}^\sigma &=& \mathbf{C}_{T}^\sigma \cup 
\left( \mathbf{C}_{D}^\sigma \setminus \mathbf{S}_{TD}^\sigma \right), \\
\mathbf{A}_{TD}^\sigma &=& \mathbf{A}_{D}^\sigma \cup 
\left( \mathbf{A}_{T}^\sigma \setminus \mathbf{S}_{TD}^\sigma \right).
\eearr
\es
Similar results can be obtained for the product $DT$. 
Using the extended a-posteriori SR, one obtains an upper bound for 
the maximal orders of $TD$ and $DT$. The commutator $\left[T,D\right]$ is ignored
finally if \reqn{eq:apriori_cri} is fulfilled.

Although the extended a-priori SR can cancel more commutators compared to
the basic a-priori SR, it has a caveat. In contrast to the basic a-priori SR, the 
extended version has to be applied individually to each element of the translation
symmetry group, which is computationally expensive. 
Therefore, in order to reach the highest efficiency we use a combination
of these a-priori SRs in practice \cite{Krull12}.

\onecolumngrid

\newpage

\section*{Supplemental Material}
\setcounter{subsection}{0}

\subsection{Matrix Elements of the Effective Hamiltonian}
In this section, the action of the various parts of the Hamiltonian~\reqn{eq:ED_ham} 
on the states~\reqn{eq:ED_states} is presented. For the sake of completeness, the effects
of $H_{2:1}$, $H_{3:2}$, and $H_{4:3}$, which are absent in a pure fermionic problem, are also included.
The results are obtained for a pure fermionic problem. However, they are in priciple
applicable to either a bosonic or a fermionic algebra or 
even to an algebra composed of both fermionic and bosonic operators. The user has 
to take care only about the fermionic sign factors which appear in front of the prefactors.

\subsubsection{$H_{1:1}$}

The action of the operator $H_{1:1}$ on the one-QP state $\ket{K}$
is given by
\be
H_{1:1} \ket{K, \sigma} = \sum_{n} \sum_{\beta} e^{iKn}
\prescript{\beta \!\!}{\sigma \!\!}{\Big{[} \mathcal{C}_{1}^{1} }{\Big{]} }^{\!n}
\ket{K,\beta}
\ee
with the definition
\be
\prescript{\beta \!\!}{\sigma \!\!}{\Big{[} \mathcal{C}_{1}^{1}}{\Big{]} }^{\!n}
=\bra{r\!-\!n,\beta} H_{1:1} \ket{r, \sigma}.
\ee
For the action of $H_{1:1}$ on the two-QP state $\ket{K, \sigma_1 ; d ,\sigma_2}$, we find
\bearr
H_{1:1} \ket{K, \sigma_1 ; d ,\sigma_2} = &+& \sum_{n>-d} \sum_{\beta_1} e^{iK\frac{n}{2}} 
\prescript{\beta_1 \!\!}{\sigma_1 \!\!}{\Big{[} \mathcal{C}_{1}^{1}}{\Big{]} }^{\!n}
\ket{K, \beta_1 ; +d+n ,\sigma_2} \nn \allowdisplaybreaks[4] \\
&-& \sum_{n<-d} \sum_{\beta_1} e^{iK\frac{n}{2}} 
\prescript{\beta_1 \!\!}{\sigma_1 \!\!}{\Big{[} \mathcal{C}_{1}^{1}}{\Big{]} }^{\!n}
\ket{K, \sigma_2 ; -d-n ,\beta_1} \nn \allowdisplaybreaks[4] \\
&+& \sum_{n<d} \sum_{\beta_2} e^{iK\frac{n}{2}} 
\prescript{\beta_2 \!\!}{\sigma_2 \!\!}{\Big{[} \mathcal{C}_{1}^{1}}{\Big{]} }^{\!n}
\ket{K, \sigma_1 ; +d-n ,\beta_2} \nn \allowdisplaybreaks[4] \\
&-& \sum_{n>d} \sum_{\beta_2} e^{iK\frac{n}{2}} 
\prescript{\beta_2 \!\!}{\sigma_2 \!\!}{\Big{[} \mathcal{C}_{1}^{1}}{\Big{]} }^{\!n}
\ket{K, \beta_2 ; -d+n ,\sigma_1}.
\eearr
The action of $H_{1:1}$ on the three-QP state $\ket{K, \sigma_1 ; d_1 ,\sigma_2 ; d_2 ,\sigma_3}$ 
leads to $9$ contributions given below
\bearr
\hspace{-1cm} H_{1:1} \ket{K, \sigma_1 ; d_1 ,\sigma_2 ; d_2 ,\sigma_3} =
&+& \sum_{\phantom{(d_1)<}n>-d_1\phantom{-+d_2}} \sum_{\beta_1} e^{iK\frac{n}{3}} 
\prescript{\beta_1 \!\!}{\sigma_1 \!\!}{\Big{[} \mathcal{C}_{1}^{1}}{\Big{]} }^{\!n}
\ket{K, \beta_1 ; +d_1+n ,\sigma_2 ; +d_2 ,\sigma_3} \nn \allowdisplaybreaks[4] \\
&-& \sum_{-(d_1+d_2)<n<-d_1} \sum_{\beta_1} e^{iK\frac{n}{3}} 
\prescript{\beta_1 \!\!}{\sigma_1 \!\!}{\Big{[} \mathcal{C}_{1}^{1}}{\Big{]} }^{\!n}
\ket{K, \sigma_2 ; -d_1-n ,\beta_1 ; d_1+d_2+n ,\sigma_3}  \nn \allowdisplaybreaks[4] \\
&+& \sum_{\phantom{d_1}n<-(d_1+d_2)\phantom{->}} \sum_{\beta_1} e^{iK\frac{n}{3}} 
\prescript{\beta_1 \!\!}{\sigma_1 \!\!}{\Big{[} \mathcal{C}_{1}^{1}}{\Big{]} }^{\!n}
\ket{K, \sigma_2 ; +d_2 ,\sigma_3 ; -d_1-d_2-n ,\beta_1} \nn \allowdisplaybreaks[4] \\
&+& \sum_{\phantom{d_1}-d_2<n<d_1\phantom{(+-)}} \sum_{\beta_2} e^{iK\frac{n}{3}} 
\prescript{\beta_2 \!\!}{\sigma_2 \!\!}{\Big{[} \mathcal{C}_{1}^{1}}{\Big{]} }^{\!n}
\ket{K, \sigma_1 ; d_1-n ,\beta_2 ; d_2+n ,\sigma_3} \nn \allowdisplaybreaks[4] \\
&-& \sum_{\phantom{(d_1)<}n<-d_2\phantom{-+d_2}} \sum_{\beta_2} e^{iK\frac{n}{3}} 
\prescript{\beta_2 \!\!}{\sigma_2 \!\!}{\Big{[} \mathcal{C}_{1}^{1}}{\Big{]} }^{\!n}
\ket{K, \sigma_1 ; d_1+d_2 ,\sigma_3 ; -d_2-n ,\beta_2} \nn \allowdisplaybreaks[4]  \\
&-& \sum_{\phantom{(-d_1)<}n>d_1\phantom{-+d_2}} \sum_{\beta_2} e^{iK\frac{n}{3}} 
\prescript{\beta_2 \!\!}{\sigma_2 \!\!}{\Big{[} \mathcal{C}_{1}^{1}}{\Big{]} }^{\!n}
\ket{K, \beta_2 ; -d_1+n ,\sigma_1 ; d_1+d_2 ,\sigma_3} \nn  \allowdisplaybreaks[4] \\
&+& \sum_{\phantom{(-d_1)<}n<d_2\phantom{-+d_2}} \sum_{\beta_3} e^{iK\frac{n}{3}} 
\prescript{\beta_3 \!\!}{\sigma_3 \!\!}{\Big{[} \mathcal{C}_{1}^{1}}{\Big{]} }^{\!n}
\ket{K, \sigma_1 ; d_1 ,\sigma_2 ; d_2-n ,\beta_3} \nn \allowdisplaybreaks[4] \\
&-& \sum_{\phantom{--}d_2<n<(d_1+d_2)} \sum_{\beta_3} e^{iK\frac{n}{3}} 
\prescript{\beta_3 \!\!}{\sigma_3 \!\!}{\Big{[} \mathcal{C}_{1}^{1}}{\Big{]} }^{\!n}
\ket{K, \sigma_1 ; d_1+d_2-n ,\beta_3 ; -d_2+n ,\sigma_2} \nn \allowdisplaybreaks[4] \\
&+& \sum_{\phantom{(-<)}n>d_1+d_2\phantom{-d_1}} \sum_{\beta_3} e^{iK\frac{n}{3}} 
\prescript{\beta_3 \!\!}{\sigma_3 \!\!}{\Big{[} \mathcal{C}_{1}^{1}}{\Big{]} }^{\!n}
\ket{K, \beta_3 ; -d_1-d_2+n ,\sigma_1 ; d_1 ,\sigma_2}.
\eearr
Finally, the application of $H_{1:1}$ on the four-QP state 
$\ket{K, \sigma_1 ; d_1 ,\sigma_2 ; d_2 ,\sigma_3 ; d_3 ,\sigma_4}$ reads
\bearr
\hspace{-1cm} H_{1:1} \ket{K, \sigma_1 ; d_1 ,\sigma_2 ; d_2 ,\sigma_3 ; d_3 ,\sigma_4} = 
&+& \sum_{\phantom{_{23}-}-d_{1}<n\phantom{<d_{12}} } \sum_{\beta_1} e^{iK\frac{n}{4}} 
\prescript{\beta_1 \!\!}{\sigma_1 \!\!}{\Big{[} \mathcal{C}_{1}^{1}}{\Big{]} }^{\!n}
\ket{K, \beta_1 ; +d_1+n ,\sigma_2 ; +d_2 ,\sigma_3 ; +d_3 ,\sigma_4} \nn \allowdisplaybreaks[4] \\
&-& \sum_{\phantom{_{3}}-d_{12}<n<-d_{1\phantom2} } \sum_{\beta_1} e^{iK\frac{n}{4}} 
\prescript{\beta_1 \!\!}{\sigma_1 \!\!}{\Big{[} \mathcal{C}_{1}^{1}}{\Big{]} }^{\!n}
\ket{K, \sigma_2 ; -d_1-n ,\beta_1 ; d_{12}+n ,\sigma_3 ; d_3 ,\sigma_4}  \nn \allowdisplaybreaks[4] \\
&+& \sum_{-d_{123} < n< -d_{12} } \sum_{\beta_1} e^{iK\frac{n}{4}} 
\prescript{\beta_1 \!\!}{\sigma_1 \!\!}{\Big{[} \mathcal{C}_{1}^{1}}{\Big{]} }^{\!n}
\ket{K, \sigma_2 ; d_2 ,\sigma_3 ; -d_{12}-n ,\beta_1 ; n+d_{123} ,\sigma_4} \nn \allowdisplaybreaks[4] \\
&-& \sum_{\phantom{-d_{12}}n< -d_{123}\phantom{<} } \sum_{\beta_1} e^{iK\frac{n}{4}} 
\prescript{\beta_1 \!\!}{\sigma_1 \!\!}{\Big{[} \mathcal{C}_{1}^{1}}{\Big{]} }^{\!n}
\ket{K, \sigma_2 ; d_2 ,\sigma_3 ; d_{3},\sigma_4 ; -n-d_{123} ,\beta_1} \nn \allowdisplaybreaks[4] \\
&-& \sum_{\phantom{-_{23}-}d_{1} <n \phantom{<d_{12}} } \sum_{\beta_2} e^{iK\frac{n}{4}} 
\prescript{\beta_2 \!\!}{\sigma_2 \!\!}{\Big{[} \mathcal{C}_{1}^{1}}{\Big{]} }^{\!n}
\ket{K, \beta_2 ; n-d_1 ,\sigma_1 ; d_{12},\sigma_3 ; d_{3} ,\sigma_4} \nn \allowdisplaybreaks[4] \\
&+& \sum_{\phantom{_{23}}-d_{2} <n <d_1{\phantom{_{2}-} } } \sum_{\beta_2} e^{iK\frac{n}{4}} 
\prescript{\beta_2 \!\!}{\sigma_2 \!\!}{\Big{[} \mathcal{C}_{1}^{1}}{\Big{]} }^{\!n}
\ket{K, \sigma_1 ; d_1-n ,\beta_2 ; d_{2}+n,\sigma_3 ; d_{3} ,\sigma_4} \nn \allowdisplaybreaks[4] \\
&-& \sum_{\phantom{_{1}}-d_{23} <n <-d_2{\phantom{_{1}} } } \sum_{\beta_2} e^{iK\frac{n}{4}} 
\prescript{\beta_2 \!\!}{\sigma_2 \!\!}{\Big{[} \mathcal{C}_{1}^{1}}{\Big{]} }^{\!n}
\ket{K, \sigma_1 ; d_{12} ,\sigma_3 ; -d_{2}-n,\beta_2 ; d_{23}+n ,\sigma_4} \nn \allowdisplaybreaks[4] \\
&+& \sum_{\phantom{d_{123}} n <-d_{23}\phantom{-<} } \sum_{\beta_2} e^{iK\frac{n}{4}} 
\prescript{\beta_2 \!\!}{\sigma_2 \!\!}{\Big{[} \mathcal{C}_{1}^{1}}{\Big{]} }^{\!n}
\ket{K, \sigma_1 ; d_{12} ,\sigma_3 ; d_{3},\sigma_4 ; -d_{23}-n ,\beta_2} \nn \allowdisplaybreaks[4] \\
&+& \sum_{\phantom{-_{3}-}d_{12} <n \phantom{<d_{23}} } \sum_{\beta_3} e^{iK\frac{n}{4}} 
\prescript{\beta_3 \!\!}{\sigma_3 \!\!}{\Big{[} \mathcal{C}_{1}^{1}}{\Big{]} }^{\!n}
\ket{K, \beta_3 ; n-d_{12} ,\sigma_1 ; d_{1},\sigma_2 ; d_{23} ,\sigma_4} \nn \allowdisplaybreaks[4] \\
&-& \sum_{\phantom{-_{13}}d_{2} <n< d_{12}\phantom{-} } \sum_{\beta_3} e^{iK\frac{n}{4}} 
\prescript{\beta_3 \!\!}{\sigma_3 \!\!}{\Big{[} \mathcal{C}_{1}^{1}}{\Big{]} }^{\!n}
\ket{K, \sigma_1 ; d_{12}-n ,\beta_3 ; n-d_{2},\sigma_2 ; d_{23} ,\sigma_4} \nn \allowdisplaybreaks[4] \\
&+& \sum_{\phantom{_{13}}-d_{3} <n< d_{2}\phantom{-_{1}} } \sum_{\beta_3} e^{iK\frac{n}{4}} 
\prescript{\beta_3 \!\!}{\sigma_3 \!\!}{\Big{[} \mathcal{C}_{1}^{1}}{\Big{]} }^{\!n}
\ket{K, \sigma_1 ; d_{1} ,\sigma_2 ; d_{2}-n,\beta_3 ; d_{3}+n ,\sigma_4} \nn \allowdisplaybreaks[4] \\
&-& \sum_{\phantom{d_{13}<}n< -d_{3}\phantom{_{12}-} } \sum_{\beta_3} e^{iK\frac{n}{4}} 
\prescript{\beta_3 \!\!}{\sigma_3 \!\!}{\Big{[} \mathcal{C}_{1}^{1}}{\Big{]} }^{\!n}
\ket{K, \sigma_1 ; d_{1} ,\sigma_2 ; d_{23},\sigma_4 ; -d_{3}-n ,\beta_3} \nn \allowdisplaybreaks[4] \\
&-& \sum_{\phantom{-<}d_{123}<n\phantom{-d_{31} }} \sum_{\beta_4} e^{iK\frac{n}{4}} 
\prescript{\beta_4 \!\!}{\sigma_4 \!\!}{\Big{[} \mathcal{C}_{1}^{1}}{\Big{]} }^{\!n}
\ket{K, \beta_4 ; n-d_{123} ,\sigma_1 ; d_{1},\sigma_2 ; d_{2} ,\sigma_3} \nn \allowdisplaybreaks[4] \\
&+& \sum_{\phantom{-}d_{23}<n<d_{123}\phantom{-} } \sum_{\beta_4} e^{iK\frac{n}{4}} 
\prescript{\beta_4 \!\!}{\sigma_4 \!\!}{\Big{[} \mathcal{C}_{1}^{1}}{\Big{]} }^{\!n}
\ket{K, \sigma_1 ; d_{123}-n ,\beta_4 ; n-d_{23},\sigma_2 ; d_{2} ,\sigma_3} \nn \allowdisplaybreaks[4] \\
&-& \sum_{\phantom{_{2}-}d_{3}<n<d_{23}\phantom{_{1}-} } \sum_{\beta_4} e^{iK\frac{n}{4}} 
\prescript{\beta_4 \!\!}{\sigma_4 \!\!}{\Big{[} \mathcal{C}_{1}^{1}}{\Big{]} }^{\!n}
\ket{K, \sigma_1 ; d_{1} ,\sigma_2 ; d_{23}-n,\beta_4 ; n-d_{3} ,\sigma_3} \nn \allowdisplaybreaks[4] \\
&+& \sum_{\phantom{d_{12}<}n<d_{3}\phantom{_{12}--} } \sum_{\beta_4} e^{iK\frac{n}{4}} 
\prescript{\beta_4 \!\!}{\sigma_4 \!\!}{\Big{[} \mathcal{C}_{1}^{1}}{\Big{]} }^{\!n}
\ket{K, \sigma_1 ; d_{1} ,\sigma_2 ; d_{2},\sigma_3 ; d_{3}-n ,\beta_4},
\eearr
where $d_{ij} := d_i + d_j$ and $d_{ijk} := d_i + d_j + d_k$.

\subsubsection{$H_{2:2}$}

The action of $H_{2:2}$ on the two-QP state is given by
\be
H_{2:2} \ket{K, \sigma_1 ; d ,\sigma_2} = -\sum_{n} \sum_{d' > 0} \sum_{\beta_1 \beta_2} e^{iK \left(n+\frac{d-d'}{2} \right)} 
\prescript{\beta_1 \! \beta_2 \!\!}{\sigma_1 \! \sigma_2 \!\!}{\Big{[} \mathcal{C}_{2}^{2}}{\Big{]} }^{\!n d'}_{d}
\ket{K, \beta_1 ; d' ,\beta_2},
\label{eq:SM:H22_on_2qp}
\ee
with the definition
\be
\prescript{\beta_1 \! \beta_2 \!\!}{\sigma_1 \! \sigma_2 \!\!}{\Big{[} \mathcal{C}_{2}^{2}}{\Big{]} }^{\!n d'}_{d}
= \bra{r\!-\!n,\beta_1; r\!-\!n\!+\!d',\beta_2} H_{2:2} \ket{r, \sigma_1;r+d,\sigma_2} .
\label{eq:SM:c22_def}
\ee
In order to fix the fermionic sign in Eqs.\ \reqn{eq:SM:H22_on_2qp} 
and~\reqn{eq:SM:c22_def}  we assume here and from now on
that in $H_{2:2}$ all annihilation
operators are put on the right of creation operators and annihilation and 
creation parts are site-ordered, separately.

For the action of $H_{2:2}$ on three-QP state, we obtain 
\bearr
\hspace{-1cm} H_{2:2} \ket{K, \sigma_1 ; d_1 ,\sigma_2 ; d_2 ,\sigma_3} &=& 
\sum_{d'>0} \Big{\{} \nn \allowdisplaybreaks[4] \\
&&- \sum_{\phantom{d_{12}<} n< -d_{12}\phantom{d'-} } \sum_{\beta_1 \beta_2} 
e^{iK \left(\frac{2n+d_1-d'}{3} \right)} 
\prescript{\beta_1 \! \beta_2 \!\!}{\sigma_1 \! \sigma_2 \!\!}{\Big{[} \mathcal{C}_{2}^{2}}{\Big{]} }^{\!n d'}_{d_1}
\ket{K, \sigma_3 ; -n-d_{12} ,\beta_1 ; d' ,\beta_2} \nn \allowdisplaybreaks[4] \\  
&&- \sum_{\phantom{<-}d'-d_{12}< n \phantom{d_{12}} } \sum_{\beta_1 \beta_2}
e^{iK \left(\frac{2n+d_1-d'}{3} \right)} 
\prescript{\beta_1 \! \beta_2 \!\!}{\sigma_1 \! \sigma_2 \!\!}{\Big{[} \mathcal{C}_{2}^{2}}{\Big{]} }^{\!n d'}_{d_1}
\ket{K, \beta_1 ; d' ,\beta_2 ; n+d_{12}-d' ,\sigma_3} \nn \allowdisplaybreaks[4] \\ 
&&+ \sum_{-d_{12}<n< d'-d_{12} } \sum_{\beta_1 \beta_2} 
e^{iK \left(\frac{2n+d_1-d'}{3} \right)} 
\prescript{\beta_1 \! \beta_2 \!\!}{\sigma_1 \! \sigma_2 \!\!}{\Big{[} \mathcal{C}_{2}^{2}}{\Big{]} }^{\!n d'}_{d_1}
\ket{K, \beta_1 ; n+d_{12} ,\sigma_3 ; d'-d_{12}-n ,\beta_2} \nn \allowdisplaybreaks[4] \\ 
&&+ \sum_{\phantom{-d_{12}}n<-d_1 \phantom{_{2}d'<} } \sum_{\beta_1 \beta_3} 
e^{iK \left(\frac{2n+d_{12}-d'}{3} \right)} 
\prescript{\beta_1 \! \beta_3 \!\!}{\sigma_1 \! \sigma_3 \!\!}{\Big{[} \mathcal{C}_{2}^{2}}{\Big{]} }^{\!n d'}_{d_{12}}
\ket{K, \sigma_2 ; -n-d_{1} ,\beta_1 ; d' ,\beta_3} \nn \allowdisplaybreaks[4] \\ 
&&- \sum_{-d_{1}<n<d'-d_1 \phantom{_{12}} } \sum_{\beta_1 \beta_3} 
e^{iK \left(\frac{2n+d_{12}-d'}{3} \right)} 
\prescript{\beta_1 \! \beta_3 \!\!}{\sigma_1 \! \sigma_3 \!\!}{\Big{[} \mathcal{C}_{2}^{2}}{\Big{]} }^{\!n d'}_{d_{12}}
\ket{K, \beta_1 ; n+d_{1} ,\sigma_2 ; d'-d_1-n ,\beta_3} \nn \allowdisplaybreaks[4] \\ 
&&+ \sum_{\phantom{<d_1}d'-d_{1}<n\phantom{_{12}-} } \sum_{\beta_1 \beta_3} 
e^{iK \left(\frac{2n+d_{12}-d'}{3} \right)} 
\prescript{\beta_1 \! \beta_3 \!\!}{\sigma_1 \! \sigma_3 \!\!}{\Big{[} \mathcal{C}_{2}^{2}}{\Big{]} }^{\!n d'}_{d_{12}}
\ket{K, \beta_1 ; d' ,\beta_3 ; n+d_1-d' ,\sigma_2} \nn \allowdisplaybreaks[4] \\ 
&&+ \sum_{\phantom{_1-}d_{1}<n<d_1+d'\phantom{_{1}} } \sum_{\beta_2 \beta_3} 
e^{iK \left(\frac{2n+d_{2}-d'}{3} \right)} 
\prescript{\beta_2 \! \beta_3 \!\!}{\sigma_2 \! \sigma_3 \!\!}{\Big{[} \mathcal{C}_{2}^{2}}{\Big{]} }^{\!n d'}_{d_{2}}
\ket{K, \beta_2 ; n-d_1 ,\sigma_1 ; d_1+d'-n ,\beta_3} \nn \allowdisplaybreaks[4] \\ 
&&- \sum_{\phantom{d_{12}-}d_1+d'<n\phantom{_{1}>} } \sum_{\beta_2 \beta_3} 
e^{iK \left(\frac{2n+d_{2}-d'}{3} \right)} 
\prescript{\beta_2 \! \beta_3 \!\!}{\sigma_2 \! \sigma_3 \!\!}{\Big{[} \mathcal{C}_{2}^{2}}{\Big{]} }^{\!n d'}_{d_{2}}
\ket{K, \beta_2 ; d' ,\beta_3 ; n-d_1-d' ,\sigma_1} \nn \allowdisplaybreaks[4] \\ 
&&- \sum_{\phantom{-d_{12}}n<d_1\phantom{-d_{1}>'} } \sum_{\beta_2 \beta_3} 
e^{iK \left(\frac{2n+d_{2}-d'}{3} \right)} 
\prescript{\beta_2 \! \beta_3 \!\!}{\sigma_2 \! \sigma_3 \!\!}{\Big{[} \mathcal{C}_{2}^{2}}{\Big{]} }^{\!n d'}_{d_{2}}
\ket{K, \sigma_1 ; d_1-n ,\beta_2 ; d' ,\beta_3}
\Big{\}}.
\eearr

The result of the application of $H_{2:2}$ to the four-QP state is given by
\bearr
H_{2:2} \ket{K, \sigma_1 ; d_1 ,\sigma_2 ; d_2 ,\sigma_3 ; d_3 ,\sigma_4} &=& 
\nn \allowdisplaybreaks[4] \\ 
&&\hspace{-4cm}-\sum_{d'>0\phantom{_1}}  \sum_{\beta_1 \beta_2}
\sum_{d'-d_{12}< n} \hspace{-0.3cm}
e^{iK \left(\frac{2n+d_1-d'}{3} \right)} 
\prescript{\beta_1 \! \beta_2 \!\!}{\sigma_1 \! \sigma_2 \!\!}{\Big{[} \mathcal{C}_{2}^{2}}{\Big{]} }^{\!n d'}_{d_1}
\ket{K, \beta_1 ; d' ,\beta_2 ; n+d_{12}-d' ,\sigma_3 ; d_3 ,\sigma_4}
\nn \allowdisplaybreaks[4] \\  
&&\hspace{-4cm}+\sum_{d'>d_3}  \sum_{\beta_1 \beta_2}
\sum_{{\substack{d'-d_{123}<n \\ d'-d_{12}>n } } } \hspace*{-0.45cm}
e^{iK \left(\frac{2n+d_1-d'}{3} \right)} 
\prescript{\beta_1 \! \beta_2 \!\!}{\sigma_1 \! \sigma_2 \!\!}{\Big{[} \mathcal{C}_{2}^{2}}{\Big{]} }^{\!n d'}_{d_1}
\ket{K, \beta_1 ; n\!+\!d_{12} ,\sigma_3 ; d'\!-\!n\!-\!d_{12} ,\beta_2 ; n\!+\!d_{123}\!-\!d' ,\sigma_4}
\nn \allowdisplaybreaks[4] \\  
&&\hspace{-4cm}-\sum_{d'>d_3}  \sum_{\beta_1 \beta_2}
\sum_{{\substack{-d_{12}<n \\ d'-d_{123}>n } } } \hspace*{-0.4cm}
e^{iK \left(\frac{2n+d_1-d'}{3} \right)} 
\prescript{\beta_1 \! \beta_2 \!\!}{\sigma_1 \! \sigma_2 \!\!}{\Big{[} \mathcal{C}_{2}^{2}}{\Big{]} }^{\!n d'}_{d_1}
\ket{K, \beta_1 ; n\!+\!d_{12} ,\sigma_3 ; d_3 ,\sigma_4 ; d'\!-\!n\!-\!d_{123} ,\beta_2}
\nn \allowdisplaybreaks[4] \\  
&&\hspace{-4cm}-\sum_{d'<d_3} \sum_{\beta_1 \beta_2}
\sum_{{\substack{d'-d_{123}<n \\ -d_{12}>n } } } \hspace*{-0.4cm}
e^{iK \left(\frac{2n+d_1-d'}{3} \right)} 
\prescript{\beta_1 \! \beta_2 \!\!}{\sigma_1 \! \sigma_2 \!\!}{\Big{[} \mathcal{C}_{2}^{2}}{\Big{]} }^{\!n d'}_{d_1}
\ket{K, \sigma_3 ; -n\!-\!d_{12} ,\beta_1 ; d' ,\beta_2 ; n\!+\!d_{123}-d' ,\sigma_4}
\nn \allowdisplaybreaks[4] \\ 
&&\hspace{-4cm}+\sum_{d'<d_3} \sum_{\beta_1 \beta_2}
\sum_{{\substack{-d_{123}<n \\ d'-d_{123}>n } } } \hspace*{-0.4cm}
e^{iK \left(\frac{2n+d_1-d'}{3} \right)} 
\prescript{\beta_1 \! \beta_2 \!\!}{\sigma_1 \! \sigma_2 \!\!}{\Big{[} \mathcal{C}_{2}^{2}}{\Big{]} }^{\!n d'}_{d_1}
\ket{K, \sigma_3 ; \!-n\!-\!d_{12} ,\beta_1 ; n\!+\!d_{123} ,\sigma_4 ; d'\!-\!n\!-\!d_{123} ,\beta_2}
\nn \allowdisplaybreaks[4] \\
&&\hspace{-4cm}-\sum_{d'>0\phantom{_3}} \sum_{\beta_1 \beta_2}
\sum_{-d_{123}>n } \hspace*{-0.2cm}
e^{iK \left(\frac{2n+d_1-d'}{3} \right)} 
\prescript{\beta_1 \! \beta_2 \!\!}{\sigma_1 \! \sigma_2 \!\!}{\Big{[} \mathcal{C}_{2}^{2}}{\Big{]} }^{\!n d'}_{d_1}
\ket{K, \sigma_3 ; d_3, \sigma_4 ; -n\!-\!d_{123} ,\beta_1 ; d' ,\beta_2}
\nn \allowdisplaybreaks[4] \\
&&\hspace{-4cm}+\sum_{d'>0\phantom{_3}} \sum_{\beta_1 \beta_3}
\sum_{d'-d_{1}<n } \hspace*{-0.2cm}
e^{iK \left(\frac{2n+d_{12}-d'}{3} \right)} 
\prescript{\beta_1 \! \beta_3 \!\!}{\sigma_1 \! \sigma_3 \!\!}{\Big{[} \mathcal{C}_{2}^{2}}{\Big{]} }^{\!n d'}_{d_{12}}
\ket{K, \beta_1 ; d', \beta_3 ; d_1\!+\!n\!-\!d' ,\sigma_2 ; d_{23} ,\sigma_4}
\nn \allowdisplaybreaks[4] \\ &&\hspace{-4cm}
-\sum_{d'>0\phantom{_3}} \sum_{\beta_1 \beta_3} \hspace*{-0.15cm}
\sum_{\substack{-d_{1}<n \\ d'-d_{123}<n \\ d'-d_1 > n}} \hspace*{-0.35cm}
e^{iK \left(\frac{2n+d_{12}-d'}{3} \right)} 
\prescript{\beta_1 \! \beta_3 \!\!}{\sigma_1 \! \sigma_3 \!\!}{\Big{[} \mathcal{C}_{2}^{2}}{\Big{]} }^{\!n d'}_{d_{12}}
\ket{K, \beta_1 ; d_1\!+\!n, \sigma_2 ; d'\!-\!d_1\!-\!n ,\beta_3 ; n\!+\!d_{123}\!-\!d' ,\sigma_4}
\nn \allowdisplaybreaks[4] \\ &&\hspace{-4cm}
+\sum_{d'>0\phantom{_3}} \sum_{\beta_1 \beta_3} \hspace*{-0.15cm}
\sum_{\substack{-d_{1}<n \\ d'-d_{123} > n}} \hspace*{-0.35cm}
e^{iK \left(\frac{2n+d_{12}-d'}{3} \right)} 
\prescript{\beta_1 \! \beta_3 \!\!}{\sigma_1 \! \sigma_3 \!\!}{\Big{[} \mathcal{C}_{2}^{2}}{\Big{]} }^{\!n d'}_{d_{12}}
\ket{K, \beta_1 ; d_1\!+\!n, \sigma_2 ; d_{23} ,\sigma_4 ; d'\!-\!d_{123}\!-\!n ,\beta_3}
\nn \allowdisplaybreaks[4] \\ &&\hspace{-4cm}
+\sum_{d'>0\phantom{_3}} \sum_{\beta_1 \beta_3} \hspace*{-0.15cm}
\sum_{\substack{-d_{1}>n \\ d'-d_{123} < n}} \hspace*{-0.35cm}
e^{iK \left(\frac{2n+d_{12}-d'}{3} \right)} 
\prescript{\beta_1 \! \beta_3 \!\!}{\sigma_1 \! \sigma_3 \!\!}{\Big{[} \mathcal{C}_{2}^{2}}{\Big{]} }^{\!n d'}_{d_{12}}
\ket{K, \sigma_2 ; -d_1\!-\!n, \beta_1 ; d' ,\beta_3 ; n\!+\!d_{123}\!-\!d' ,\sigma_4}
\nn \allowdisplaybreaks[4] \\ &&\hspace{-4cm}
- \hspace*{-0.05cm} \sum_{d'<d_{23} } \sum_{\beta_1 \beta_3} \hspace*{-0.15cm}
\sum_{\substack{-d_{123}<n \\ d'-d_{123} > n}} \hspace*{-0.4cm}
e^{iK \left(\frac{2n+d_{12}-d'}{3} \right)} 
\prescript{\beta_1 \! \beta_3 \!\!}{\sigma_1 \! \sigma_3 \!\!}{\Big{[} \mathcal{C}_{2}^{2}}{\Big{]} }^{\!n d'}_{d_{12}}
\ket{K, \sigma_2 ; -d_1\!-\!n, \beta_1 ; n+d_{123} ,\sigma_4 ; d'\!-\!n\!-\!d_{123} ,\beta_3}
\nn \allowdisplaybreaks[4] \\ &&\hspace{-4cm}
+ \sum_{d'>0\phantom{_1}} \sum_{\beta_1 \beta_3} \hspace*{0cm}
\sum_{-d_{123}>n } \hspace*{-0.2cm}
e^{iK \left(\frac{2n+d_{12}-d'}{3} \right)} 
\prescript{\beta_1 \! \beta_3 \!\!}{\sigma_1 \! \sigma_3 \!\!}{\Big{[} \mathcal{C}_{2}^{2}}{\Big{]} }^{\!n d'}_{d_{12}}
\ket{K, \sigma_2 ; d_{23}, \sigma_4 ; -n-d_{123} ,\beta_1 ; d' ,\beta_3}
\nn \allowdisplaybreaks[4] \\ &&\hspace{-4cm}
- \sum_{d'>0\phantom{_1}} \sum_{\beta_1 \beta_4} \hspace*{0cm}
\sum_{d'-d_{1}<n } \hspace*{-0.2cm}
e^{iK \left(\frac{2n+d_{123}-d'}{3} \right)} 
\prescript{\beta_1 \! \beta_4 \!\!}{\sigma_1 \! \sigma_4 \!\!}{\Big{[} \mathcal{C}_{2}^{2}}{\Big{]} }^{\!n d'}_{d_{123}}
\ket{K, \beta_1 ; d', \beta_4 ; d_1\!+\!n\!-\!d' ,\sigma_2 ; d_2 ,\sigma_3}
\nn \allowdisplaybreaks[4] \\ &&\hspace{-4cm}
+\sum_{d'>d_2} \sum_{\beta_1 \beta_4} \hspace*{0cm}
\sum_{\substack{d'-d_1>n \\ d'-d_{12}<n} } \hspace*{-0.35cm}
e^{iK \left(\frac{2n+d_{123}-d'}{3} \right)} 
\prescript{\beta_1 \! \beta_4 \!\!}{\sigma_1 \! \sigma_4 \!\!}{\Big{[} \mathcal{C}_{2}^{2}}{\Big{]} }^{\!n d'}_{d_{123}}
\ket{K, \beta_1 ; d_1\!+\!n, \sigma_2 ; d'\!-\!d_1\!-\!n ,\beta_4 ; n\!+\!d_{12}\!-\!d' ,\sigma_3}
\nn \allowdisplaybreaks[4] \\ &&\hspace{-4cm}
-\sum_{d'>0\phantom{_2} } \sum_{\beta_1 \beta_4} \hspace*{0cm}
\sum_{\substack{-d_1<n \\ d'-d_{12}>n} } \hspace*{-0.35cm}
e^{iK \left(\frac{2n+d_{123}-d'}{3} \right)} 
\prescript{\beta_1 \! \beta_4 \!\!}{\sigma_1 \! \sigma_4 \!\!}{\Big{[} \mathcal{C}_{2}^{2}}{\Big{]} }^{\!n d'}_{d_{123}}
\ket{K, \beta_1 ; d_1\!+\!n, \sigma_2 ; d_2 ,\sigma_3 ; d'\!-\!n\!-\!d_{12} ,\beta_4}
\nn \allowdisplaybreaks[4] \\ &&\hspace{-4cm}
-\sum_{d'>0\phantom{_2} } \sum_{\beta_1 \beta_4} \hspace*{0cm}
\sum_{\substack{-d_1>n \\ d'-d_{12}<n} } \hspace*{-0.35cm}
e^{iK \left(\frac{2n+d_{123}-d'}{3} \right)} 
\prescript{\beta_1 \! \beta_4 \!\!}{\sigma_1 \! \sigma_4 \!\!}{\Big{[} \mathcal{C}_{2}^{2}}{\Big{]} }^{\!n d'}_{d_{123}}
\ket{K, \sigma_2 ; -d_1\!-\!n, \beta_1 ; d' ,\beta_4 ; n\!+\!d_{12}\!-\!d' ,\sigma_3}
\nn \allowdisplaybreaks[4] \\ &&\hspace{-4cm}
+\sum_{d'<d_2} \sum_{\beta_1 \beta_4} \hspace*{0cm}
\sum_{\substack{-d_{12}<n \\ d'-d_{12}>n} } \hspace*{-0.35cm}
e^{iK \left(\frac{2n+d_{123}-d'}{3} \right)} 
\prescript{\beta_1 \! \beta_4 \!\!}{\sigma_1 \! \sigma_4 \!\!}{\Big{[} \mathcal{C}_{2}^{2}}{\Big{]} }^{\!n d'}_{d_{123}}
\ket{K, \sigma_2 ; \!-d_1\!-\!n, \beta_1 ; n\!+\!d_{12} ,\sigma_3 ; d'\!-\!n\!-\!d_{12} ,\beta_4 }
\nn \allowdisplaybreaks[4] \\ &&\hspace{-4cm}
-\sum_{d'>0\phantom{_2}} \sum_{\beta_1 \beta_4} \hspace*{0cm}
\sum_{-d_{12}>n } \hspace*{-0.15cm}
e^{iK \left(\frac{2n+d_{123}-d'}{3} \right)} 
\prescript{\beta_1 \! \beta_4 \!\!}{\sigma_1 \! \sigma_4 \!\!}{\Big{[} \mathcal{C}_{2}^{2}}{\Big{]} }^{\!n d'}_{d_{123}}
\ket{K, \sigma_2 ; d_2, \sigma_3 ; -n\!-\!d_{12} ,\beta_1 ; d' ,\beta_4 }
\nn \allowdisplaybreaks[4] \\ &&\hspace{-4cm}
-\sum_{d'>0\phantom{_2}} \sum_{\beta_2 \beta_3} \hspace*{0cm}
\sum_{\substack{d_{1}>n \\  d'-d_{23}<n}} \hspace*{-0.3cm}
e^{iK \left(\frac{2n+d_{2}-d'}{3} \right)} 
\prescript{\beta_2 \! \beta_3 \!\!}{\sigma_2 \! \sigma_3 \!\!}{\Big{[} \mathcal{C}_{2}^{2}}{\Big{]} }^{\!n d'}_{d_{2}}
\ket{K, \sigma_1 ; d_1\!-\!n, \beta_2 ; d' ,\beta_3 ; n\!+\!d_{23}\!-\!d' ,\sigma_4 }
\nn \allowdisplaybreaks[4] \\ &&\hspace{-4cm}
+\sum_{d'>0\phantom{_2}} \sum_{\beta_2 \beta_3} \hspace*{0cm}
\sum_{\substack{d_{1}<n \\  d_1+d'>n \\ d'-d_{23}<n}} \hspace*{-0.3cm}
e^{iK \left(\frac{2n+d_{2}-d'}{3} \right)} 
\prescript{\beta_2 \! \beta_3 \!\!}{\sigma_2 \! \sigma_3 \!\!}{\Big{[} \mathcal{C}_{2}^{2}}{\Big{]} }^{\!n d'}_{d_{2}}
\ket{K, \beta_2 ; n\!-\!d_1, \sigma_1 ; d_1\!+\!d'\!-\!n ,\beta_3 ; n\!+\!d_{23}\!-\!d' ,\sigma_4 }
\nn \allowdisplaybreaks[4] \\ &&\hspace{-4cm}
-\sum_{d'>0\phantom{_2}} \sum_{\beta_2 \beta_3} \hspace*{-0.1cm}
\sum_{\substack{d_{1}<n  \\ d'-d_{123}>n}} \hspace*{-0.3cm}
e^{iK \left(\frac{2n+d_{2}-d'}{3} \right)} 
\prescript{\beta_2 \! \beta_3 \!\!}{\sigma_2 \! \sigma_3 \!\!}{\Big{[} \mathcal{C}_{2}^{2}}{\Big{]} }^{\!n d'}_{d_{2}}
\ket{K, \beta_2 ; n\!-\!d_1, \sigma_1 ; d_{123} ,\sigma_4 ; d'\!-\!n\!-\!d_{23} ,\beta_3 }
\nn \allowdisplaybreaks[4] \\ &&\hspace{-4cm}
-\sum_{d'>0\phantom{_2}} \sum_{\beta_2 \beta_3} \hspace*{-0cm}
\sum_{d_{1}+d'<n } \hspace*{-0.2cm}
e^{iK \left(\frac{2n+d_{2}-d'}{3} \right)} 
\prescript{\beta_2 \! \beta_3 \!\!}{\sigma_2 \! \sigma_3 \!\!}{\Big{[} \mathcal{C}_{2}^{2}}{\Big{]} }^{\!n d'}_{d_{2}}
\ket{K, \beta_2 ; d', \beta_3 ; n\!-\!d_{1}\!-\!d' ,\sigma_1 ; d_{123} ,\sigma_4 }
\nn \allowdisplaybreaks[4] \\ &&\hspace{-4cm}
+\sum_{d'>0\phantom{_2}} \sum_{\beta_2 \beta_3} \hspace*{-0cm}
\sum_{\substack{ d_{1}>n \\ -d_{23}<n \\ d'-d_{23}>n}} \hspace*{-0.25cm}
e^{iK \left(\frac{2n+d_{2}-d'}{3} \right)} 
\prescript{\beta_2 \! \beta_3 \!\!}{\sigma_2 \! \sigma_3 \!\!}{\Big{[} \mathcal{C}_{2}^{2}}{\Big{]} }^{\!n d'}_{d_{2}}
\ket{K, \sigma_1 ; d_1\!-\!n, \beta_2 ; n\!+\!d_{23} ,\sigma_4 ; d'\!-\!n\!-\!d_{23} ,\beta_3 }
\nn \allowdisplaybreaks[4] \\ &&\hspace{-4cm}
-\sum_{d'>0\phantom{_2}} \sum_{\beta_2 \beta_3} \hspace*{-0cm}
\sum_{-d_{23}>n} \hspace*{-0.15cm}
e^{iK \left(\frac{2n+d_{2}-d'}{3} \right)} 
\prescript{\beta_2 \! \beta_3 \!\!}{\sigma_2 \! \sigma_3 \!\!}{\Big{[} \mathcal{C}_{2}^{2}}{\Big{]} }^{\!n d'}_{d_{2}}
\ket{K, \sigma_1 ; d_{123} , \sigma_4 ; -n\!-\!d_{23} ,\beta_2 ; d' ,\beta_3 }
\nn \allowdisplaybreaks[4] \\ &&\hspace{-4cm}
+\sum_{d'>0\phantom{_2}} \sum_{\beta_2 \beta_4} \hspace*{-0cm}
\sum_{d_{1}+d'<n} \hspace*{-0.2cm}
e^{iK \left(\frac{2n+d_{23}-d'}{3} \right)} 
\prescript{\beta_2 \! \beta_4 \!\!}{\sigma_2 \! \sigma_4 \!\!}{\Big{[} \mathcal{C}_{2}^{2}}{\Big{]} }^{\!n d'}_{d_{23}}
\ket{K, \beta_2 ; d' , \beta_4 ; n\!-\!d_{1}\!-\!d' ,\sigma_1 ; d_{12} ,\sigma_3 }
\nn \allowdisplaybreaks[4] \\ &&\hspace{-4cm}
-\sum_{d'>0\phantom{_2}} \sum_{\beta_2 \beta_4} \hspace*{-0cm}
\sum_{\substack{d_{1}<n \\ d_{1}+d'>n \\ d'-d_{2}<n} } \hspace*{-0.2cm}
e^{iK \left(\frac{2n+d_{23}-d'}{3} \right)} 
\prescript{\beta_2 \! \beta_4 \!\!}{\sigma_2 \! \sigma_4 \!\!}{\Big{[} \mathcal{C}_{2}^{2}}{\Big{]} }^{\!n d'}_{d_{23}}
\ket{K, \beta_2 ; n\!-\!d_1 , \sigma_1 ; d_{1}\!+\!d'\!-\!n ,\beta_4 ; n\!+\!d_{2}\!-\!d' ,\sigma_3 }
\nn \allowdisplaybreaks[4] \\ &&\hspace{-4cm}
+\sum_{d'>0\phantom{_2}} \sum_{\beta_2 \beta_4} \hspace*{-0cm}
\sum_{\substack{d_{1}<n \\ d'-d_{2}>n} } \hspace*{-0.2cm}
e^{iK \left(\frac{2n+d_{23}-d'}{3} \right)} 
\prescript{\beta_2 \! \beta_4 \!\!}{\sigma_2 \! \sigma_4 \!\!}{\Big{[} \mathcal{C}_{2}^{2}}{\Big{]} }^{\!n d'}_{d_{23}}
\ket{K, \beta_2 ; n\!-\!d_1 , \sigma_1 ; d_{12} ,\sigma_3 ; d'\!-\!n\!-\!d_{2} ,\beta_4 }
\nn \allowdisplaybreaks[4] \\ &&\hspace{-4cm}
+\sum_{d'>0\phantom{_2}} \sum_{\beta_2 \beta_4} \hspace*{-0cm}
\sum_{\substack{d_{1}>n \\ d'-d_{2}<n} } \hspace*{-0.2cm}
e^{iK \left(\frac{2n+d_{23}-d'}{3} \right)} 
\prescript{\beta_2 \! \beta_4 \!\!}{\sigma_2 \! \sigma_4 \!\!}{\Big{[} \mathcal{C}_{2}^{2}}{\Big{]} }^{\!n d'}_{d_{23}}
\ket{K, \sigma_1 ; d_1\!-\!n , \beta_2 ; d' ,\beta_4 ; n\!+\!d_{2}\!-\!d' ,\sigma_3 }
\nn \allowdisplaybreaks[4] \\ &&\hspace{-4cm}
-\sum_{d'>0\phantom{_2}} \sum_{\beta_2 \beta_4} \hspace*{-0cm}
\sum_{\substack{d_{1}>n \\ -d_{2}<n \\ d'-d_{2}>n } } \hspace*{-0.2cm}
e^{iK \left(\frac{2n+d_{23}-d'}{3} \right)} 
\prescript{\beta_2 \! \beta_4 \!\!}{\sigma_2 \! \sigma_4 \!\!}{\Big{[} \mathcal{C}_{2}^{2}}{\Big{]} }^{\!n d'}_{d_{23}}
\ket{K, \sigma_1 ; d_1\!-\!n , \beta_2 ; n\!+\!d_2 ,\sigma_3 ; d'\!-\!n\!-\!d_{2} ,\beta_4 }
\nn \allowdisplaybreaks[4] \\ &&\hspace{-4cm}
+\sum_{d'>0\phantom{_2}} \sum_{\beta_2 \beta_4} \hspace*{0.1cm}
\sum_{-d_{2}>n} \hspace*{-0.1cm}
e^{iK \left(\frac{2n+d_{23}-d'}{3} \right)} 
\prescript{\beta_2 \! \beta_4 \!\!}{\sigma_2 \! \sigma_4 \!\!}{\Big{[} \mathcal{C}_{2}^{2}}{\Big{]} }^{\!n d'}_{d_{23}}
\ket{K, \sigma_1 ; d_{12} , \sigma_3 ; -n\!-\!d_2 ,\beta_2 ; d' ,\beta_4 }
\nn \allowdisplaybreaks[4] \\ &&\hspace{-4cm}
-\sum_{d'>0\phantom{_2}} \sum_{\beta_3 \beta_4} \hspace*{0.1cm}
\sum_{d'+d_{12}<n} \hspace*{-0.3cm}
e^{iK \left(\frac{2n+d_{3}-d'}{3} \right)} 
\prescript{\beta_3 \! \beta_4 \!\!}{\sigma_3 \! \sigma_4 \!\!}{\Big{[} \mathcal{C}_{2}^{2}}{\Big{]} }^{\!n d'}_{d_{3}}
\ket{K, \beta_3 ; d' , \beta_4 ; n\!-\!d_{12}\!-\!d' ,\sigma_1 ; d_1 ,\sigma_2 }
\nn \allowdisplaybreaks[4] \\ &&\hspace{-4cm}
+\sum_{d'>0\phantom{_2}} \sum_{\beta_3 \beta_4} \hspace*{0cm}
\sum_{\substack{d_{12}<n \\ d'+d_{12}>n \\ d'+d_{2}<n  } } \hspace*{-0.3cm}
e^{iK \left(\frac{2n+d_{3}-d'}{3} \right)} 
\prescript{\beta_3 \! \beta_4 \!\!}{\sigma_3 \! \sigma_4 \!\!}{\Big{[} \mathcal{C}_{2}^{2}}{\Big{]} }^{\!n d'}_{d_{3}}
\ket{K, \beta_3 ; n\!-\!d_{12} , \sigma_1 ; d_{12}\!+\!d'\!-\!n ,\beta_4 ; n\!-\!d_2\!-\!d' ,\sigma_2 }
\nn \allowdisplaybreaks[4] \\ &&\hspace{-4cm}
-\sum_{d'>0\phantom{_2}} \sum_{\beta_3 \beta_4} \hspace*{0cm}
\sum_{\substack{d_{12}<n \\ d'+d_{2}>n} } \hspace*{-0.2cm}
e^{iK \left(\frac{2n+d_{3}-d'}{3} \right)} 
\prescript{\beta_3 \! \beta_4 \!\!}{\sigma_3 \! \sigma_4 \!\!}{\Big{[} \mathcal{C}_{2}^{2}}{\Big{]} }^{\!n d'}_{d_{3}}
\ket{K, \beta_3 ; n\!-\!d_{12} , \sigma_1 ; d_{1} ,\sigma_2 ; d_2\!+\!d'\!-\!n ,\beta_4 }
\nn \allowdisplaybreaks[4] \\ &&\hspace{-4cm}
-\sum_{d'>0\phantom{_2}} \sum_{\beta_3 \beta_4} \hspace*{0cm}
\sum_{\substack{d_{12}>n \\ d'+d_{2}<n} } \hspace*{-0.2cm}
e^{iK \left(\frac{2n+d_{3}-d'}{3} \right)} 
\prescript{\beta_3 \! \beta_4 \!\!}{\sigma_3 \! \sigma_4 \!\!}{\Big{[} \mathcal{C}_{2}^{2}}{\Big{]} }^{\!n d'}_{d_{3}}
\ket{K, \sigma_1 ; d_{12}\!-\!n , \beta_3 ; d' ,\beta_4 ; n\!-\!d'\!-\!d_2 ,\sigma_2 }
\nn \allowdisplaybreaks[4] \\ &&\hspace{-4cm}
+\sum_{d'>0\phantom{_2}} \sum_{\beta_3 \beta_4} \hspace*{0cm}
\sum_{\substack{d_{12}>n \\ d_{2}<n \\ d_{2}+d'>n} } \hspace*{-0.2cm}
e^{iK \left(\frac{2n+d_{3}-d'}{3} \right)} 
\prescript{\beta_3 \! \beta_4 \!\!}{\sigma_3 \! \sigma_4 \!\!}{\Big{[} \mathcal{C}_{2}^{2}}{\Big{]} }^{\!n d'}_{d_{3}}
\ket{K, \sigma_1 ; d_{12}\!-\!n , \beta_3 ; n\!-\!d_2 ,\sigma_2 ; d_2\!+\!d'\!-\!n ,\beta_4 }
\nn \allowdisplaybreaks[4] \\ &&\hspace{-4cm}
-\sum_{d'>0\phantom{_2}} \sum_{\beta_3 \beta_4} \hspace*{0.1cm}
\sum_{d_{2}>n} \hspace*{-0cm}
e^{iK \left(\frac{2n+d_{3}-d'}{3} \right)} 
\prescript{\beta_3 \! \beta_4 \!\!}{\sigma_3 \! \sigma_4 \!\!}{\Big{[} \mathcal{C}_{2}^{2}}{\Big{]} }^{\!n d'}_{d_{3}}
\ket{K, \sigma_1 ; d_{1} , \sigma_2 ; d_2\!-\!n ,\beta_3 ; d' ,\beta_4 }.
\eearr

\subsubsection{$H_{3:3}$}

The effect of $H_{3:3}$ on the three-QP state is given by
\bearr
H_{3:3} \ket{K, \sigma_1 ; d_1 ,\sigma_2 ; d_2 ,\sigma_3} = 
-\sum_{d'_1 d'_2}  \sum_{\beta_1 \beta_2 \beta_3}
\sum_{n} \hspace{-0cm}
e^{iK \left(n+\frac{2(d_1-d'_1)+(d_2-d'_2)}{3} \right)} 
\prescript{\beta_1 \! \beta_2 \! \beta_3 \!\!}{\sigma_1 \! \sigma_2 \! \sigma_3 \!\!}
{\Big{[} \mathcal{C}_{3}^{3}}{\Big{]} }^{\!n d'_1 \!d'_2}_{\!d_1 \!d_2}
\ket{K, \beta_1 ; d'_1 ,\beta_2 ; d'_2 ,\beta_3},
\eearr
where $d'_1$ and $d'_2$ take  positive integer values. We have also defined
\bearr
\prescript{\beta_1 \! \beta_2 \! \beta_3 \!\!}{\sigma_1 \! \sigma_2 \! \sigma_3 \!\!}
{\Big{[} \mathcal{C}_{3}^{3}}{\Big{]} }^{\!n d'_1 \!d'_2}_{\!d_1 \!d_2} := 
\bra{r\!-\!n, \beta_1 ; r\!-\!n\!+\!d'_1 ,\beta_2 ; r\!-\!n\!+\!d'_{12} ,\beta_3}
H_{3:3} \ket{r, \sigma_1 ; r\!+\!d_1 ,\sigma_2 ; r\!+\!d_{12} ,\sigma_3}, \nn \\
\eearr
where $d'_{ij}=d'_i+d'_j$. 

The action of $H_{3:3}$ on the four-QP state reads
\bearr
H_{3:3} \ket{K, \sigma_1 ; d_1 ,\sigma_2 ; d_2 ,\sigma_3 ; d_3 ,\sigma_4} &=& 
\sum_{d'_1 d'_2 > 0} \Big{\{}
\nn \allowdisplaybreaks[4] \\ && \hspace{-2cm}
-\hspace{-0.4cm} \sum_{n>d'_{12}-d_{123}} \hspace{-0.5cm}
e^{iK \left(\frac{D_{33}}{4} \right)}  \sum_{\beta_1 \beta_2 \beta_3} 
\prescript{\beta_1 \! \beta_2 \! \beta_3 \!\!}{\sigma_1 \! \sigma_2 \! \sigma_3 \!\!}
{\Big{[} \mathcal{C}_{3}^{3}}{\Big{]} }^{\!n d'_1 \!d'_2}_{\!d_1 \!d_2}
\ket{K, \beta_1 ; d'_1 ,\beta_2 ; d'_2 ,\beta_3 ; n\!+\!d_{123}\!-\!d'_{12} ,\sigma_4}
\nn \allowdisplaybreaks[4] \\ && \hspace{-2cm}
+\hspace{-0.4cm} \sum_{\substack{n<d'_{12}-d_{123} \\ n>d'_{1}-d_{123}} } \hspace{-0.5cm}
e^{iK \left(\frac{D_{33}}{4} \right)}  \sum_{\beta_1 \beta_2 \beta_3} 
\prescript{\beta_1 \! \beta_2 \! \beta_3 \!\!}{\sigma_1 \! \sigma_2 \! \sigma_3 \!\!}
{\Big{[} \mathcal{C}_{3}^{3}}{\Big{]} }^{\!n d'_1 \!d'_2}_{\!d_1 \!d_2}
\ket{K, \beta_1 ; d'_1, \beta_2 ; n\!+\!d_{123}\!-\!d'_{1} ,\sigma_4 ; d'_{12}\!-\!n\!-\!d_{123} ,\beta_3}
\nn \allowdisplaybreaks[4] \\ && \hspace{-2cm}
-\hspace{-0.4cm} \sum_{\substack{n<d'_{1}-d_{123} \\ n>-d_{123}} } \hspace{-0.4cm}
e^{iK \left(\frac{D_{33}}{4} \right)}  \sum_{\beta_1 \beta_2 \beta_3} 
\prescript{\beta_1 \! \beta_2 \! \beta_3 \!\!}{\sigma_1 \! \sigma_2 \! \sigma_3 \!\!}
{\Big{[} \mathcal{C}_{3}^{3}}{\Big{]} }^{\!n d'_1 \!d'_2}_{\!d_1 \!d_2}
\ket{K, \beta_1 ; n\!+\!d_{123} ,\sigma_4 ; d'_{1}\!-\!n\!-\!d_{123} ,\beta_2 ; d'_2, \beta_3}
\nn \allowdisplaybreaks[4] \\ &&  \hspace{-2cm}
+\hspace{-0.2cm} \sum_{ n<-d_{123} } \hspace{-0.3cm}
e^{iK \left(\frac{D_{33}}{4} \right)}  \sum_{\beta_1 \beta_2 \beta_3} 
\prescript{\beta_1 \! \beta_2 \! \beta_3 \!\!}{\sigma_1 \! \sigma_2 \! \sigma_3 \!\!}
{\Big{[} \mathcal{C}_{3}^{3}}{\Big{]} }^{\!n d'_1 \!d'_2}_{\!d_1 \!d_2}
\ket{K, \sigma_4 ; -\!n\!-\!d_{123} ,\beta_1 ; d'_{1} ,\beta_2 ; d'_2, \beta_3}
\nn \allowdisplaybreaks[4] \\ && \hspace{-2cm}
+\hspace{-0.4cm} \sum_{ n>d'_{12}-d_{12} } \hspace{-0.4cm}
e^{iK \left(\frac{D_{33}+d_3}{4} \right)}  \sum_{\beta_1 \beta_2 \beta_4} 
\prescript{\beta_1 \! \beta_2 \! \beta_4 \!\!}{\sigma_1 \! \sigma_2 \! \sigma_4 \!\!}
{\Big{[} \mathcal{C}_{3}^{3}}{\Big{]} }^{\!n d'_1 \!d'_2}_{\!d_1 \!d_{23}}
\ket{K, \beta_1 ; d'_{1} ,\beta_2 ; d'_2, \beta_4 ; n\!+\!d_{12}\!-\!d'_{12} ,\sigma_3}
\nn \allowdisplaybreaks[4] \\ && \hspace{-2cm}
-\hspace{-0.4cm} \sum_{\substack{ n<d'_{12}-d_{12} \\ n>d'_{1}-d_{12}} } \hspace{-0.4cm}
e^{iK \left(\frac{D_{33}+d_3}{4} \right)} \hspace{-0.1cm}  \sum_{\beta_1 \beta_2 \beta_4} 
\prescript{\beta_1 \! \beta_2 \! \beta_4 \!\!}{\sigma_1 \! \sigma_2 \! \sigma_4 \!\!}
{\Big{[} \mathcal{C}_{3}^{3}}{\Big{]} }^{\!n d'_1 \!d'_2}_{\!d_1 \!d_{23}}
\ket{K, \beta_1 ; d'_{1} ,\beta_2 ; n\!+\!d_{12}\!-\!d'_{1} ,\sigma_3 ; d'_{12}\!-\!n\!-\!d_{12} ,\beta_4}
\nn \allowdisplaybreaks[4] \\ && \hspace{-2cm}
+\hspace{-0.4cm} \sum_{\substack{ n<d'_{1}-d_{12} \\ n>-d_{12}} } \hspace{-0.3cm}
e^{iK \left(\frac{D_{33}+d_3}{4} \right)} \hspace{-0.1cm}  \sum_{\beta_1 \beta_2 \beta_4} 
\prescript{\beta_1 \! \beta_2 \! \beta_4 \!\!}{\sigma_1 \! \sigma_2 \! \sigma_4 \!\!}
{\Big{[} \mathcal{C}_{3}^{3}}{\Big{]} }^{\!n d'_1 \!d'_2}_{\!d_1 \!d_{23}}
\ket{K, \beta_1 ; n\!+\!d_{12} ,\sigma_3 ; d'_{1}\!-\!n\!-\!d_{12} ,\beta_2 ; d'_{2} ,\beta_4}
\nn \allowdisplaybreaks[4] \\ && \hspace{-2cm}
-\hspace{-0.3cm} \sum_{ n<-d_{12} } \hspace{-0.1cm}
e^{iK \left(\frac{D_{33}+d_3}{4} \right)} \hspace{-0.1cm}  \sum_{\beta_1 \beta_2 \beta_4} 
\prescript{\beta_1 \! \beta_2 \! \beta_4 \!\!}{\sigma_1 \! \sigma_2 \! \sigma_4 \!\!}
{\Big{[} \mathcal{C}_{3}^{3}}{\Big{]} }^{\!n d'_1 \!d'_2}_{\!d_1 \!d_{23}}
\ket{K, \sigma_3 ; -\!n\!-\!d_{12} ,\beta_1 ; d'_{1} ,\beta_2 ; d'_{2} ,\beta_4}
\nn \allowdisplaybreaks[4] \\ && \hspace{-2cm}
-\hspace{-0.5cm} \sum_{ n>d'_{12}-d_{1} } \hspace{-0.3cm}
e^{iK \left(\frac{D_{33}+d_{23}}{4} \right)} \hspace{-0.1cm}  \sum_{\beta_1 \beta_3 \beta_4} 
\prescript{\beta_1 \! \beta_3 \! \beta_4 \!\!}{\sigma_1 \! \sigma_3 \! \sigma_4 \!\!}
{\Big{[} \mathcal{C}_{3}^{3}}{\Big{]} }^{\!n d'_1 \!d'_2}_{\!d_{12} \!d_{3}}
\ket{K, \beta_1 ; d'_{1} ,\beta_3 ; d'_{2} ,\beta_4 ; n\!+\!d_{1}\!-\!d'_{12} ,\sigma_2}
\nn \allowdisplaybreaks[4] \\ && \hspace{-2cm}
+\hspace{-0.5cm} \sum_{ \substack{n<d'_{12}-d_{1} \\ n>d'_{1}-d_{1} } } \hspace{-0.3cm}
e^{iK \left(\frac{D_{33}+d_{23}}{4} \right)} \hspace{-0.1cm}  \sum_{\beta_1 \beta_3 \beta_4} 
\prescript{\beta_1 \! \beta_3 \! \beta_4 \!\!}{\sigma_1 \! \sigma_3 \! \sigma_4 \!\!}
{\Big{[} \mathcal{C}_{3}^{3}}{\Big{]} }^{\!n d'_1 \!d'_2}_{\!d_{12} \!d_{3}}
\ket{K, \beta_1 ; d'_{1} ,\beta_3 ; n\!+\!d_{1}\!-\!d'_1 ,\sigma_2 ; d'_{12}\!-\!n\!-\!d_{1} ,\beta_4}
\nn \allowdisplaybreaks[4] \\ && \hspace{-2cm}
-\hspace{-0.4cm} \sum_{ \substack{n<d'_{1}-d_{1} \\ n>-d_{1} } } \hspace{-0.2cm}
e^{iK \left(\frac{D_{33}+d_{23}}{4} \right)} \hspace{-0.1cm}  \sum_{\beta_1 \beta_3 \beta_4} 
\prescript{\beta_1 \! \beta_3 \! \beta_4 \!\!}{\sigma_1 \! \sigma_3 \! \sigma_4 \!\!}
{\Big{[} \mathcal{C}_{3}^{3}}{\Big{]} }^{\!n d'_1 \!d'_2}_{\!d_{12} \!d_{3}}
\ket{K, \beta_1 ; n\!+\!d_1 ,\sigma_2 ; d'_1\!-\!n\!+\!d_{1} ,\beta_3 ; d'_{2} ,\beta_4}
\nn \allowdisplaybreaks[4] \\ && \hspace{-2cm}
+\hspace{-0.2cm} \sum_{  n<-d_{1} } \hspace{-0.1cm}
e^{iK \left(\frac{D_{33}+d_{23}}{4} \right)} \hspace{-0.1cm}  \sum_{\beta_1 \beta_3 \beta_4} 
\prescript{\beta_1 \! \beta_3 \! \beta_4 \!\!}{\sigma_1 \! \sigma_3 \! \sigma_4 \!\!}
{\Big{[} \mathcal{C}_{3}^{3}}{\Big{]} }^{\!n d'_1 \!d'_2}_{\!d_{12} \!d_{3}}
\ket{K, \sigma_2 ; -\!n\!-\!d_1 ,\beta_1 ; d'_1 ,\beta_3 ; d'_{2} ,\beta_4}
\nn \allowdisplaybreaks[4] \\ && \hspace{-2cm}
+\hspace{-0.4cm} \sum_{  n>d_{1}+d'_{12} } \hspace{-0.2cm}
e^{iK \left(\frac{D_{33}+d_{23}-2d_1}{4} \right)} \hspace{-0.1cm}  \sum_{\beta_2 \beta_3 \beta_4} 
\prescript{\beta_2 \! \beta_3 \! \beta_4 \!\!}{\sigma_2 \! \sigma_3 \! \sigma_4 \!\!}
{\Big{[} \mathcal{C}_{3}^{3}}{\Big{]} }^{\!n d'_1 \!d'_2}_{\!d_{2} \!d_{3}}
\ket{K, \beta_2 ; d'_1 ,\beta_3 ; d'_{2} ,\beta_4 ; \!n\!-\!d_1-\!d'_{12} ,\sigma_1}
\nn \allowdisplaybreaks[4] \\ && \hspace{-2cm}
-\hspace{-0.4cm} \sum_{ \substack{ n>d_{1}+d'_{1} \\ n<d_{1}+d'_{12} } } \hspace{-0.2cm}
e^{iK \left(\frac{D_{33}+d_{23}-2d_1}{4} \right)} \hspace{-0.1cm}  \sum_{\beta_2 \beta_3 \beta_4} 
\prescript{\beta_2 \! \beta_3 \! \beta_4 \!\!}{\sigma_2 \! \sigma_3 \! \sigma_4 \!\!}
{\Big{[} \mathcal{C}_{3}^{3}}{\Big{]} }^{\!n d'_1 \!d'_2}_{\!d_{2} \!d_{3}}
\ket{K, \beta_2 ; d'_1 ,\beta_3 ; \!n\!-\!d_1\!-\!d'_{1} ,\sigma_1 ; d_1\!+\!d'_{12}\!-\!n ,\beta_4}
\nn \allowdisplaybreaks[4] \\ && \hspace{-2cm}
+\hspace{-0.3cm} \sum_{ \substack{ n<d_{1}+d'_{1} \\ n>d_{1} } } \hspace{-0.2cm}
e^{iK \left(\frac{D_{33}+d_{23}-2d_1}{4} \right)} \hspace{-0.1cm}  \sum_{\beta_2 \beta_3 \beta_4} 
\prescript{\beta_2 \! \beta_3 \! \beta_4 \!\!}{\sigma_2 \! \sigma_3 \! \sigma_4 \!\!}
{\Big{[} \mathcal{C}_{3}^{3}}{\Big{]} }^{\!n d'_1 \!d'_2}_{\!d_{2} \!d_{3}}
\ket{K, \beta_2 ; n\!-\!d_1 ,\sigma_1 ; d_1\!+\!d'_{1}\!-\!n ,\beta_3 ; d'_2 ,\beta_4}
\nn \allowdisplaybreaks[4] \\ && \hspace{-2cm}
-\hspace{-0cm} \sum_{ n<d_{1} } \hspace{-0cm}
e^{iK \left(\frac{D_{33}+d_{23}-2d_1}{4} \right)} \hspace{-0.1cm}  \sum_{\beta_2 \beta_3 \beta_4} 
\prescript{\beta_2 \! \beta_3 \! \beta_4 \!\!}{\sigma_2 \! \sigma_3 \! \sigma_4 \!\!}
{\Big{[} \mathcal{C}_{3}^{3}}{\Big{]} }^{\!n d'_1 \!d'_2}_{\!d_{2} \!d_{3}}
\ket{K, \sigma_1 ; d_1\!-\!n ,\beta_2 ; d'_{1} ,\beta_3 ; d'_2 ,\beta_4} \Big{\}},
\eearr
where $D_{33} := 2d_1+2d'_1+d_2+d'_2$.

\subsubsection{$H_{4:4}$}

The action of $H_{4:4}$ on the four-QP state is given by
\bearr
H_{4:4} \ket{K, \sigma_1 ; d_1 ,\sigma_2 ; d_2 ,\sigma_3 ; d_3 ,\sigma_4} = 
+\!\!\!\!\!\sum_{d'_1 d'_2 d'_3 > 0}  \sum_{ \substack{\beta_1 \beta_2 \\ \beta_3 \beta_4}}
\sum_{n} \hspace{-0cm}
e^{iK \left(n+\frac{D_{44}}{4} \right)} 
\prescript{\beta_1 \! \beta_2 \! \beta_3 \! \beta_4 \!\!}{\sigma_1 \! \sigma_2 \! \sigma_3 \! \sigma_4 \!\!}
{\Big{[} \mathcal{C}_{4}^{4}}{\Big{]} }^{\!n d'_1 \!d'_2 \!d'_3}_{\!d_1 \!d_2 \!d_3}
\ket{K, \beta_1 ; d'_1 ,\beta_2 ; d'_2 ,\beta_3 ; d'_3 ,\beta_4},
\eearr
where $D_{44}=3(d_1-d'_1)+2(d_2-d'_2)+(d_3-d'_3)$. We have also defined
\bearr
\prescript{\beta_1 \! \beta_2 \! \beta_3 \! \beta_4 \!\!}{\sigma_1 \! \sigma_2 \! \sigma_3 \! \sigma_4 \!\!}
{\Big{[} \mathcal{C}_{4}^{4}}{\Big{]} }^{\!n d'_1 \!d'_2 \!d'_3}_{\!d_1 \!d_2 \!d_3} := 
\bra{r', \beta_1 ; r'\!+\!d'_1 ,\beta_2 ; r'\!+\!d'_{12} ,\beta_3 ; r'\!+\!d'_{123} ,\beta_4}
H_{3:3} \ket{r, \sigma_1 ; r\!+\!d_1 ,\sigma_2 ; r\!+\!d_{12} ,\sigma_3 ; r\!+\!d_{123} ,\sigma_4},
\eearr
where $r':= r\!-\!n$ and $d'_{ijk} := d'_i+d'_j+d'_k$.

\subsubsection{$H_{2:1}$}

The action of the off-diagonal interaction $H_{2:1}$ on the one-QP state
is given by
\bearr
H_{2:1} \ket{K, \sigma} = 
+\sum_{d'>0} \sum_{n} \sum_{\beta_1 \beta_2}
e^{iK \left(n-\frac{d'}{2} \right)} 
\prescript{\beta_1 \! \beta_2 \!\!}{\sigma \!\!}
{\Big{[} \mathcal{C}_{1}^{2}}{\Big{]} }^{\!n d'}_{}
\ket{K, \beta_1 ; d' ,\beta_2 },
\eearr
where 
\bearr
\prescript{\beta_1 \! \beta_2 \!\!}{\sigma \!\!}
{\Big{[} \mathcal{C}_{1}^{2}}{\Big{]} }^{\!n d'}_{} := 
\bra{r\!-\!n, \beta_1 ; r\!-\!n\!+\!d' ,\beta_2}
H_{2:1} \ket{r, \sigma}.
\eearr
The action of $H_{2:1}$ on the two-QP state leads to $6$ contributions
given here
\bearr
H_{2:1} \ket{K, \sigma_1; d_1, \sigma_2} = \sum_{d'_1>0} \Big{\{}
&+& \hspace{-0.2cm }\sum_{d'_1-d_1<n} \sum_{\beta_1 \beta_3}
e^{iK \left( \frac{4n+d_1-2d'_1}{6} \right)} 
\prescript{\beta_1 \! \beta_3 \!\!}{\sigma_1 \!\!}
{\Big{[} \mathcal{C}_{1}^{2}}{\Big{]} }^{\!n d'_1}_{}
\ket{K, \beta_1 ; d'_1 ,\beta_3 ; n\!+\!d_1\!-\!d'_1 ,\sigma_2 }
\nn \allowdisplaybreaks[4] \\ 
&-& \hspace{-0.2cm }\sum_{\substack{ d'_1-d_1>n \\ -d_1<n } } \sum_{\beta_1 \beta_3}
e^{iK \left( \frac{4n+d_1-2d'_1}{6} \right)} 
\prescript{\beta_1 \! \beta_3 \!\!}{\sigma_1 \!\!}
{\Big{[} \mathcal{C}_{1}^{2}}{\Big{]} }^{\!n d'_1}_{}
\ket{K, \beta_1 ; n\!+\!d_1 ,\sigma_2 ; d'_1\!-\!n\!-\!d_1 ,\beta_3 }
\nn \allowdisplaybreaks[4] \\ 
&+& \hspace{-0.05cm }\sum_{ -d_1>n } \hspace{0.15cm } \sum_{\beta_1 \beta_3} 
e^{iK \left( \frac{4n+d_1-2d'_1}{6} \right)} 
\prescript{\beta_1 \! \beta_3 \!\!}{\sigma_1 \!\!}
{\Big{[} \mathcal{C}_{1}^{2}}{\Big{]} }^{\!n d'_1}_{}
\ket{K, \sigma_2 ; -n\!-\!d_1 ,\beta_1 ; d'_1 ,\beta_3 }
\nn \allowdisplaybreaks[4] \\ 
&-& \hspace{-0cm }\sum_{ d_1>n } \hspace{0.3cm } \sum_{\beta_2 \beta_3} 
e^{iK \left( \frac{4n-d_1-2d'_1}{6} \right)} 
\prescript{\beta_2 \! \beta_3 \!\!}{\sigma_2 \!\!}
{\Big{[} \mathcal{C}_{1}^{2}}{\Big{]} }^{\!n d'_1}_{}
\ket{K, \sigma_1 ; d_1\!-\!n ,\beta_2 ; d'_1 ,\beta_3 }
\nn \allowdisplaybreaks[4] \\ 
&+& \hspace{-0.2cm }\sum_{ \substack{ d_1+d'_1>n \\ d_1<n } } \hspace{0cm } \sum_{\beta_2 \beta_3} 
e^{iK \left( \frac{4n-d_1-2d'_1}{6} \right)} 
\prescript{\beta_2 \! \beta_3 \!\!}{\sigma_2 \!\!}
{\Big{[} \mathcal{C}_{1}^{2}}{\Big{]} }^{\!n d'_1}_{}
\ket{K, \beta_2 ; n\!-\!d_1 ,\sigma_1 ; d_1\!+\!d'_1\!-\!n ,\beta_3 }
\nn \allowdisplaybreaks[4] \\ 
&-& \hspace{-0.2cm }\sum_{d_1+d'_1<n  } \hspace{0cm } \sum_{\beta_2 \beta_3} 
e^{iK \left( \frac{4n-d_1-2d'_1}{6} \right)} 
\prescript{\beta_2 \! \beta_3 \!\!}{\sigma_2 \!\!}
{\Big{[} \mathcal{C}_{1}^{2}}{\Big{]} }^{\!n d'_1}_{}
\ket{K, \beta_2 ; d'_1 ,\beta_3 ; n\!-\!d_1\!-\!d'_1 ,\sigma_1 } \Big{\}}.
\eearr

The application of $H_{2:1}$ on the three-QP state is given by
\bearr
H_{2:1} \ket{K, \sigma_1; d_1, \sigma_2; d_2, \sigma_3} &=& \sum_{d'_1>0} \Big{\{}
+ \hspace{-0.2cm }\sum_{d'_1-d_1<n} \sum_{\beta_1 \beta_4}
e^{iK \left( \frac{D_{21}}{12} \right)} 
\prescript{\beta_1 \! \beta_4 \!\!}{\sigma_1 \!\!}
{\Big{[} \mathcal{C}_{1}^{2}}{\Big{]} }^{\!n d'_1}_{}
\ket{K, \beta_1 ; d'_1 ,\beta_4 ; n\!+\!d_1\!-\!d'_1 ,\sigma_2 ; d_2 ,\sigma_3 }
\nn \allowdisplaybreaks[4] \\ && \hspace{-1.5cm }
- \hspace{-0.3cm }\sum_{ \substack{ d'_1-d_1>n \\ -d_1<n \\ d'_1-d_{12}<n } } \sum_{\beta_1 \beta_4}
e^{iK \left( \frac{D_{21}}{12} \right)} 
\prescript{\beta_1 \! \beta_4 \!\!}{\sigma_1 \!\!}
{\Big{[} \mathcal{C}_{1}^{2}}{\Big{]} }^{\!n d'_1}_{}
\ket{K, \beta_1 ; n\!+\!d_1 ,\sigma_2 ; d'_1\!-\!n\!-\!d_1 ,\beta_4 ; n\!+\!d_{12}\!-\!d'_1 ,\sigma_3 }
\nn \allowdisplaybreaks[4] \\ && \hspace{-1.5cm }
+ \hspace{-0.3cm }\sum_{ \substack{ d'_1-d_{12}>n \\ -d_1<n} } \sum_{\beta_1 \beta_4}
e^{iK \left( \frac{D_{21}}{12} \right)} 
\prescript{\beta_1 \! \beta_4 \!\!}{\sigma_1 \!\!}
{\Big{[} \mathcal{C}_{1}^{2}}{\Big{]} }^{\!n d'_1}_{}
\ket{K, \beta_1 ; n\!+\!d_1 ,\sigma_2 ; d_2, \sigma_3 ; d'_1\!-\!n\!-\!d_{12} ,\beta_4 }
\nn \allowdisplaybreaks[4] \\ && \hspace{-1.5cm }
+ \hspace{-0.3cm }\sum_{ \substack{ d'_1-d_{12}<n \\ -d_1>n} } \sum_{\beta_1 \beta_4}
e^{iK \left( \frac{D_{21}}{12} \right)} 
\prescript{\beta_1 \! \beta_4 \!\!}{\sigma_1 \!\!}
{\Big{[} \mathcal{C}_{1}^{2}}{\Big{]} }^{\!n d'_1}_{}
\ket{K, \sigma_2 ; -n\!-\!d_1 ,\beta_1 ; d'_1, \beta_4 ; n\!+\!d_{12}\!-\!d'_1 ,\sigma_3 }
\nn \allowdisplaybreaks[4] \\ && \hspace{-1.5cm }
- \hspace{-0.3cm }\sum_{ \substack{ d'_1-d_{12}>n \\ -d_1>n \\ -d_{12}<n} } \sum_{\beta_1 \beta_4}
e^{iK \left( \frac{D_{21}}{12} \right)} 
\prescript{\beta_1 \! \beta_4 \!\!}{\sigma_1 \!\!}
{\Big{[} \mathcal{C}_{1}^{2}}{\Big{]} }^{\!n d'_1}_{}
\ket{K, \sigma_2 ; -n\!-\!d_1 ,\beta_1 ; n\!+\!d_{12}, \sigma_3 ; d'_1\!-\!d_{12}\!-\!n ,\beta_4 }
\nn \allowdisplaybreaks[4] \\ && \hspace{-1.5cm }
+ \hspace{-0.2cm }\sum_{  -d_{12}>n} \hspace{0.2cm } \sum_{\beta_1 \beta_4}
e^{iK \left( \frac{D_{21}}{12} \right)} 
\prescript{\beta_1 \! \beta_4 \!\!}{\sigma_1 \!\!}
{\Big{[} \mathcal{C}_{1}^{2}}{\Big{]} }^{\!n d'_1}_{}
\ket{K, \sigma_2 ; d_2 ,\sigma_3 ; -n\!-\!d_{12}, \beta_1 ; d'_1 ,\beta_4 }
\nn \allowdisplaybreaks[4] \\ && \hspace{-1.5cm }
- \hspace{-0.3cm }\sum_{  d_{1}+d'_1<n} \hspace{0cm } \sum_{\beta_2 \beta_4}
e^{iK \left( \frac{D_{21}-3d_1}{12} \right)} 
\prescript{\beta_2 \! \beta_4 \!\!}{\sigma_2 \!\!}
{\Big{[} \mathcal{C}_{1}^{2}}{\Big{]} }^{\!n d'_1}_{}
\ket{K, \beta_2 ; d'_1 ,\beta_4 ; n\!-\!d_{1}\!-\!d'_{1}, \sigma_1 ; d_{12} ,\sigma_3 }
\nn \allowdisplaybreaks[4] \\ && \hspace{-1.5cm }
+ \hspace{-0.3cm }\sum_{ \substack{ d_{1}+d'_1>n \\ d'_{1}-d_2<n \\ d_1<n}} \hspace{0cm } \sum_{\beta_2 \beta_4}
e^{iK \left( \frac{D_{21}-3d_1}{12} \right)} 
\prescript{\beta_2 \! \beta_4 \!\!}{\sigma_2 \!\!}
{\Big{[} \mathcal{C}_{1}^{2}}{\Big{]} }^{\!n d'_1}_{}
\ket{K, \beta_2 ; n\!-\!d_1 ,\sigma_1 ; d'_1\!+\!d_{1}\!-\!n, \beta_4 ; n\!+\!d_{2}\!-\!d'_1 ,\sigma_3 }
\nn \allowdisplaybreaks[4] \\ && \hspace{-1.5cm }
- \hspace{-0.3cm }\sum_{ \substack{ d'_{1}-d_2>n \\ d_1<n}} \hspace{0cm } \sum_{\beta_2 \beta_4}
e^{iK \left( \frac{D_{21}-3d_1}{12} \right)} 
\prescript{\beta_2 \! \beta_4 \!\!}{\sigma_2 \!\!}
{\Big{[} \mathcal{C}_{1}^{2}}{\Big{]} }^{\!n d'_1}_{}
\ket{K, \beta_2 ; n\!-\!d_1 ,\sigma_1 ; d_{12}, \sigma_3 ; d'_1\!-\!d_{2}\!-\!n ,\beta_4 }
\nn \allowdisplaybreaks[4] \\ && \hspace{-1.5cm }
- \hspace{-0.3cm }\sum_{ \substack{ d'_{1}-d_2<n \\ d_1>n}} \hspace{0cm } \sum_{\beta_2 \beta_4}
e^{iK \left( \frac{D_{21}-3d_1}{12} \right)} 
\prescript{\beta_2 \! \beta_4 \!\!}{\sigma_2 \!\!}
{\Big{[} \mathcal{C}_{1}^{2}}{\Big{]} }^{\!n d'_1}_{}
\ket{K, \sigma_1 ; d_1\!-\!n ,\beta_2 ; d'_{1}, \beta_4 ; n\!+\!d_{2}\!-\!d'_1 ,\sigma_3 }
\nn \allowdisplaybreaks[4] \\ && \hspace{-1.5cm }
+ \hspace{-0.3cm }\sum_{ \substack{ d'_{1}-d_2>n \\ d_1>n \\ -d_2<n } } \hspace{0cm } \sum_{\beta_2 \beta_4}
e^{iK \left( \frac{D_{21}-3d_1}{12} \right)} 
\prescript{\beta_2 \! \beta_4 \!\!}{\sigma_2 \!\!}
{\Big{[} \mathcal{C}_{1}^{2}}{\Big{]} }^{\!n d'_1}_{}
\ket{K, \sigma_1 ; d_1\!-\!n ,\beta_2 ; d_{2}\!+\!n, \sigma_3 ; d'_1\!-\!n\!-\!d_2 ,\beta_4 }
\nn \allowdisplaybreaks[4] \\ && \hspace{-1.5cm }
- \hspace{-0.2cm }\sum_{ -d_2>n } \hspace{0.2cm } \sum_{\beta_2 \beta_4}
e^{iK \left( \frac{D_{21}-3d_1}{12} \right)} 
\prescript{\beta_2 \! \beta_4 \!\!}{\sigma_2 \!\!}
{\Big{[} \mathcal{C}_{1}^{2}}{\Big{]} }^{\!n d'_1}_{}
\ket{K, \sigma_1 ; d_{12} ,\sigma_3 ; -d_{2}\!-\!n, \beta_2 ; d'_1 ,\beta_4 }
\nn \allowdisplaybreaks[4] \\ && \hspace{-1.5cm }
+ \hspace{-0.4cm }\sum_{ d_{12}+d'_1<n } \hspace{0cm } \sum_{\beta_3 \beta_4}
e^{iK \left( \frac{D_{21}-3d_{12} }{12} \right)} 
\prescript{\beta_3 \! \beta_4 \!\!}{\sigma_3 \!\!}
{\Big{[} \mathcal{C}_{1}^{2}}{\Big{]} }^{\!n d'_1}_{}
\ket{K, \beta_3 ; d'_{1} ,\beta_4 ; n\!-\!d_{12}\!-\!d'_1, \sigma_1 ; d_1 ,\sigma_2 }
\nn \allowdisplaybreaks[4] \\ && \hspace{-1.5cm }
- \hspace{-0.4cm }\sum_{ \substack{ d_{12}+d'_1>n \\ d_{12}<n \\ d'_1+d_2<n } } \hspace{0cm } \sum_{\beta_3 \beta_4}
e^{iK \left( \frac{D_{21}-3d_{12} }{12} \right)} 
\prescript{\beta_3 \! \beta_4 \!\!}{\sigma_3 \!\!}
{\Big{[} \mathcal{C}_{1}^{2}}{\Big{]} }^{\!n d'_1}_{}
\ket{K, \beta_3 ; n\!-\!d_{12} ,\sigma_1 ; d_{12}\!+\!d'_{1}\!-\!n, \beta_4 ; n\!-\!d_2\!-\!d'_1 ,\sigma_2 }
\nn \allowdisplaybreaks[4] \\ && \hspace{-1.5cm }
+ \hspace{-0.3cm }\sum_{ \substack{ d_{2}+d'_1>n \\ d_{12}<n } } \hspace{0cm } \sum_{\beta_3 \beta_4}
e^{iK \left( \frac{D_{21}-3d_{12} }{12} \right)} 
\prescript{\beta_3 \! \beta_4 \!\!}{\sigma_3 \!\!}
{\Big{[} \mathcal{C}_{1}^{2}}{\Big{]} }^{\!n d'_1}_{}
\ket{K, \beta_3 ; n\!-\!d_{12} ,\sigma_1 ; d_{1} , \sigma_2 ; d_2\!+\!d'_1\!-\!n ,\beta_4 }
\nn \allowdisplaybreaks[4] \\ && \hspace{-1.5cm }
+ \hspace{-0.3cm }\sum_{ \substack{ d_{2}+d'_1<n \\ d_{12}>n } } \hspace{0cm } \sum_{\beta_3 \beta_4}
e^{iK \left( \frac{D_{21}-3d_{12} }{12} \right)} 
\prescript{\beta_3 \! \beta_4 \!\!}{\sigma_3 \!\!}
{\Big{[} \mathcal{C}_{1}^{2}}{\Big{]} }^{\!n d'_1}_{}
\ket{K, \sigma_1 ; d_{12}\!-\!n ,\beta_3 ; d'_{1} , \beta_4 ; n\!-\!d'_1\!-\!d_2 ,\sigma_2 }
\nn \allowdisplaybreaks[4] \\ && \hspace{-1.5cm }
- \hspace{-0.3cm }\sum_{ \substack{ d_{2}+d'_1>n \\ d_{12}>n \\ d_{2}<n } } \hspace{0cm } \sum_{\beta_3 \beta_4}
e^{iK \left( \frac{D_{21}-3d_{12} }{12} \right)} 
\prescript{\beta_3 \! \beta_4 \!\!}{\sigma_3 \!\!}
{\Big{[} \mathcal{C}_{1}^{2}}{\Big{]} }^{\!n d'_1}_{}
\ket{K, \sigma_1 ; d_{12}\!-\!n ,\beta_3 ; n\!-\!d_{2} , \sigma_2 ; d_2\!+\!d'_1\!-\!n ,\beta_4 }
\nn \allowdisplaybreaks[4] \\ && \hspace{-1.5cm }
+ \hspace{-0cm }\sum_{ d_{2}>n } \hspace{0.2cm } \sum_{\beta_3 \beta_4}
e^{iK \left( \frac{D_{21}-3d_{12} }{12} \right)} 
\prescript{\beta_3 \! \beta_4 \!\!}{\sigma_3 \!\!}
{\Big{[} \mathcal{C}_{1}^{2}}{\Big{]} }^{\!n d'_1}_{}
\ket{K, \sigma_1 ; d_{1} ,\sigma_2 ; d_2\!-\!n , \beta_3 ; d'_1 ,\beta_4 } \Big{\} },
\eearr
with $D_{21} := 6n+2d_1+d_2-3d'_1$.

\subsubsection{$H_{3:1}$}

The action of $H_{3:1}$ on the one-QP state is given by
\bearr
H_{3:1} \ket{K, \sigma} = 
+\sum_{d'_1 d'_2} \sum_{n} \sum_{\beta_1 \beta_2  \beta_3}
e^{iK \left(n-\frac{2d'_1+d'_2}{3} \right)} 
\prescript{\beta_1 \! \beta_2 \! \beta_3 \!\!}{\sigma \!\!}
{\Big{[} \mathcal{C}_{1}^{3}}{\Big{]} }^{\!n d'_1 d'_2}_{}
\ket{K, \beta_1 ; d'_1 ,\beta_2 ; d'_2 ,\beta_3 },
\eearr
where 
\bearr
\prescript{\beta_1 \! \beta_2 \! \beta_3 \!\!}{\sigma \!\!}
{\Big{[} \mathcal{C}_{1}^{3}}{\Big{]} }^{\!n d'_1 d'_2}_{} := 
\bra{r\!-\!n, \beta_1 ; r\!-\!n\!+\!d'_1 ,\beta_2 ; r\!-\!n\!+\!d'_{12} ,\beta_3}
H_{3:1} \ket{r, \sigma}.
\eearr

The action of $H_{3:1}$ on the two-QP state is given by
\bearr
H_{3:1} \ket{K, \sigma_1 ; d_1 ,\sigma_2} &=& 
\sum_{d'_1 d'_2} \Big{\{}
+\hspace{-0.4cm} \sum_{n>d'_{12}-d_{1}} \hspace{-0.3cm}
e^{iK \left(\frac{D_{31}}{4} \right)}  \sum_{\beta_1 \beta_3 \beta_4} 
\prescript{\beta_1 \! \beta_3 \! \beta_4 \!\!}{\sigma_1 \!\!}
{\Big{[} \mathcal{C}_{1}^{3}}{\Big{]} }^{\!n d'_1 \!d'_2}_{}
\ket{K, \beta_1 ; d'_1 ,\beta_3 ; d'_2 ,\beta_4 ; n\!+\!d_{1}\!-\!d'_{12} ,\sigma_2}
\nn \allowdisplaybreaks[4] \\ &&
-\hspace{-0.4cm} \sum_{ \substack{n<d'_{12}-d_{1} \\ n>d'_{1}-d_{1} } } \hspace{-0.3cm}
e^{iK \left(\frac{D_{31}}{4} \right)}  \sum_{\beta_1 \beta_3 \beta_4} 
\prescript{\beta_1 \! \beta_3 \! \beta_4 \!\!}{\sigma_1 \!\!}
{\Big{[} \mathcal{C}_{1}^{3}}{\Big{]} }^{\!n d'_1 \!d'_2}_{}
\ket{K, \beta_1 ; d'_1 ,\beta_3 ; n\!+\!d_{1}\!-\!d'_{1} ,\sigma_2 ; d'_{12}\!-\!n\!-\!d_{1} ,\beta_4}
\nn \allowdisplaybreaks[4] \\ &&
+\hspace{-0.3cm} \sum_{ \substack{n<d'_{1}-d_{1} \\ n>-d_{1} } } \hspace{-0.2cm}
e^{iK \left(\frac{D_{31}}{4} \right)}  \sum_{\beta_1 \beta_3 \beta_4} 
\prescript{\beta_1 \! \beta_3 \! \beta_4 \!\!}{\sigma_1 \!\!}
{\Big{[} \mathcal{C}_{1}^{3}}{\Big{]} }^{\!n d'_1 \!d'_2}_{}
\ket{K, \beta_1 ; d_1\!+\!n ,\sigma_2 ; d'_{1}\!-\!n\!-\!d_{1} ,\beta_3 ; d'_{2} ,\beta_4}
\nn \allowdisplaybreaks[4] \\ &&
-\hspace{-0.1cm} \sum_{ n<-d_{1} } \hspace{-0.2cm}
e^{iK \left(\frac{D_{31}}{4} \right)}  \sum_{\beta_1 \beta_3 \beta_4} 
\prescript{\beta_1 \! \beta_3 \! \beta_4 \!\!}{\sigma_1 \!\!}
{\Big{[} \mathcal{C}_{1}^{3}}{\Big{]} }^{\!n d'_1 \!d'_2}_{}
\ket{K, \sigma_2 ; -d_1\!-\!n ,\beta_1 ; d'_{1} ,\beta_3 ; d'_{2} ,\beta_4}
\nn \allowdisplaybreaks[4] \\ &&
+\hspace{-0cm} \sum_{ n<d_{1} } \hspace{-0cm}
e^{iK \left(\frac{D_{31}-2d_1}{4} \right)}  \sum_{\beta_2 \beta_3 \beta_4} 
\prescript{\beta_2 \! \beta_3 \! \beta_4 \!\!}{\sigma_2 \!\!}
{\Big{[} \mathcal{C}_{1}^{3}}{\Big{]} }^{\!n d'_1 \!d'_2}_{}
\ket{K, \sigma_1 ; d_1\!-\!n ,\beta_2 ; d'_{1} ,\beta_3 ; d'_{2} ,\beta_4}
\nn \allowdisplaybreaks[4] \\ &&
-\hspace{-0.3cm} \sum_{ \substack{n>d_{1} \\ n<d_{1}+d'_1 } } \hspace{-0.3cm}
e^{iK \left(\frac{D_{31}-2d_1}{4} \right)}  \sum_{\beta_2 \beta_3 \beta_4} 
\prescript{\beta_2 \! \beta_3 \! \beta_4 \!\!}{\sigma_2 \!\!}
{\Big{[} \mathcal{C}_{1}^{3}}{\Big{]} }^{\!n d'_1 \!d'_2}_{}
\ket{K, \beta_2 ; n\!-\!d_1 ,\sigma_1 ; d_{1}\!+\!d'_1\!-\!n ,\beta_3 ; d'_{2} ,\beta_4}
\nn \allowdisplaybreaks[4] \\ &&
+\hspace{-0.4cm} \sum_{ \substack{n>d_{1}+d'_1 \\ n<d_{1}+d'_{12} } } \hspace{-0.3cm}
e^{iK \left(\frac{D_{31}-2d_1}{4} \right)}  \sum_{\beta_2 \beta_3 \beta_4} 
\prescript{\beta_2 \! \beta_3 \! \beta_4 \!\!}{\sigma_2 \!\!}
{\Big{[} \mathcal{C}_{1}^{3}}{\Big{]} }^{\!n d'_1 \!d'_2}_{}
\ket{K, \beta_2 ; d'_{1} ,\beta_3 ; n\!-\!d_1\!-\!d'_1 ,\sigma_1 ; d_{1}\!+\!d'_{12}\!-\!n ,\beta_4}
\nn \allowdisplaybreaks[4] \\ &&
-\hspace{-0.45cm} \sum_{ n>d_{1}+d'_{12} } \hspace{-0.3cm}
e^{iK \left(\frac{D_{31}-2d_1}{4} \right)}  \sum_{\beta_2 \beta_3 \beta_4} 
\prescript{\beta_2 \! \beta_3 \! \beta_4 \!\!}{\sigma_2 \!\!}
{\Big{[} \mathcal{C}_{1}^{3}}{\Big{]} }^{\!n d'_1 \!d'_2}_{}
\ket{K, \beta_2 ; d'_{1} ,\beta_3 ; d'_2 ,\beta_4 ; n\!-\!d_{1}\!-\!d'_{12} ,\sigma_1} \Big{\}},
\eearr
where $D_{31} := 3n+d_1-2d'_1-d'_2$.

\subsubsection{$H_{4:1}$}
The action of $H_{4:1}$ on the one-QP state is given by
\bearr
H_{4:1} \ket{K, \sigma} = 
+\sum_{d'_1 d'_2 d'_3} \sum_{n} \sum_{\substack{\beta_1 \beta_2 \\ \beta_3 \beta_4}}
e^{iK \left(n-\frac{3d'_1+2d'_2+d'_3}{4} \right)} 
\prescript{\beta_1 \! \beta_2 \! \beta_3 \! \beta_4 \!\!}{\sigma \!\!}
{\Big{[} \mathcal{C}_{1}^{4}}{\Big{]} }^{\!n d'_1 d'_2 d'_3}_{}
\ket{K, \beta_1 ; d'_1 ,\beta_2 ; d'_2 ,\beta_3 ; d'_3 ,\beta_4 },
\eearr
where 
\bearr
\prescript{\beta_1 \! \beta_2 \! \beta_3 \! \beta_4 \!\!}{\sigma \!\!}
{\Big{[} \mathcal{C}_{1}^{4}}{\Big{]} }^{\!n d'_1 d'_2 d'_3}_{} := 
\bra{r\!-\!n, \beta_1 ; r\!-\!n\!+\!d'_1 ,\beta_2 ; r\!-\!n\!+\!d'_{12} ,\beta_3 
; r\!-\!n\!+\!d'_{123} ,\beta_4}
H_{4:1} \ket{r, \sigma}.
\eearr

\subsubsection{$H_{3:2}$}

The action of $H_{3:2}$ on the two-QP state is given by
\bearr
H_{3:2} \ket{K, \sigma_1 ; d_1 , \sigma_2} = 
-\sum_{d'_1 d'_2} \sum_{n} \sum_{\beta_1 \beta_2  \beta_3} \hspace{-0.2cm}
e^{iK \left(n+\frac{d_1}{2}-\frac{2d'_1+d'_2}{3} \right)} 
\prescript{\beta_1 \! \beta_2 \! \beta_3 \!\!}{\sigma_1 \! \sigma_2 \!\!}
{\Big{[} \mathcal{C}_{2}^{3}}{\Big{]} }^{\!n d'_1 d'_2}_{d_1}
\ket{K, \beta_1 ; d'_1 ,\beta_2 ; d'_2 ,\beta_3 },
\eearr
where 
\bearr
\prescript{\beta_1 \! \beta_2 \! \beta_3 \!\!}{\sigma_1 \sigma_2 \!\!}
{\Big{[} \mathcal{C}_{2}^{3}}{\Big{]} }^{\!n d'_1 d'_2}_{d_1} := 
\bra{r\!-\!n, \beta_1 ; r\!-\!n\!+\!d'_1 ,\beta_2 ; r\!-\!n\!+\!d'_{12} ,\beta_3}
H_{3:1} \ket{r, \sigma_1 ; r\!+\!d_1, \sigma_2}. 
\eearr

For the application of $H_{3:2}$ on the three-QP state, we find
\bearr
H_{3:2} \ket{K, \sigma_1 ; d_1 , \sigma_2 ; d_2 , \sigma_3} &=& \sum_{d'_1 d'_2} \Big{\{}
+ \hspace{-0cm} \sum_{-d_{12}>n}  \hspace{-0cm}
e^{iK \left( \frac{D_{32}}{12} \right)} \sum_{\beta_1 \beta_2  \beta_4}
\prescript{\beta_1 \! \beta_2 \! \beta_4 \!\!}{\sigma_1 \! \sigma_2 \!\!}
{\Big{[} \mathcal{C}_{2}^{3}}{\Big{]} }^{\!n d'_1 d'_2}_{d_1}
\ket{K, \sigma_3 ; -n\!-\!d_{12} ,\beta_1 ; d'_1 ,\beta_2 ; d'_2 ,\beta_4 }
\nn \allowdisplaybreaks[4] \\ && \hspace{-2cm}
- \hspace{-0.2cm} \sum_{\substack{ d'_1-d_{12}>n \\  -d_{12}<n } }  \hspace{-0.1cm}
e^{iK \left( \frac{D_{32}}{12} \right)} \sum_{\beta_1 \beta_2  \beta_4}
\prescript{\beta_1 \! \beta_2 \! \beta_4 \!\!}{\sigma_1 \! \sigma_2 \!\!}
{\Big{[} \mathcal{C}_{2}^{3}}{\Big{]} }^{\!n d'_1 d'_2}_{d_1}
\ket{K, \beta_1 ; n\!+\!d_{12} ,\sigma_3 ; d'_1\!-\!n\!-\!d_{12} ,\beta_2 ; d'_2 ,\beta_4 }
\nn \allowdisplaybreaks[4] \\ && \hspace{-2cm}
+ \hspace{-0.3cm} \sum_{\substack{ d'_1-d_{12}<n \\  d'_{12}-d_{12}>n } }  \hspace{-0.15cm}
e^{iK \left( \frac{D_{32}}{12} \right)} \sum_{\beta_1 \beta_2  \beta_4}
\prescript{\beta_1 \! \beta_2 \! \beta_4 \!\!}{\sigma_1 \! \sigma_2 \!\!}
{\Big{[} \mathcal{C}_{2}^{3}}{\Big{]} }^{\!n d'_1 d'_2}_{d_1}
\ket{K, \beta_1 ; d'_1 ,\beta_2 ; n\!+\!d_{12}\!-\!d'_1 ,\sigma_3 ; d'_{12}\!-\!n\!-\!d_{12} ,\beta_4 }
\nn \allowdisplaybreaks[4] \\ && \hspace{-2cm}
- \hspace{-0.3cm} \sum_{ d'_{12}-d_{12}<n }  \hspace{-0.15cm}
e^{iK \left( \frac{D_{32}}{12} \right)} \sum_{\beta_1 \beta_2  \beta_4}
\prescript{\beta_1 \! \beta_2 \! \beta_4 \!\!}{\sigma_1 \! \sigma_2 \!\!}
{\Big{[} \mathcal{C}_{2}^{3}}{\Big{]} }^{\!n d'_1 d'_2}_{d_1}
\ket{K, \beta_1 ; d'_1 ,\beta_2 ; d'_2 ,\beta_4 ; n\!+\!d_{12}\!-\!d'_{12} ,\sigma_3 }
\nn \allowdisplaybreaks[4] \\ && \hspace{-2cm}
- \hspace{-0.0cm} \sum_{ -d_{1}>n }  \hspace{-0cm}
e^{iK \left( \frac{D_{32}+3d_2}{12} \right)} \sum_{\beta_1 \beta_3  \beta_4}
\prescript{\beta_1 \! \beta_3 \! \beta_4 \!\!}{\sigma_1 \! \sigma_3 \!\!}
{\Big{[} \mathcal{C}_{2}^{3}}{\Big{]} }^{\!n d'_1 d'_2}_{d_{12}}
\ket{K, \sigma_2 ; -n\!-\!d_1 ,\beta_1 ; d'_1 ,\beta_3 ; d'_{2} ,\beta_4 }
\nn \allowdisplaybreaks[4] \\ && \hspace{-2cm}
+ \hspace{-0.2cm} \sum_{ \substack{d'_1-d_{1}>n \\ -d_{1}<n } }  \hspace{-0.1cm}
e^{iK \left( \frac{D_{32}+3d_2}{12} \right)} \sum_{\beta_1 \beta_3  \beta_4}
\prescript{\beta_1 \! \beta_3 \! \beta_4 \!\!}{\sigma_1 \! \sigma_3 \!\!}
{\Big{[} \mathcal{C}_{2}^{3}}{\Big{]} }^{\!n d'_1 d'_2}_{d_{12}}
\ket{K, \beta_1 ; n\!+\!d_1 ,\sigma_2 ; d'_1\!-\!n\!-\!d_1 ,\beta_3 ; d'_{2} ,\beta_4 }
\nn \allowdisplaybreaks[4] \\ && \hspace{-2cm}
- \hspace{-0.25cm} \sum_{ \substack{d'_{12}-d_{1}>n \\ d'_1-d_{1}<n } }  \hspace{-0.2cm}
e^{iK \left( \frac{D_{32}+3d_2}{12} \right)} \sum_{\beta_1 \beta_3  \beta_4}
\prescript{\beta_1 \! \beta_3 \! \beta_4 \!\!}{\sigma_1 \! \sigma_3 \!\!}
{\Big{[} \mathcal{C}_{2}^{3}}{\Big{]} }^{\!n d'_1 d'_2}_{d_{12}}
\ket{K, \beta_1 ; d'_{1} ,\beta_3 ; n\!+\!d_1\!-\!d'_1 ,\sigma_2 ; d'_{12}\!-\!n\!-\!d_1 ,\beta_4 }
\nn \allowdisplaybreaks[4] \\ && \hspace{-2cm}
+ \hspace{-0.25cm} \sum_{ d'_{12}-d_{1}<n }  \hspace{-0.2cm}
e^{iK \left( \frac{D_{32}+3d_2}{12} \right)} \sum_{\beta_1 \beta_3  \beta_4}
\prescript{\beta_1 \! \beta_3 \! \beta_4 \!\!}{\sigma_1 \! \sigma_3 \!\!}
{\Big{[} \mathcal{C}_{2}^{3}}{\Big{]} }^{\!n d'_1 d'_2}_{d_{12}}
\ket{K, \beta_1 ; d'_{1} ,\beta_3 ; d'_{2} ,\beta_4 ; n\!+\!d_1\!-\!d'_{12} ,\sigma_2 }
\nn \allowdisplaybreaks[4] \\ && \hspace{-2cm}
- \hspace{-0cm} \sum_{ d_{1}>n }  \hspace{-0cm}
e^{iK \left( \frac{D_{32}+3d_2-6d_1}{12} \right)} \sum_{\beta_2 \beta_3  \beta_4}
\prescript{\beta_2 \! \beta_3 \! \beta_4 \!\!}{\sigma_2 \! \sigma_3 \!\!}
{\Big{[} \mathcal{C}_{2}^{3}}{\Big{]} }^{\!n d'_1 d'_2}_{d_{2}}
\ket{K, \sigma_1 ; d_{1}\!-\!n ,\beta_2 ; d'_{1} ,\beta_3 ; d'_{2} ,\beta_4 }
\nn \allowdisplaybreaks[4] \\ && \hspace{-2cm}
+ \hspace{-0.25cm} \sum_{ \substack{ d_{1}<n \\ d_{1}+d'_{1}>n } }  \hspace{-0.25cm}
e^{iK \left( \frac{D_{32}+3d_2-6d_1}{12} \right)} \sum_{\beta_2 \beta_3  \beta_4}
\prescript{\beta_2 \! \beta_3 \! \beta_4 \!\!}{\sigma_2 \! \sigma_3 \!\!}
{\Big{[} \mathcal{C}_{2}^{3}}{\Big{]} }^{\!n d'_1 d'_2}_{d_{2}}
\ket{K, \beta_2 ; n\!-\!d_{1} ,\sigma_1 ; d'_{1}\!+\!d_{1}\!-\!n ,\beta_3 ; d'_{2} ,\beta_4 }
\nn \allowdisplaybreaks[4] \\ && \hspace{-2cm}
- \hspace{-0.3cm} \sum_{ \substack{ d_{1}+d'_{1}<n \\ d_{1}+d'_{12}>n } }  \hspace{-0.2cm}
e^{iK \left( \frac{D_{32}+3d_2-6d_1}{12} \right)} \sum_{\beta_2 \beta_3  \beta_4}
\prescript{\beta_2 \! \beta_3 \! \beta_4 \!\!}{\sigma_2 \! \sigma_3 \!\!}
{\Big{[} \mathcal{C}_{2}^{3}}{\Big{]} }^{\!n d'_1 d'_2}_{d_{2}}
\ket{K, \beta_2 ; d'_{1} ,\beta_3 ; n\!-\!d_{1}\!-\!d'_{1} ,\sigma_1 ; d'_{12}\!+\!d_1\!-\!n ,\beta_4 }
\nn \allowdisplaybreaks[4] \\ && \hspace{-2cm}
+ \hspace{-0.3cm} \sum_{ d_{1}+d'_{12}<n }  \hspace{-0.2cm}
e^{iK \left( \frac{D_{32}+3d_2-6d_1}{12} \right)} \sum_{\beta_2 \beta_3  \beta_4}
\prescript{\beta_2 \! \beta_3 \! \beta_4 \!\!}{\sigma_2 \! \sigma_3 \!\!}
{\Big{[} \mathcal{C}_{2}^{3}}{\Big{]} }^{\!n d'_1 d'_2}_{d_{2}}
\ket{K, \beta_2 ; d'_{1} ,\beta_3 ; d'_{2} ,\beta_4 ; n\!-\!d_{1}\!-\!d'_{12} ,\sigma_1 } \Big{\}},
\eearr
with $D_{32} := 9n+5d_1+d_2-6d'_1-3d'_2$.

\subsubsection{$H_{4:2}$}

The action of $H_{4:2}$ on the two-QP state is given by
\bearr
H_{4:2} \ket{K, \sigma_1 ; d_1 , \sigma_2} = 
-\sum_{d'_1 d'_2 d'_3} \sum_{n} \sum_{\substack{\beta_1 \beta_2 \\ \beta_3 \beta_4} } \hspace{-0cm}
e^{iK \left(n+\frac{d_1}{2}-\frac{3d'_1+2d'_2+d'_3}{4} \right)} 
\prescript{\beta_1 \! \beta_2 \! \beta_3 \! \beta_4 \!\!}{\sigma_1 \! \sigma_2 \!\!}
{\Big{[} \mathcal{C}_{2}^{4}}{\Big{]} }^{\!n d'_1 d'_2 d'_3}_{d_1}
\ket{K, \beta_1 ; d'_1 ,\beta_2 ; d'_2 ,\beta_3 ; d'_3 ,\beta_4 },
\eearr
where we have defined
\bearr
\prescript{\beta_1 \! \beta_2 \! \beta_3 \! \beta_4 \!\!}{\sigma_1 \! \sigma_2 \!\!}
{\Big{[} \mathcal{C}_{2}^{4}}{\Big{]} }^{\!n d'_1 d'_2 d'_3}_{d_1} := 
\bra{r', \beta_1 ; r'\!+\!d'_1 ,\beta_2 ; r'\!+\!d'_{12} ,\beta_3 ; r'\!+\!d'_{123} ,\beta_4}
H_{4:2} \ket{r, \sigma_1 ; r\!+\!d_1, \sigma_2}, 
\eearr
with $r' := r\!-\!n$.

\subsubsection{$H_{4:3}$}

The action of $H_{4:3}$ on the three-QP state is given by
\bearr
H_{4:3} \ket{K, \sigma_1 ; d_1 , \sigma_2 ; d_2 , \sigma_3} =
-\!\!\!\sum_{d'_1 d'_2 d'_3} \sum_{n} \sum_{\substack{\beta_1 \beta_2 \\ \beta_3 \beta_4} } \hspace{-0cm}
e^{iK \left(n+\frac{2d_1+d_2}{3}-\frac{3d'_1+2d'_2+d'_3}{4} \right)} 
\prescript{\beta_1 \! \beta_2 \! \beta_3 \! \beta_4 \!\!}{\sigma_1 \! \sigma_2 \! \sigma_3 \!\!}
{\Big{[} \mathcal{C}_{3}^{4}}{\Big{]} }^{\!n d'_1 d'_2 d'_3}_{d_1 d_2}
\ket{K, \beta_1 ; d'_1 ,\beta_2 ; d'_2 ,\beta_3 ; d'_3 ,\beta_4 }, \nn \\
\eearr
where we have defined
\bearr
\prescript{\beta_1 \! \beta_2 \! \beta_3 \! \beta_4 \!\!}{\sigma_1 \! \sigma_2 \! \sigma_3 \!\!}
{\Big{[} \mathcal{C}_{3}^{4}}{\Big{]} }^{\!n d'_1 d'_2 d'_3}_{d_1 d_2} :=  
\bra{r', \beta_1 ; r'\!+\!d'_1 ,\beta_2 ; r'\!+\!d'_{12} ,\beta_3 ; r'\!+\!d'_{123} ,\beta_4}
H_{4:3} \ket{r, \sigma_1 ; r\!+\!d_1, \sigma_2 ; r\!+\!d_{12}, \sigma_3}, 
\eearr
with $r' := r\!-\!n$.

\subsection{Spin Eigen States}

The electron-hole transformation~\reqn{eq:ph_trans} on the odd sites of the IHM affects
the spin eigen states and makes them deviate from their standard form. 
For instance, the two-QP state $\ket{K;d}$ with total spin zero reads
\be
\ket{K;d}^{S=0,M=0} =
\begin{cases}
\frac{1}{\sqrt{2}} \big{(} \ket{K,+;d,+} + \ket{K,-;d,-} \big{)} & {\rm for} ~ d \in {\rm odd} \\
\frac{1}{\sqrt{2}} \big{(} \ket{K,+;d,-} - \ket{K,-;d,+} \big{)} & {\rm for} ~ d \in {\rm even} \\
\end{cases}
\label{app:eq:2qp_S0}
\ee
where $d$ is the distance between the two QPs. For constructing the Hamiltonian matrix in the ED, 
it is more suitable to apply the transformation $g^\dagger_{i,\sigma} \rightarrow g^\dagger_{i,\bar{\sigma}}$ to 
the odd sites of the effective Hamiltonian derived from the deepCUT calculations. The aim 
of this transformation is to make the spin eigen states independent of relative distances up to
a sign factor.

The one-, two-, three-, and four-QP states with total spins $S=0,\frac{1}{2},1$ and corresponding
total magnetic numbers 
$M=0,\frac{1}{2}, 1$ are presented in Table~\ref{app:tab:SS}. These relations are used for the 
construction of the Hamiltonian matrix in the ED treatment.

\begin{table}
\begin{tabular}[b]{ | l | c | }
\hline 
& \\
$\ket{K}^{S=1/2, M=1/2}$ & $\ket{K+}$ \\[0.5cm] 
$\ket{K;d_1}^{S=1, M=1}$ & $ \ket{K+;d+}$ \\[0.5cm] 
$\ket{K;d_1}^{S=0, M=0}$ & $ \frac{1}{\sqrt{2}}
\Big{(} 
\ket{K+;d_1-}-(-1)^{d_1} \ket{K-;d_1+} 
\Big{)}$ \\[0.7cm] 
$\ket{K;d_1;d_2}^{S=1/2, M=1/2}_a$ & 
$ \frac{1}{\sqrt{2}} 
\Big{(} 
\ket{K+;d_1-;d_2+}-(-1)^{d_1} \ket{K-;d_1+;d_2+} 
\Big{)}$ \\[0.8cm] 
$\ket{K;d_1;d_2}^{S=1/2, M=1/2}_b$ & 
$
\begin{array}{lcl}
\frac{1}{\sqrt{6}} 
\Big{(} 
\ket{K-;d_1+;d_2+}&+&(-1)^{d_1} \ket{K+;d_1-;d_2+}  \\ 
&-&2(-1)^{d_1+d_2} \ket{K+;d_1+;d_2-}
\Big{)}
\end{array}
$
\\[1cm] 
$\ket{K;d_1;d_2;d_3}^{S=1, M=1}_a$ & 
$
\frac{1}{\sqrt{2}} 
\Big{(} 
\ket{K+;d_1+;d_2+;d_3-} -
(-1)^{d_3}\ket{K+;d_1+;d_2-;d_3+}
\Big{)}
$
\\[0.9cm] 
$\ket{K;d_1;d_2;d_3}^{S=1, M=1}_b$ & 
$
\frac{1}{\sqrt{2}} 
\Big{(} 
\ket{K+;d_1-;d_2+;d_3+} -
(-1)^{d_1}\ket{K-;d_1+;d_2+;d_3+}
\Big{)}
$
\\[1cm] 
$\ket{K;d_1;d_2;d_3}^{S=1, M=1}_c$ & 
$
\begin{array}{lcl}
\frac{1}{2} 
\Big{(} 
\ket{K+;d_1+;d_2+;d_3-} &+&
(-1)^{d_3}\ket{K+;d_1+;d_2-;d_3+} \\
&-&
(-1)^{d_{2}}\ket{K-;d_1+;d_2+;d_3+} \\
&-&
(-1)^{d_{1}+d_2}\ket{K+;d_1-;d_2+;d_3+}
\Big{)}
\end{array}
$
\\[1.2cm] 
$\ket{K;d_1;d_2;d_3}^{S=0, M=0}_a$ & 
$
\begin{array}{lcl}
\frac{1}{2} 
\Big{(} 
\ket{K+;d_1-;d_2+;d_3-} &+&
(-1)^{d_1+d_3}\ket{K-;d_1+;d_2-;d_3+} \\
&-&
(-1)^{d_{3}}\ket{K+;d_1-;d_2-;d_3+} \\
&-&
(-1)^{d_{1}}\ket{K-;d_1+;d_2+;d_3-}
\Big{)}
\end{array}
$
\\[1.2cm] 
$\ket{K;d_1;d_2;d_3}^{S=0, M=0}_b$ & 
$
\begin{array}{lcl}
\frac{1}{2\sqrt{3}} 
\Big{(} 
\ket{K-;d_1+;d_2+;d_3-} &-&
2(-1)^{d_2}\ket{K+;d_1+;d_2-;d_3-} \\
&-&
2(-1)^{d_2}\ket{K-;d_1-;d_2+;d_3+} \\
&+&
(-1)^{d_{1}}\ket{K+;d_1-;d_2+;d_3-}\\
&+&
(-1)^{d_{3}}\ket{K-;d_1+;d_2-;d_3+}  \\
&+&
(-1)^{d_{1}+d_3}\ket{K+;d_1-;d_2-;d_3+}
\Big{)}
\end{array}
$
\\ & \\
\hline
\end{tabular}
\caption{One-, two-, three-, and four-QP states with total spin $S$ and total magnetic number 
$M$ employed in the ED analysis for constructing the Hamiltonian matrix in different sectors of total $S$ and 
$M$.}
\label{app:tab:SS}
\end{table}

\end{document}